\documentclass[a4paper,12pt]{article}

\usepackage{amsmath}
\usepackage[backend=biber,style=numeric-comp,bibstyle=numeric,sorting=none,url=false,natbib=true,maxbibnames=99,firstinits=true]{biblatex}

\usepackage{setspace}

\usepackage{hyperref}
\usepackage[a4paper,left=2.5cm,right=2.5cm,top=3cm,bottom=2.5cm]{geometry}
\usepackage{authblk}
\usepackage{amsmath,bm}
\usepackage{graphicx}
\usepackage{mathrsfs}
\usepackage{amssymb}
\usepackage{physics}
\usepackage{tensor}
\usepackage{xcolor}
\usepackage{tikz}
\usepackage{pgfplots}
\usetikzlibrary{spy}
\usetikzlibrary{external}

\pgfplotsset{grid style={line width=.1pt,dashed,lightgray},compat=newest}
\usepackage{caption}
\usepackage{subcaption}

\DeclareCaptionLabelFormat{subfigure}{#2)}
\usepackage{amsthm}
\usepackage{booktabs,dcolumn,caption}
\usepackage{multirow}
\usepackage[english]{babel}
\usepackage{soul}
\usepackage{mathtools}

\usepackage[title]{appendix}

\usepackage{xstring}

\newtheoremstyle{remark}
{}
{}
{\textnormal}
{}
{\bfseries}
{.}
{ }
{\thmname{#1}\thmnumber{ #2}}%

\theoremstyle{plain}

\title{Efficient snap-to-contact computations for van der Waals interacting fibers}

\author[1,2]{A.~Borković}
\author[1]{M.~H.~Gfrerer}
\author[3,4,5]{R.~A.~Sauer}
\author[1]{B. Marussig}

\affil[1]{Institute of Applied Mechanics, Graz University of Technology, Technikerstraße 4/II, 8010 Graz, Austria, aleksandar.borkovic@aggf.unibl.org, aborkovic@tugraz.at}
\affil[2]{University of Banja Luka, Faculty of Architecture, Civil Engineering and Geodesy, Department of Mechanics and Theory of Structures, 78000 Banja Luka, Bosnia and Herzegovina}
\affil[3]{Institute for Structural Mechanics, Ruhr University Bochum, Universitätsstraße 150, 44801 Bochum }
\affil[4]{Department of Structural Mechanics, Gdansk University of Technology, ul. Narutowicza 11/12, 80-233 Gdansk, Poland} 
\affil[5]{Dept. of Mechanical Engineering, Indian Institute of Technology Guwahati, Assam 781039, India }
\date{}                     
\begin{document}
	
\newcommand{\ssub}[2]{{#1}_{#2}} 
\newcommand{\vsub}[2]{\textbf{#1}_{#2}} 
\newcommand{\ssup}[2]{{#1}^{#2}} 
\newcommand{\vsup}[2]{\textbf{#1}^{#2}} 
\newcommand{\ssupsub}[3]{{#1}^{#2}_{#3}} 
\newcommand{\vsupsub}[3]{\textbf{#1}^{#2}_{#3}} 

\newcommand{\veq}[1]{\bar{\boldsymbol{#1}}} 
\newcommand{\veqn}[1]{\bar{\textbf{#1}}} 
\newcommand{\seq}[1]{\bar{#1}} 
\newcommand{\ve}[1]{\boldsymbol{#1}} 
\newcommand{\ven}[1]{\textbf{#1}} 
\newcommand{\vepre}[1]{\boldsymbol{#1}^\sharp} %
\newcommand{\sdef}[1]{#1^*} 
\newcommand{\vdef}[1]{{\boldsymbol{#1}}^*} 
\newcommand{\vdefeq}[1]{{\bar{\boldsymbol{#1}}}^*} 
\newcommand{\trans}[1]{\boldsymbol{#1}^\mathsf{T}} 
\newcommand{\transn}[1]{\textbf{#1}^\mathsf{T}} 
\newcommand{\transmd}[1]{\dot{\boldsymbol{#1}}^\mathsf{T}} 
\newcommand{\mdvdef}[1]{\dot{\boldsymbol{#1}}^*} 
\newcommand{\mdsdef}[1]{\dot{#1}^*} 
\newcommand{\mdv}[1]{\dot{\bm{#1}}} 
\newcommand{\mdvni}[1]{\dot{\textbf{#1}}} 
\newcommand{\mds}[1]{\dot{#1}} 

\newcommand{\loc}[1]{\hat{#1}} 
\newcommand{\iloc}[3]{\hat{#1}^{#2}_{#3}} 
\newcommand{\ilocmd}[3]{\dot{\hat{#1}}^{#2}_{#3}} 
\newcommand{\md}[1]{\dot{#1}} 

\newcommand{\ii}[3]{{#1}^{#2}_{#3}} 
\newcommand{\iv}[3]{\boldsymbol{#1}^{#2}_{#3}} 
\newcommand{\idef}[3]{{#1}^{* #2}_{#3}} 
\newcommand{\ivdef}[3]{\boldsymbol{#1}^{* #2}_{#3}} 
\newcommand{\ipre}[3]{{#1}^{\sharp #2}_{#3}} 
\newcommand{\ivpre}[3]{\textbf{#1}^{\sharp #2}_{#3}} 

\newcommand{\ieq}[3]{\bar{#1}^{#2}_{#3}} 
\newcommand{\ic}[3]{\tilde{#1}^{#2}_{#3}} 
\newcommand{\icdef}[3]{\tilde{#1}^{* #2}_{#3}} 
\newcommand{\icpre}[3]{\tilde{#1}^{\sharp #2}_{#3}} 
\newcommand{\iveq}[3]{\bar{\textbf{#1}}^{#2}_{#3}} 
\newcommand{\ieqdef}[3]{\bar{#1}^{* #2}_{#3}} 
\newcommand{\iveqdef}[3]{\bar{\textbf{#1}}^{* #2}_{#3}} 
\newcommand{\ieqmddef}[3]{\dot{\bar{#1}}^{* #2}_{#3}} 
\newcommand{\icmddef}[3]{\dot{\tilde{#1}}^{* #2}_{#3}} 
\newcommand{\iveqmddef}[3]{\dot{\bar{\textbf{#1}}}^{* #2}_{#3}} 
\newcommand{\iveqmdddef}[3]{\ddot{\bar{\textbf{#1}}}^{* #2}_{#3}} 

\newcommand{\ieqpre}[3]{\bar{#1}^{\sharp #2}_{#3}} 
\newcommand{\iveqpre}[3]{\bar{\textbf{#1}}^{\sharp #2}_{#3}} 
\newcommand{\ieqmdpre}[3]{\dot{\bar{#1}}^{\sharp #2}_{#3}} 
\newcommand{\icmdpre}[3]{\dot{\tilde{#1}}^{\sharp #2}_{#3}} 
\newcommand{\iveqmdpre}[3]{\dot{\bar{\textbf{#1}}}^{\sharp #2}_{#3}} 

\newcommand{\ieqmd}[3]{\dot{\bar{#1}}^{#2}_{#3}} 
\newcommand{\icmd}[3]{\dot{\tilde{#1}}^{#2}_{#3}} 
\newcommand{\iveqmd}[3]{\dot{\bar{\boldsymbol{#1}}}^{#2}_{#3}} 
\newcommand{\iveqmdd}[3]{\ddot{\bar{\boldsymbol{#1}}}^{#2}_{#3}} 

\newcommand{\imddef}[3]{\dot{#1}^{* #2}_{#3}} 
\newcommand{\ivmddef}[3]{\dot{\textbf{#1}}^{* #2}_{#3}} 
\newcommand{\ivmdddef}[3]{\ddot{\textbf{#1}}^{* #2}_{#3}} 

\newcommand{\imdpre}[3]{\dot{#1}^{\sharp #2}_{#3}} 
\newcommand{\ivmdpre}[3]{\dot{\textbf{#1}}^{\sharp #2}_{#3}} 

\newcommand{\imd}[3]{\dot{#1}^{#2}_{#3}} 
\newcommand{\imdd}[3]{\ddot{#1}^{#2}_{#3}} 

\newcommand{\ivmd}[3]{\dot{\textbf{#1}}^{#2}_{#3}} 

\newcommand{\iii}[5]{^{#2}_{#3}{#1}^{#4}_{#5}} 
\newcommand{\iiv}[5]{^{#2}_{#3}{\boldsymbol{#1}}^{#4}_{#5}} 
\newcommand{\iivn}[5]{^{#2}_{#3}{\tilde{\boldsymbol{#1}}}^{#4}_{#5}} 
\newcommand{\iiieq}[5]{^{#2}_{#3}{\bar{#1}}^{#4}_{#5}} 
\newcommand{\iiieqt}[5]{^{#2}_{#3}{\tilde{#1}}^{#4}_{#5}} 

\newcommand{\eqqref}[1]{Eq.~\eqref{#1}} 
\newcommand{\fref}[1]{Fig.~\ref{#1}} 
	
\maketitle
	
\section*{Abstract}

We consider van der Waals interactions between in-plane fibers, where the computational model employs the Lennard-Jones potential and the coarse-grained approach. The involved 6D integral over two interacting fibers is split into a 4D analytical pre-integration over cross sections and the remaining 2D numerical integration along the fibers' axes. Two section-section interaction laws are implemented, refined, and compared. Fibers are modeled using the Bernoulli-Euler beam theory and spatially discretized with isogeometric finite elements. We derive and solve the weak form of both quasi-static and dynamic boundary value problems. Four numerical examples involving highly nonlinear and dynamic snap-to-contact phenomena are scrutinized. We observe that the coarse-graining and pre-integration of interaction potentials enable the efficient modeling of complex phenomena at small length scales.

\textbf{Keywords}: interaction potential; van der Waals attraction; coarse-grained approach; contact mechanics; snap-to-contact; fiber-fiber interaction

\section{Introduction}


The interactions between molecular assemblies that resemble shapes of fibers are the underlying cause of many macroscopic phenomena \cite{2011israelachvili}. These interactions are important in industry for the development of glass fibers \cite{2008alavinasab}, silicon nanotubes \cite{2012yoo}, carbon nanotubes \cite{2021čanadija}, and nano-electromechanical systems (NEMS) such as nano-beam actuators \cite{2024khan}. Furthermore, interactions at small length scales govern the behavior of many biological fiber-like macromolecules, e.g.~proteins, such as filamentous actin \cite{2012murrell} and collagen \cite{2021slepukhin}, nucleic acids, such as DNA and RNA \cite{2018franquelim}, cellulose \cite{2018nishiyama}, and hyphae \cite{2017islama}. Modeling of interactions between these fibers is challenging due to the interplay of many forces and the involved time and length scales at nano- and micro-levels. Among the many general types of interactions between fibers, an important example is adhesion that is preceded by a snap-to-contact phenomenon. The understanding of this is central to many important applications in coating, bonding, and adhesion technology. One of the main challenges in these applications is the abrupt change of stress that occurs at the instance of snap-to-contact.

Van der Waals (vdW) attraction is the most significant intermolecular interaction, playing a central role in nearly all phenomena involving intermolecular forces. This interaction occurs even between neutral molecules due to fluctuating charge distributions and is relevant across both small and large separations. Accurately modeling vdW interactions is challenging, as they result from multiple contributing factors, depend on retardation effects, and exhibit non-additivity \cite{2005parsegian}. A common approximation for calculating vdW interactions between bodies is the pairwise summation approach, which assumes that the total interaction is a mere sum of individual point-pair interactions. The accuracy of this approach significantly depends on the type and shape of molecules \cite{2017venkataram, 2017ambrosetti}. In this paper, we focus on the simplified case of non-retarded vdW interactions, where the pairwise summation approach is valid and can be effectively modeled using an inverse-power law with an exponent of 6.

The numerical modeling of intermolecular interactions is often based on molecular dynamics or Monte Carlo simulations \cite{2011israelachvili}. Another approach, based on homogenization and coarse-graining of the molecular model, is the \emph{coarse-grained} model. This approach provides a good balance between accuracy and efficiency \cite{1997argento, 2007sauer, 2008sauer, 2016fan} by utilizing the physics of molecular interactions with the efficacy of continuum contact formulations \cite{2006wriggers}. The interactions are separated into those that occur within the body (intrasolid) and those between bodies (intersolid), which allows us to represent the interaction potential between two bodies as a function of the gap vector.

Solving the boundary value problem of potential-based interactions between bodies at involved time and length scales is computationally demanding due to high gradients of configuration-dependent interaction forces at small separations. For large separations, fibers can be treated as lines and the implementation simplifies \cite{2024wang}. The coarse-grained model can be implemented into structural beam and shell theories \cite{2014sauerb, 2020grill, 2024mokhalingam}, resulting in a good balance of computational efficiency and accuracy. A section-section approach is introduced in \cite{2020grill,2023meier} and utilized for short-range interactions between deformable planar beams in \cite{2021grilla,2024borkovićb,2024borkovićg,2025borković}. Finding an appropriate section-section law for short-range interactions is prohibitive for spatial beams, and other approaches are considered, such as a section-beam model \cite{2023grill,2024grill}.

Bodies interacting via intermolecular forces can suddenly \emph{snap-to-contact} when they are moved closer towards to each other, or \emph{snap-off-contact} when they are pulled apart. The snap-to-contact of fibers interacting via electrostatic force is modeled by dynamic analysis in \cite{2020grill,2021grilla}. Although these snapping phenomena are inherently dynamic, a quasi-static model can be suitable for cases when inertial effect are not significant or when the steady-state response corresponds to the static equilibrium. A well-known approach for the quasi-static analysis of systems that exhibit instabilities is the arc-length continuation method. Initially developed to track equilibrium paths with load limit points, such as shallow shells and steel diaphragms \cite{1981crisfield}, it has been successfully applied for modeling delamination of composite structures \cite{2002crisfield} and solving vdW interactions between deformable bodies \cite{2000feng,2006sauer,2006wu,2022roy,2023roy}.

Motivated by the lack of thorough computational modeling of intermolecular interactions between slender deformable bodies in a complex snap-to-contact setting, this research investigates the behavior of fibers that abruptly adhere to each other. The modeling of intermolecular interactions between fibers is inherently complex and therefore simplified by the following assumptions:
\begin{itemize}
	\item The total body-body interaction equals the pairwise summation (integration) of point-pair interactions.
	\item The point-point interaction potential is modeled as an inverse-power law of the point-pair distance.
	\item Only two-body interaction is considered and many-body effects are neglected.
	\item Any influence of a surrounding medium is neglected.
	\item There is no redistribution of particles or charges inside the bodies; that is, we are dealing with dielectric or nonconducting materials. 
	\item The density distributions of particles and physical constants over the interacting bodies are homogeneous at initial configuration.
\end{itemize}

The main contributions of the present paper are twofold: (i) implementation of two section-section laws within the computational formulation that enables accurate and efficient modeling of highly nonlinear quasi-static and dynamic interactions between in-plane fibers, and (ii) thorough numerical analysis of non-trivial examples involving snap-to-contact behavior due to the interaction potentials. In particular, we present four novel numerical experiments of fiber-fiber interactions: (i) modeling snap-to-contact between two cantilever fibers using both static and dynamic analysis, (ii) finding the equilibrium configuration between two free deformable fibers, (iii) modeling snap-to-contact between deformable fibers involving strong collision, (iv) simulating the deformation of a fiber due to the bending of an adhered fiber.

The remaining paper is organized as follows: The problem of potential-based interactions between fibers is discussed in the next section. Our computational model is scrutinized in Section 3, where we focus on a beam model, section-section laws, solvers, and integration approach. The four proposed numerical experiments are presented in Section 4, which is followed by conclusions in Section 5.

\section{Statement of the problem}


The concept of point-pair interaction potentials and their integration over the two interacting fibers using the coarse-grained method are revised in this section. The basic idea of the section-section approach is introduced, and the general form of the equation of motion is presented.
	
Let us consider an interaction between particles $i$ and $j$. They interact via some potential field modeled as an inverse power law w.r.t.~their distance $r_{ij}$. At infinite separation, the interaction is zero. A point-point interaction potential of $m^\text{th}$ order, ${\Pi}^m_{\operatorname{P-P}}$, is the energy required to move these particles from distance $r_{ij}$ to infinite separation, i.e.,
	\begin{equation}
		\label{eq: ip01}
		\begin{aligned}
			{\Pi}_{\operatorname{P-P}}^m = k_m \, r_{ij}^{-m},
		\end{aligned}
	\end{equation}
where $k_m$ is a physical constant. The interaction force is obtained as the gradient of this potential w.r.t.~the distance; it acts on both particles with the same intensity but in the opposite direction. By the pairwise summation concept \cite{2011israelachvili}, a volume interaction potential between bodies $X$ and $Y$ is
\begin{equation}
	\label{eq: ip01sum}
	\begin{aligned}
		\Pi_{\operatorname{{B-B}_{PW}}}^{m} &=  \sum_{i \in X}^{} \sum_{j \in Y}^{} \Pi_{\operatorname{{P-P}}}^m (r_{ij}).
	\end{aligned}
\end{equation}
The coarse-graining procedure, introduced in \cite{2007sauer}, approximates the pairwise summation as a volume integral over both bodies,
\begin{equation}
	\label{eq: ip01x}
	\begin{aligned}
		\Pi_{\operatorname{{B-B}_{PW}}}^{m} \approx \Pi_{\operatorname{B-B}}^{m} =  \int_{\idef{V}{}{x}} \int_{\idef{V}{}{y}} \beta^*_x  \beta^*_y {\Pi}_{\operatorname{P-P}}^m \dd{V^*_y} \dd{V^*_x},
	\end{aligned}
\end{equation}
where $V_k^*$ are volumes, while $\beta_k^*$ are particle volume densities; both at the current configuration. Since we assume that the interacting property of an elementary volume is conserved during deformation, we have $\beta_{k} \dd{V}_k = \beta^*_{k} \dd{V^*_k}$.
This fact allows us to calculate an interaction potential at the current configuration by integrating over the reference volume $V_k$, using the reference particle densities $\beta_{k}$, i.e.,
\begin{equation}
	\label{eq: ip01xhh}
	\begin{aligned}
		\Pi_{\operatorname{B-B}}^{m} = \int_{\idef{V}{}{x}} \int_{\idef{V}{}{y}} \beta^*_x \,  \beta^*_y \, k_m \, r^{-m}  \dd{V^*_y} \dd{V^*_x} = \int_{V_x} \int_{V_y} \beta_x  \, \beta_y \, k_m \, r^{-m} \dd{V_y} \dd{V_x}.
	\end{aligned}
\end{equation}
%
It turns out that the straightforward calculation of this integral for practical time and space resolutions is highly demanding and often impossible. Therefore, we aim to employ a more efficient computational model.

For the interactions between fibers, the concept of the section-section potential is introduced in \cite{2020grill}. The idea is to pre-integrate \eqref{eq: ip01xhh} over cross sections to obtain the section-section potential, $\Pi_{\operatorname{S-S}}^m$, and then to numerically integrate $\Pi_{\operatorname{S-S}}^m$ along the beams' axes, i.e.,
\begin{equation}
	\label{eq: ip01xhah}
	\begin{aligned}
		\Pi_{\operatorname{B-B}}^{m} &\approx\int_{L_x} \int_{L_y} \int_{A_x} \int_{A_y} \beta_x  \, \beta_y \, k_m \, r^{-m} \dd{A_y} \dd{A_x}\dd{s_y} \dd{s_x}=\int_{L_x} \int_{L_y} \beta_x  \, \beta_y k_m  \Pi_{\operatorname{S-S}}^m\dd{s_y} \dd{s_x}, \\
		\Pi_{\operatorname{S-S}}^m&:=\int_{A_x} \int_{A_y}  \, \, r^{-m} \dd{A_y} \dd{A_x}.
	\end{aligned}
\end{equation}
Although an exact analytical integration of interaction potentials over cross sections has not been obtained yet, two laws that provide a good balance between accuracy and efficiency for in-plane fibers are proposed in \cite{2024borkovićb,2025borković} and revised in Subsection \ref{sec:sec-sec}. 

Let us emphasize that the closest point-pairs, which are usually only a few, dominate the interaction between bodies in close proximity when $m>3$. This short-range effect leads to the complex competition of repulsive and attractive forces at the contact interface. The computational modeling of this behavior is both crucial and demanding, which has motivated the development of short-range section-section models \cite{2020grill, 2024borkovićb}. On the other hand, modeling long- and moderate-range effects is computationally less involved but can be equally important since these forces can bring the objects together. For this reason, we developed a section-section law \cite{2025borković} that exhibits good results for both short- and long-ranges.

For the modeling of adhesion due to vdW forces, it is necessary to consider repulsive effects as well. The repulsion develops between bodies in close vicinity due to overlapping electron clouds. This repulsion effect is observed as \emph{contact} from a macroscopic point of view. vdW and steric interactions exist for practically all bodies, making them one of the most common forces in nature. By modeling the repulsive steric potential with an inverse-power law with $m=12$, and adding it to the vdW potential, we obtain the well-known Lennard-Jones (LJ) potential between two particles
	\begin{equation}
		\begin{aligned}
			\Pi_{\operatorname{P-P}}^{\operatorname{LJ}}=4\epsilon \left[\left(\frac{\sigma}{r}\right)^{12}-\left(\frac{\sigma}{r}\right)^6\right] = k_6 r^{-6} + k_{12} r^{-12},
		\end{aligned}
		\label{eqLJex}
	\end{equation}
where $\sigma$ is the distance at which the potential is zero, while $\epsilon$ is the minimum value of the potential. 
In our numerical analysis, we exclusively consider LJ fiber-fiber interactions that stem from the integration of the point-point LJ potential \eqref{eqLJex}.

	
The computational modeling of potential-based interactions between bodies requires solving an appropriate (initial) boundary value problem. The strong form of this problem consists of the equation of motion (balance of linear momentum), boundary and initial conditions, and a constitutive relation. To derive the weak form, we define the total potential energy of a system involving elastic bodies, $\Pi_{\mathrm{tot}}$. It consists of the strain energy, $\Pi_{\mathrm{int}}$, the kinetic energy, $\Pi_{\mathrm{kin}}$, the work of external forces, $\Pi_{\mathrm{ext}}$, and an interaction potential. By the principle of stationary total potential energy, the weak form of the boundary value problem then follows from
\begin{equation}
	\label{eq: poten}
	\begin{aligned}
		\delta \Pi_{\mathrm{tot}} = \delta \Pi_{\mathrm{int}} +\delta \Pi_{\mathrm{kin}}+\delta \Pi_{\mathrm{ext}}+\delta \Pi_{\operatorname{B-B}}^m=0.
	\end{aligned}
\end{equation}
In Section \ref{sec:met}, a methodology for the derivation and approximate solution of this equation is briefly discussed.

	\section{Computational model}
	\label{sec:met}
In this section, we present the beam model that allows an efficient integration of the total potential energy of interacting fibers. Details on the section-section laws and the beam-beam approach are provided. For completeness, three algorithms for solving the nonlinear equations are briefly presented and the numerical integration of the section-section interaction along the beam's axis is discussed.
	
	\subsection{Bernoulli-Euler beam model}
	
	\label{besec}
	
A beam is an infinite set of plane figures (cross sections) that are attached at their centroids to a finite-length smooth curve (beam axis). According to the Bernoulli-Euler (BE) hypothesis, we assume that cross sections are rigid and perpendicular to the beam axis in all configurations. This approach is well-suited for the computational modeling of arbitrarily curved slender bodies, such as fibers. Although we consider an interaction between two beams, $X$ and $Y$, it is sufficient to focus on one beam in this subsection.  
	
\begin{figure}[h]
		\includegraphics[width=\linewidth]{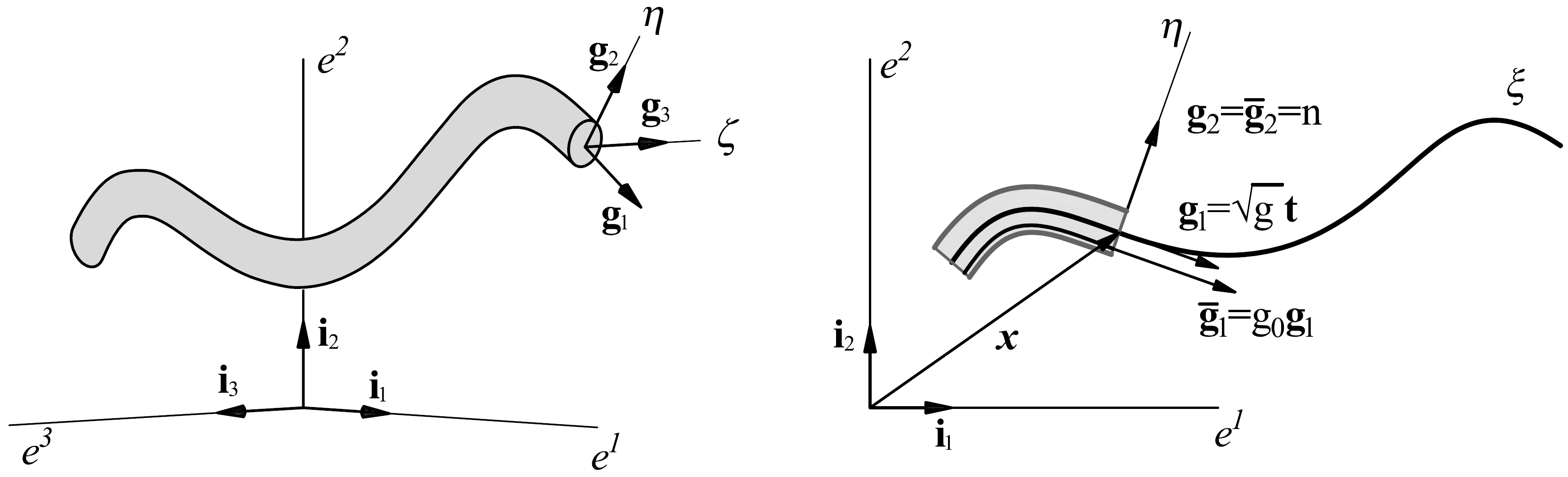}
		\caption{Degeneration of a planar 3D beam into an in-plane 1D beam model. The base vectors at the centroid and at an equidistant line are depicted.}
		\label{fig:Figure 1}
	\end{figure}

Due to the BE assumptions, the continuum of an in-plane beam can be described solely by the position of the beam axis, see \fref{fig:Figure 1}. 
We parameterize the axis with both the arc-length coordinate $s$ and a parametric coordinate $\xi$. 
	%
	%
	%
	The position vector of the beam axis is $\ve{x}$, while the tangent base vectors of a beam axis, w.r.t.~parametric and arc-length coordinates are
	\begin{equation}
		\begin{aligned}
			\iv{g}{}{1}&=\frac{\partial \ve{x}}{\partial \xi} = \iv{x}{}{,1}, \\
			\ve{t} &=\frac{\partial \ve{x}}{\partial s} = \iv{x}{}{,s}=\iv{g}{}{1}/\sqrt{g}, \quad \sqrt{g}=\norm{\iv{g}{}{1}} = \sqrt{\iv{g}{}{1}\cdot\iv{g}{}{1}}.	
		\end{aligned}	
	\end{equation}
The other base vector, $\iv{g}{}{2}=\ve{n}$, has unit length and is defined here through an anti-clockwise rotation of the vector $\iv{t}{}{}$ w.r.t.~cross section direction $\zeta$, i.e.
\begin{equation}
	\iv{g}{}{2}= \iv{n}{}{} = \pmb \Lambda \iv{t}{}{}, \quad \pmb 	\Lambda =
	\begin{bmatrix}
		0 & -1\\
		1 & 0
	\end{bmatrix}.
\end{equation}
 The derivatives of the basis vectors are defined via Christoffel symbols, 
	\begin{equation}
		\label{eq: def: derivatives of base vectors}
		\begin{bmatrix}
			\iv{g}{}{1,1}\\
			\iv{g}{}{2,1}
		\end{bmatrix}
		=
		\begin{bmatrix}
			\ii{\Gamma}{1}{11} & \ii{\Gamma}{2}{11}\\
			\ii{\Gamma}{1}{21} & \ii{\Gamma}{2}{21}
		\end{bmatrix}
		\begin{bmatrix}
			\iv{g}{}{1}\\
			\iv{g}{}{2}
		\end{bmatrix}
		=
		\begin{bmatrix}
			\ii{\Gamma}{1}{11} & \ic{K}{}{}\\
			-K & 0
		\end{bmatrix}
		\begin{bmatrix}
			\iv{g}{}{1}\\
			\iv{g}{}{2}
		\end{bmatrix},
	\end{equation}
where $K$ and $\ic{K}{}{} = g K$ are the so-called \textit{signed curvatures} w.r.t.~arc-length and parametric coordinates, respectively.

The position and base vectors at an arbitrary point of the beam continuum are
\begin{align}
	\label{eq:def:r_eq}
	\begin{aligned}
		\veq{x} &= \ve{x} + \eta \, \ve{g}_{2}, \\
		\veq{g}_{1} &= \veq{x}_{,1} = \ve{g}_{1} - \eta \, K \,  \iv{g}{}{1}, \\
		\veq{g}_{2} &=  \ve{g}_{2} = \ve{n},
	\end{aligned}
\end{align}
and we designate them with an overbar. In this way, the metric of a beam in the reference configuration is defined by the metric of the beam axis. 
Since the current configuration is obtained by adding the displacement field to the initial position, $\vdef{x}=\ve{x}+\ve{u}$, the relations given in \eqqref{eq:def:r_eq} are valid for every configuration \cite{2023borković}. 

Due to the BE assumptions, the only remaining component of the Green-Lagrange strain tensor is the axial strain
	\begin{equation}
		\label{eq:def:strain}
		\ieq{\epsilon}{}{11} = \frac{1}{2} \left( \ieqdef{g}{}{11} - \ieq{g}{}{11} \right),\quad \ieq{g}{}{11}= \veq{g}_{1} \cdot \veq{g}_{1}. 
	\end{equation}
By neglecting the influence of higher-order terms w.r.t.~the $\eta$ coordinate and the initial curvature, the axial strain at an arbitrary point is 
	\begin{equation}
		\label{eq:e11eq}
		\begin{aligned}
			\ieq{\epsilon}{}{11}&=  \ii{\epsilon}{}{11}+ \eta \kappa, \\
			\ii{\epsilon}{}{11} &:= \frac{1}{2} \left( \idef{g}{}{11} - \ii{g}{}{11} \right), \quad \ii{\kappa}{}{} :=\icdef{K}{}{} - \ic{K}{}{}. 
		\end{aligned}
	\end{equation}
where $\ii{\epsilon}{}{11}$ is the axial strain of the beam axis, while $\ii{\kappa}{}{}$ is the change of curvature of the beam axis w.r.t.~the parametric convective coordinate \cite{2023borković}.
	%
	

We employ the hyperelastic St.~Venant--Kirchhoff material model which allows us to write the second Piola-Kirchhoff stress component as
	\begin{equation}
		\label{eq:const}
		\begin{aligned}
			\ieq{S}{11}{} = E \ieq{g}{11}{} \ieq{g}{11}{} \ieq{\epsilon}{}{11}\,,
		\end{aligned}
	\end{equation}
where $E$ is the Young's modulus of elasticity. Now, the strain energy of a beam is
	\begin{equation}
		\label{eq:internalener}
		\begin{aligned}
			\Pi_{\mathrm{int}} = \frac{1}{2} \int_{V}^{} \ieq{S}{11}{}  \ieq{\epsilon}{}{11} \dd{\ieq{V}{}{}}.
		\end{aligned}
	\end{equation}
	%
By introducing the velocity vector, $\mdv{x}$, the kinetic energy of a BE beam is
\begin{equation}
	\label{eq:kinetic}
	\begin{aligned}
		\Pi_{\mathrm{kin}} = \frac{1}{2} \int_{V}^{} \rho \iveqmd{x}{}{} \cdot \iveqmd{x}{}{} \dd{\ieq{V}{}{}},
	\end{aligned}
\end{equation}
where $\rho$ is the mass density. Moreover, the work of external forces can be defined as
\begin{equation}
	\label{eq:external}
	\begin{aligned}
		\Pi_{\mathrm{ext}} = -\int_{V}^{} \veq{p} \cdot \veq{u} \dd{\ieq{V}{}{}},
	\end{aligned}
\end{equation}
where $\veq{p}$ is a body force at an equidistant point of the beam. 
	%

For further details on the variation of the potential and the IGA-based spatial discretization of unknown fields, we refer to \cite{2019borkovica, 2023borković}.

\subsection{Section-section interaction potential}
\label{sec:sec-sec}

The interaction potential between two in-plane disks, $\Pi_{\operatorname{D-D_{IP}}}$, depends only on the smallest distance between them, referred to as the normal gap $q_2$. When the disks belong to two parallel planes with an offset $q_1$, the disk-disk potential, $\Pi_{\operatorname{D-D}}$, becomes a function of both the gap and the offset. Our \emph{improved section-section interaction potential} (ISSIP) \cite{2024borkovićb} addresses this additional dimension by improving the integration approach for in-plane disks \cite{1972langbein}. The ISSIP is represented as a product of an approximate disk-disk in-plane potential, $\hat{\Pi}_{\operatorname{D-D_{IP}}}^m$, derived in \cite{1972langbein}, and a hypergeometric function, i.e.,
\begin{equation}
	\label{appDD1}
	\begin{aligned}
		\Pi_{\mathrm{ISSIP}}^m &= \hat{\Pi}_{\operatorname{D-D_{IP}}}^m \, _2F_1 \left(\frac{2m-7}{4},\frac{2m-5}{4};\frac{m}{2};-\frac{q_1^2}{q_2^2}\right), \\
		\hat{\Pi}_{\operatorname{D-D_{IP}}}^m &= 2^{\frac{5}{2}-m} \pi^\frac{3}{2} \sqrt{\frac{R_x R_y}{R_x + R_y}} \frac{\Gamma (m-\frac{7}{2})}{\Gamma (\frac{m}{2})^2} q_2^{\frac{7}{2}-m},
	\end{aligned}
\end{equation}
where $R_x$ and $R_y$ are the radii of interacting cross sections. Here, $_2F_1 \left(a,b;c;z\right)$ is the Gaussian hypergeometric function
\begin{equation}
	\label{eq: ip13hyp}
	\begin{aligned}
		_2F_1 \left(a,b;c;z\right)= \sum_{k=0}^{\infty} \frac{\left(a\right)_k \left(b\right)_k}{\left(c\right)_k} \frac{z^k}{k!}
	\end{aligned}
\end{equation}
and $\left(a\right)_k$ is the Pochhammer symbol
\begin{equation}
	\label{eq: ip13poch}
	\begin{aligned}
		\left(a\right)_k=\frac{\Gamma \left(a+k\right)}{\Gamma \left(k\right)},
	\end{aligned}
\end{equation}
where $\Gamma\left(z\right)= \int_{0}^{\infty} p^{z-1} e^{-w} \dd{w}$ is the gamma function. For even $m$, the law \eqref{appDD1} evaluates to elliptic integrals of the first and second kind, while for odd $m$ we obtain rational functions. 

To further ameliorate the ISSIP law, we have proposed replacing the approximate in-plane potential $\hat{\Pi}_{\operatorname{D-D_{IP}}}^m$ in \eqref{appDD1} with the exact one for even $m$, ${\Pi}_{\operatorname{D-D_{IP}}}^m$, \cite{2025borković}. This law is named \emph{the approximate disk-disk law}, $\operatorname{D-D_{app}}$, and it is defined as
\begin{equation}
	\label{appDD2}
	\begin{aligned}
		\Pi_{\operatorname{D-D_{app}}}^m = \Pi_{\operatorname{D-D_{IP}}}^m \, _2F_1 \left(\frac{2m-7}{4},\frac{2m-5}{4};\frac{m}{2};-\frac{q_1^2}{q_2^2}\right)\; \text{for} \; m=4,6,8,...
	\end{aligned}
\end{equation}
Using the exact $\operatorname{D-D_{IP}}$ makes the $\operatorname{D-D_{app}}$ law significantly more complicated than the ISSIP one. Expressions for the exact in-plane vdW and steric potentials are given in Notebook 2 of \cite{2025borkovićb}.

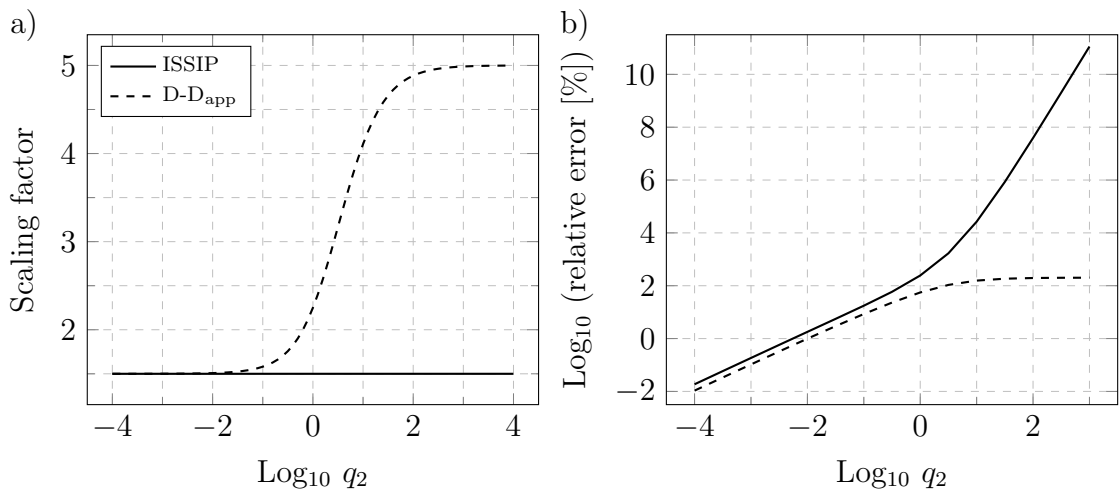
\begin{figure}[h!]
	\centering
	\begin{tikzpicture}
		\begin{axis}[
			xlabel = {Log$_{10}$ $q_2$},
			ylabel = {Scaling factor},
			ylabel near ticks,
			legend pos=north west,
			legend cell align=left,
			legend style={font=\scriptsize},
			width=0.47\textwidth,
			xmin=-4.5,xmax=4.5,
			minor y tick num = 1,
			minor x tick num = 1,
			ytick distance=1,
			xtick distance=2,
			clip=false,grid=both];
			\node [text width=1em,anchor=north west] at (rel axis cs: -0.195,1.1){a)};
			\addplot[black,thick,restrict x to domain=-4:4] table [col sep=comma] {data/dataDCissipScaling.csv};
			\addplot[black,thick,dashed,restrict x to domain=-4:4] table [col sep=comma] {data/dataDCddappScaling.csv};
			\legend{ISSIP,$\operatorname{D-D_{app}}$} 
		\end{axis}
	\end{tikzpicture}
\begin{tikzpicture}
	\begin{axis}[
		xlabel = {Log$_{10}$ $q_2$},
		ylabel = {Log$_{10}$ (relative error $[\%]$)},
		ylabel near ticks,
		legend pos=north west,
		legend cell align=left,
		legend style={font=\scriptsize,nodes={scale=0.91, transform shape}},
		width=0.47\textwidth,
		yminorticks=true,
		grid=major,
		xmin=-4.5,xmax=3.5,
		ymin=-2.5,ymax=11.5,
		minor x tick num = 1,
		ytick distance=2,
		xtick distance=2,
		clip=false,grid=both
		];
		\addplot[black,thick,restrict x to domain=-10000:6] table [col sep=comma] {data/dataDCissipError.csv};
		\addplot[black,dashed,thick] table [col sep=comma] {data/dataDCddappError.csv};
		\node [text width=1em,anchor=north west] at (rel axis cs: -0.26,1.1){b)};
	\end{axis}
\end{tikzpicture}
	\caption{Comparison of the $\operatorname{D-D_{app}}$ and ISSIP laws for disk-infinite cylinder vdW interaction ($R_x=R_y=1$). a) Scaling factor function. b) Relative error w.r.t.~a highly accurate numerical result. 
	}
	\label{fig:LJpoten3}
\end{figure}

Let us compare the accuracy of these two laws for the vdW case ($m=6$). Since we are interested in modeling interactions between beams, the interaction force acting on one section of the beam $X$ is obtained by integrating the interaction force over the length of the beam $Y$. Therefore, for small and moderate separations, it is more informative to observe the disk-infinite cylinder law, than the disk-disk law. We can find a disk-infinite cylinder law by integrating ISSIP \eqref{appDD1} and $\operatorname{D-D_{app}}$ \eqref{appDD2} along $q_1$ from $-\infty$ to $\infty$. Since only the hypergeometric function depends on $q_1$ in \eqref{appDD1} and \eqref{appDD2}, the disk-infinite cylinder potentials for even $m$ are:
\begin{equation}
	\label{appDD1s}
	\begin{aligned}
		\Pi_{\operatorname{D-C,ISSIP}} &= \int_{-\infty}^{\infty} \Pi_{\mathrm{ISSIP}}  \dd{q_1}= \hat{\Pi}_{\operatorname{D-D_{IP}}}^m \frac{4 \sqrt{\pi} q_2 \Gamma(m/2)}{(2m-9) \Gamma(\frac{m-1}{2})}, \\
	\Pi_{\operatorname{D-C,D-D_{app}}} &= \int_{-\infty}^{\infty} \Pi_{\operatorname{D-D_{app}}}  \dd{q_1}= \Pi_{\operatorname{D-D_{IP}}}^m \frac{4 \sqrt{\pi} q_2 \Gamma(m/2)}{(2m-9) \Gamma(\frac{m-1}{2})}.
	\end{aligned}
\end{equation}
An important feature of the interaction potential is its scaling w.r.t.~the gap, for which we utilize a scaling factor function \cite{2025borković}. The scaling factor functions of the obtained disk-infinite cylinder potentials vs.~the gap $q_2$ for $m=6$ are shown in Fig.~\ref{fig:LJpoten3}a.
These results confirm that the $\operatorname{D-D_{app}}$ law returns the correct scaling, while the ISSIP fails in this regard. Therefore, the ISSIP law is appropriate only for small separations.

Next, let us further scrutinize the accuracy of these two approximations by comparing them with a numerical reference solution. The corresponding relative errors of the ISSIP and $\operatorname{D-D_{app}}$ laws for disk-infinite cylinder vdW interaction are plotted in Fig.~\ref{fig:LJpoten3}b. The highly accurate reference solution is obtained by numerically integrating the exact point-infinite cylinder law over the disk area. The $\operatorname{D-D_{app}}$ is generally more accurate, yet both laws are in good agreement for small separations. As the gap increases, however, the error of the ISSIP blows up, while the error of the $\operatorname{D-D_{app}}$ is bounded. Although this relative error is close to 200 \%, the absolute value is very small due to the large separation \cite{2025borković}.



\begin{figure}[h!]
	\centering
		\begin{tikzpicture}
		\begin{axis}[
			xlabel = {$\bar q_2=q_2/R$},
			ylabel = {Disk-cylinder force 
			},
			ylabel near ticks,
			legend pos=south east,
			legend cell align=left,
			legend style={font=\scriptsize},
			width=0.45\textwidth,
			ymin = -0.195, ymax = 0.055,
			xmin=0.025,xmax=0.205,
			minor y tick num = 1,
			minor x tick num = 1,
			ytick distance=0.05,
			grid=both,clip=false,
			xticklabel style={/pgf/number format/fixed, /pgf/number format/precision=2},
			yticklabel style={/pgf/number format/fixed, /pgf/number format/precision=2}]
			\addplot[black,thick] table [col sep=comma] {data/dataNumExampleLJF2DDappSigma1.csv};
			\addplot[red,dashed,thick] table [col sep=comma] {data/dataNumExampleLJF2ISSIPSigma1.csv};
			\addplot[blue,thick,dashed] table [col sep=comma] {data/dataNumExampleLJF2DDExactDistx1.csv};
			\node [text width=1em,anchor=north west] at (rel axis cs: -0.36,1.1){a)};
			\legend{$\operatorname{D-D_{app}}$,ISSIP,Numeric} 
		\end{axis}
	\end{tikzpicture}
	\begin{tikzpicture}
		\begin{axis}[
			xlabel = {$\bar q_2=q_2/R$},
			ylabel = {Log$_{10}$ (absolute error)},
			ylabel near ticks,
			legend pos=north east,
			legend cell align=left,
			legend style={font=\scriptsize,nodes={scale=0.91, transform shape}},
			width=0.45\textwidth,
			minor y tick num = 1,
			minor x tick num = 1,
			grid=both,
			xmin=0.025,xmax=0.205,
			ymin = -3.75, ymax = -1,
			clip=false,
			xticklabel style={/pgf/number format/fixed, /pgf/number format/precision=2}
		]
			\addplot[black,thick,restrict x to domain=0:0.2] table [col sep=comma] {data/dataLawsTestErrorDDapp.csv};
			\addplot[red,dashed,thick,restrict x to domain=0:0.2] table [col sep=comma] {data/dataLawsTestErrorISSIP.csv};
			\node [text width=1em,anchor=north west] at (rel axis cs: -0.26,1.1){b)};
			\legend{$\operatorname{D-D_{app}}$,ISSIP
			} 
		\end{axis}
	\end{tikzpicture}
	\caption{Comparison of disk-infinite cylinder LJ force functions for $R=R_x=R_y=0.02$, $k_6=-10^{-7}$, $k_{12}=5 \times 10^{-25}$. Highly accurate disk-infinite cylinder force values are obtained by numerically integrating the exact point-infinite cylinder law. a) LJ force vs.~normalized gap $\bar q_2$. b) Scaled absolute error of $\operatorname{D-D_{app}}$ and ISSIP laws w.r.t.~a highly accurate numerical result vs.~the normalized gap $\bar q_2$. }
	\label{fig:LJpoten3s}
\end{figure}
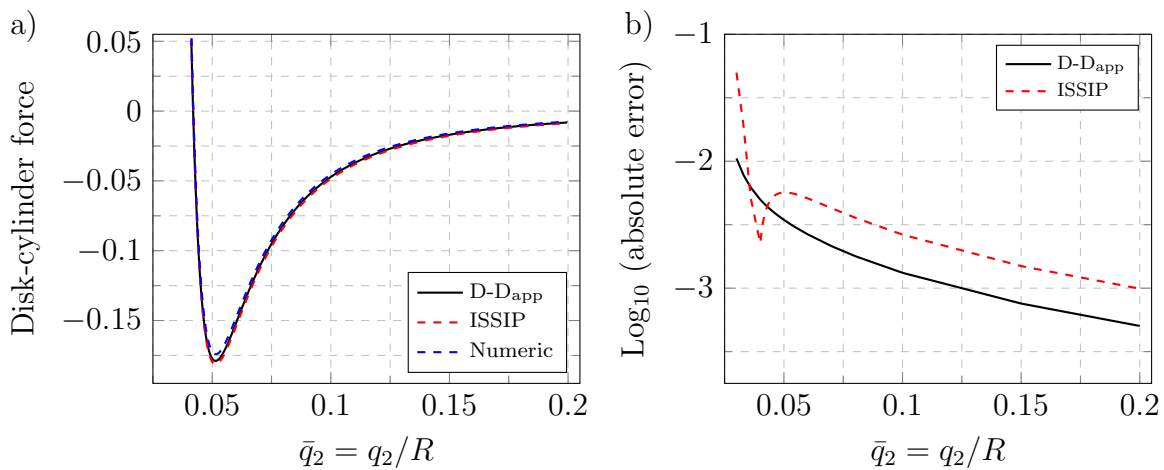

To illustrate this decay of force with distance, the disk-infinite cylinder LJ force is displayed in Fig.~\ref{fig:LJpoten3s}a for both laws, and compared with a highly accurate numerical solution.
By inspecting these force plots, the differences between the numerical, $\operatorname{D-D_{app}}$, and ISSIP results are noticeable but not significant. However, this difference strongly depends on the adopted physical constants \cite{2025borković}. 

The absolute error of the two considered laws is plotted in Fig.~\ref{fig:LJpoten3s}b. Although there is a small range of gaps where the ISSIP law returns more accurate results, the $\operatorname{D-D_{app}}$ law is generally more accurate and consistent. This plot shows that the absolute error decreases with the gap, although the relative one is increasing, as evidenced in Fig.~\ref{fig:LJpoten3}b for the vdW case.


Finally, let us consider the equilibrium gap between a disk and an infinite cylinder in parallel orientation, $q_\text{2,eq}$. An analytical expression for $q_\text{2,eq}$ can be found by employing the ISSIP law. Starting from \eqref{appDD1s}, we define a disk-infinite cylinder ISSIP LJ potential, and find its minimum which corresponds to the force equilibrium, i.e.,
\begin{equation}
	\frac{\partial \Pi_{\operatorname{D-C,ISSIP}}^{LJ}}{\partial q_2} \bigg\rvert_{ q_2=q_\text{2,eq}}= 0 \;\; \rightarrow \;\; q_\text{2,eq}=\left(-\frac{143 \: k_{12}}{2048 \: k_{6}}\right)^{1/6}.
	\label{eq:2}
\end{equation}
It can be shown that this is the same result as derived in \cite{2020grill}. 
For the $\operatorname{D-D_{app}}$ law, however, an analytical equilibrium disk-infinite cylinder distance is more challenging to find and its existence is an open question.

\subsection{Beam-beam interaction formulation}
\label{sec:beam-beam}



\begin{figure}[h!]
	\begin{center}
		\includegraphics[width=0.9\textwidth]{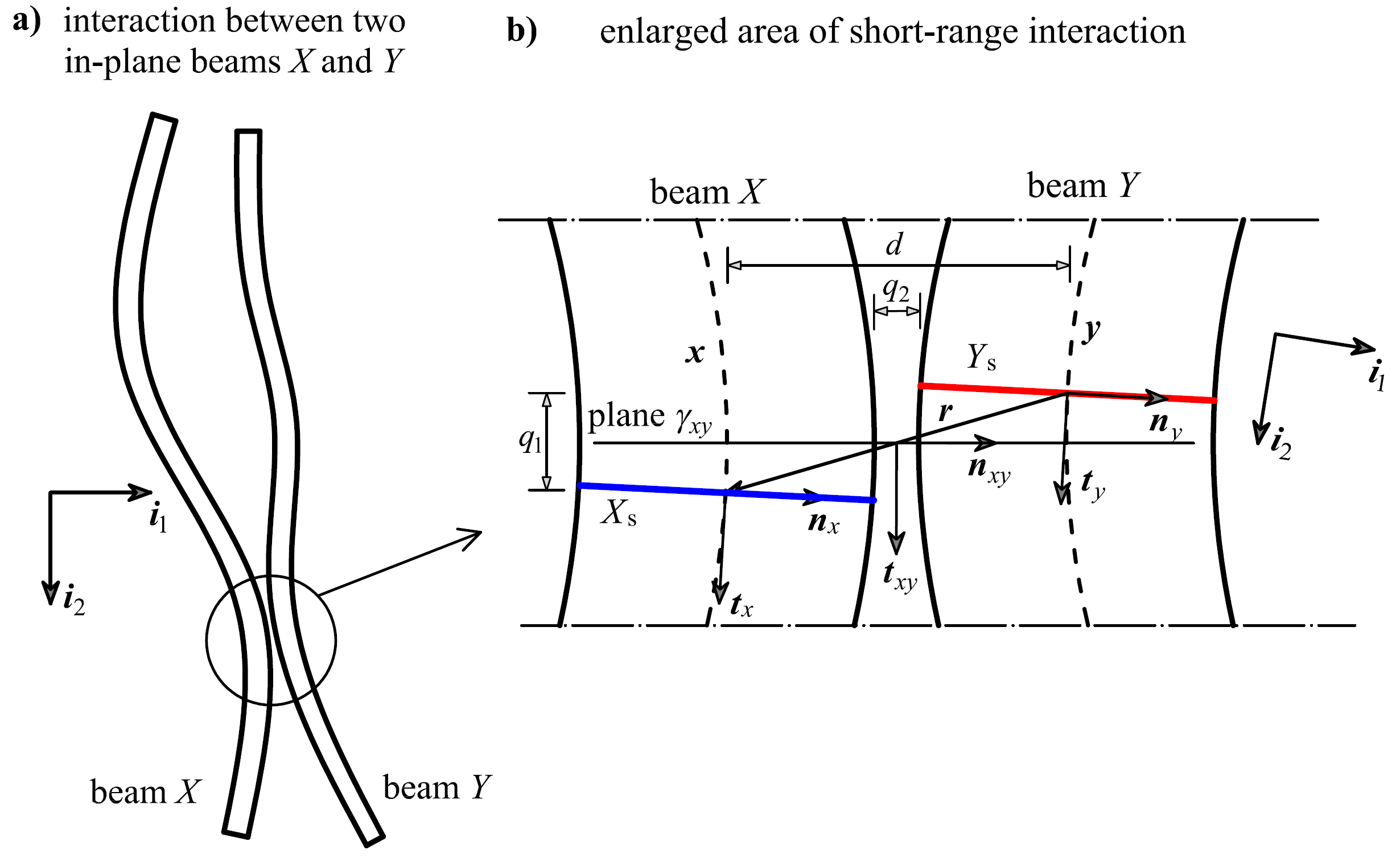}
		\caption{a) Interaction of two in-plane beams. b) Enlarged area of short-range interaction. \cite{2024borkovićb}}
		\label{fig:cfa}
	\end{center}
\end{figure}

We compute an interaction potential between two beams by integrating their section-section potentials along the beams' axes. Fig.~\ref{fig:cfa}a illustrates two in-plane beams, $X$ and $Y$, that are interacting via some interaction potential.
%
By taking a closer look at the area of beam-beam contact, we focus on the interaction between two arbitrary cross sections in Fig.~\ref{fig:cfa}b. 
Although these cross sections can be in an arbitrary orientation, we assume that they are parallel, i.e., $\ve{t}_x \parallel \ve{t}_y$, and employ disk-disk laws from Subsection \ref{sec:sec-sec}.

To define $q_1$ and $q_2$, we need to adopt a reference local coordinate system (LCS). Choosing the LCS of either beam as the reference LCS leads to a bias. To avoid this issue, we adopt an averaged reference LCS $\left(\hat{\ve{t}}_{xy},\hat{\ve{n}}_{xy}\right)$ by adding and normalizing the basis vectors of both beams, i.e.,
\begin{equation}
	\label{eq: g1x2}
	\begin{aligned}
		\hat{\ve{t}}_{xy} &= \frac{\ve{t}_{xy}}{\sqrt{\ve{t}_{xy} \cdot \ve{t}_{xy}}}, \quad \ve{t}_{xy} = \ve{t}_x+\ve{t}_y, \\
		\hat{\ve{n}}_{xy} &= \ve{\Lambda} \hat{\ve{t}}_{xy}.
	\end{aligned}
\end{equation}
Now, we define the offset $q_1$ and gap $q_2$ by projecting the distance vector to the axes of the new reference LCS. This gives
\begin{equation}
	\label{eq: g1xa}
	\begin{aligned}
		q_1 &= \ve{r}\cdot \hat{\ve{t}}_{xy},   \\
		q_2 &= d - (R_x + R_y) \cos \beta , \quad d = \abs{\ve{r}\cdot \hat{\ve{n}}_{xy}} = s_{\alpha} \left(\ve{r}\cdot \hat{\ve{n}}_{xy}\right), \quad s_\alpha = \textrm{sgn} \left(\ve{r} \cdot \hat{\ve{n}}_{xy} \right),\\
		\cos \beta &= \hat{\ve{n}}_{xy} \cdot \ve{n}_x=\hat{\ve{n}}_{xy} \cdot \ve{n}_y,
	\end{aligned}
\end{equation}
where we also project the beam radii via angle $\beta$. This step proves crucial for some cases of small separations, depending on the orientation of sections.

\begin{figure}[h!]
	\begin{center}
		\includegraphics[width=0.95\textwidth]{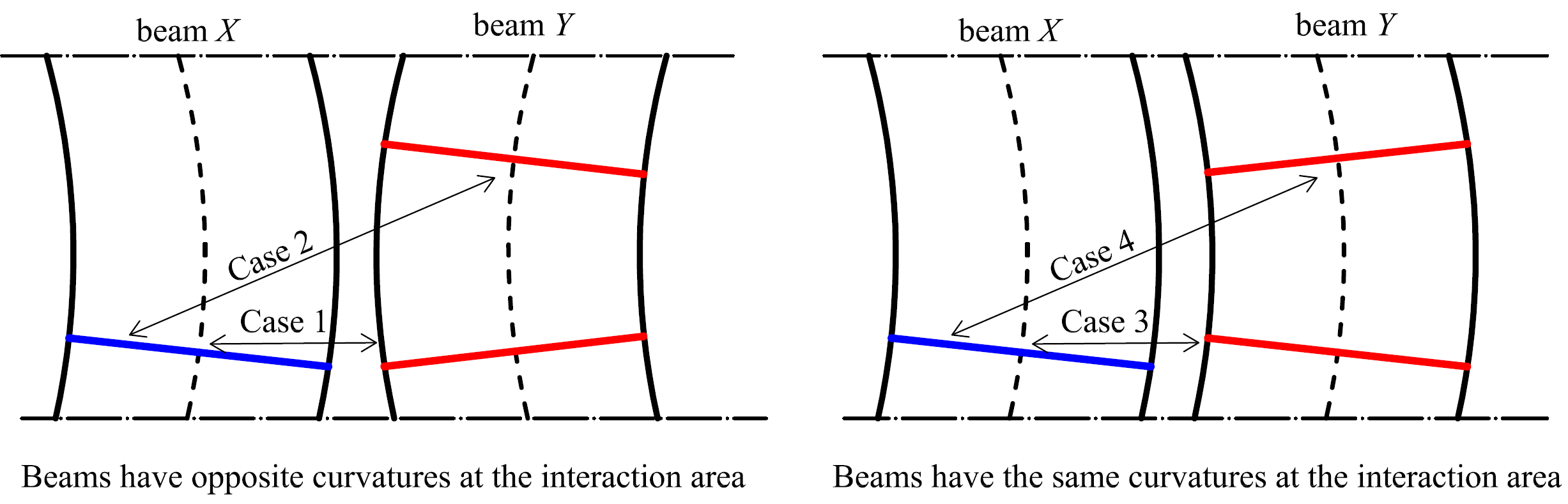}
		\caption{Four characteristic cases of beam-beam and section-section interactions.}
		\label{fig:cases}
	\end{center}
\end{figure}

Let us consider four characteristic cases of section-section orientation shown in Fig.~\ref{fig:cases}.
Case 2 and Case 3 are very accurately described with our disk-disk laws, since there we assume interaction between parallel cross sections. For Cases 1 and 4, this assumption is violated, and the error depends on the gap, the offset, and the angle between the cross sections. In our implementation, the definition of $q_2$ in Eq.~\eqref{eq: g1xa} improves the accuracy for Cases 1 and 4, compared to the definition in \cite{2024borkovićb}.

Furthermore, for a large offset in Case 4, the projection to the averaged LCS may lead to an overlap, i.e., the gap $q_2$ becomes negative. 
Hence, we implement a minimum positive limit value for the gap, $q_\text{2,lim}=10^{-8}$. Our calculations show that this practically does not influence the accuracy, since such configurations occur for relatively large $q_1$ when the contribution from this section-section interaction is negligible.

With definitions \eqref{eq: g1xa} at hand, the gradients of the gap and offset w.r.t.~positions of both beams are
\begin{equation}
	\label{eq: full3x66}
	\begin{aligned}
		\nabla_{\ve{x}} q_1 &=  \nabla_{\ve{x}}  \left(\ve r \cdot \hat{\ve{t}}_{xy}\right) =   \hat{\ve{t}}_{xy} = - \nabla_{\ve{y}} q_1, \\
		\nabla_{\ve{x}} q_2 &= \nabla_{\ve{x}}   \abs{\ve {r} \cdot \hat{\ve{n}}_{xy}} = s_{\alpha} \hat{\ve{n}}_{xy} = -\nabla_{\ve{y}} q_2.
	\end{aligned}
\end{equation}
Finally, the variations of the interaction potential between two circular cross sections, $\Pi_{\operatorname{D-D}}^m$, neglecting the interaction moment, are
\begin{equation}
	\label{eq: full3x}
	\begin{aligned}
		\delta_{x} \Pi_{\operatorname{D-D}}^m &= \nabla_{\ve{x}} \Pi_{\operatorname{D-D}}^m \cdot \delta \iv{u}{}{x}= \ve{f}_x \cdot \delta \iv{u}{}{x}=\ve{f} \cdot \delta \iv{u}{}{x} , \\
		\delta_{y} \Pi_{\operatorname{D-D}}^m &= \nabla_{\ve{y}} \Pi_{\operatorname{D-D}}^m \cdot \delta \iv{u}{}{y}=\ve{f}_y \cdot \delta \iv{u}{}{y}= -\ve{f} \cdot \delta \iv{u}{}{y} ,
	\end{aligned}
\end{equation}
where the section-section interaction force is
\begin{equation}
	\label{eq: full3xff}
	\begin{aligned}
		\ve{f} &=  \frac{\partial \Pi_{\operatorname{D-D}}^m }{\partial q_1} \hat{\ve{t}}_{xy} + \frac{\partial \Pi_{\operatorname{D-D}}^m }{\partial q_2}  s_\alpha \hat{\ve{n}}_{xy} = f_1  \hat{\ve{t}}_{xy} + f_2  \hat{\ve{n}}_{xy} .
	\end{aligned}
\end{equation}
Further details on the beam-beam formulation are given in \cite{2024borkovićb}.

	\subsection{Solution algorithms}
	
For the sake of completeness, we briefly discuss solution techniques for the set of spatially discretized equations of equilibrium.

Since the time-continuous analytical solution of Eq.~\eqref{eq: poten} rarely exists, we aim to find its solution at a discrete set of time instances $t_i$, $i=1,2,...,M$. In order to find this set, we start from some equilibrium configuration, defined with the vector of nodal displacements $\textbf{q}^i$, and aim to find the next one, $\textbf{q}^{i+1}$. Therefore, at instance $t_{i+1}$, the balance equation is
\begin{equation}
	 \textbf{M} \ddot{\textbf{q}}^{i+1}+\textbf{F}(\textbf{q}^{i+1}) -\textbf{Q}(\textbf{q}^{i+1})-\pmb{\Phi}(\textbf{q}^{i+1})=\textbf{R}(\textbf{q}^{i+1})\overset{!}{=} 
	\textbf{0},
	\label{eq:equil}
\end{equation}
where the vector of inertial nodal forces is represented as a product of the mass matrix $\textbf{M}$ and the nodal acceleration vector $\ddot{\textbf{q}}$, $\textbf{F}$ is the vector of internal nodal forces, $\textbf{Q}$ is the vector of external nodal forces, $\pmb{\Phi}$ is the vector of nodal interaction forces, and $\textbf{R}$ is the vector of residual nodal forces, which should be equal to the zero vector $\textbf{0}$ at equilibrium.

\subsubsection{Static analysis with the Newton-Raphson algorithm}

In this subsection, we consider the quasi-static case where inertia is neglected. By the Taylor expansion w.r.t.~the configuration at $t_{i}$, equilibrium equation \eqref{eq:equil} becomes
\begin{equation}
	\label{eq:tayloreq}
	\textbf{R} (\textbf{q}^{i+1})= \textbf{R}(\textbf{q}^i) + \nabla \textbf{R} (\textbf{q}^i)  \Delta \textbf{q}^{i+1} + \operatorname{h.o.t.}=\textbf{K}_\text{T}^i \Delta \textbf{q}^{i+1} + \operatorname{h.o.t.}\overset{!}{=}\textbf{0}
\end{equation}
where $\textbf{K}_\text{T}^i=\nabla \textbf{R} (\textbf{q}^i)$ is the tangent stiffness matrix. Since this equation is nonlinear w.r.t.~the unknown increment $\Delta \textbf{q}$, we aim to solve it by an iterative Newton-Raphson (NR) procedure. For brevity, the time index is omitted in the remainder of this subsection since all quantities refer to the current time step $i+1$. If the equilibrium is not reached at the $j^{th}$ iteration, we calculate the new iterative displacement increment from
	\begin{equation}
		{}^j \textbf{K}_\text{T}^{} \Delta {}^{j+1} \textbf{q}^{} = {}^j\textbf{R}^{}={}^j\textbf{F}^{}-{}^j\textbf{Q}^{}-{}^j\pmb{\Phi}^{},
		\label{eq:linearized}
	\end{equation}
	and update the configuration
	\begin{equation}
		{}^{j+1}\textbf{q}^{}={}^j\textbf{q}^{}+\Delta {}^{j+1}\textbf{q}^{}.
	\end{equation}
Then we calculate the residual at the updated configuration and check if it satisfies a convergence criteria that is here defined as
\begin{equation}
	\frac{\norm{{}^{j+1}\textbf{R}^{}}}{\norm{{}^{j+1}\textbf{Q}^{}+{}^{j+1}\pmb{\Phi}^{}}}\le \epsilon.
\end{equation}
In all of our simulations in Section \ref{sec:num}, the convergence threshold is set to $\epsilon=10^{-6}$.

\subsubsection{Static analysis with the arc-length method}

The NR method from the previous subsection is applicable for the examples that do not exhibit snap-through behavior, meaning that there is no load limit points and the stiffness matrix is regular in all configurations. If there is a load limit point on the equilibrium path, as in Subsection \ref{sec:num1}, we use the arc-length (AL) method. There are many different implementations of the AL method developed for adhesion problems \cite{2006sauer,2022roy,2023roy}. In this paper, we use the modified Riks method, as employed in Abaqus \cite{2009smith}. 

	
The main idea of the AL method is to introduce a load proportionality factor $\lambda$ as an additional unknown in \eqref{eq:tayloreq}. This factor scales the total value of the external load, or the total interaction force, or both. Without losing generality, let us assume that there is only an external load acting on the system. The load proportionality factor $\lambda$ allows us to overcome load limit points by increasing or decreasing the external load during the simulation, i.e.,
\begin{equation}
		\label{eq eq for numeric1}
		\textbf{Q}^i = \ii{\lambda}{i}{} \textbf{Q}^M.
\end{equation}
%

Let us start from some known, converged, configuration defined with $(\textbf{q}^{i}, \ii{\lambda}{i}{})$, and search for the next equilibrium point $(\textbf{q}^{i+1}, \ii{\lambda}{i+1}{})$. The first step is to find a predictor solution, called the predictor tangential displacement vector, $\textbf{q}_\text{T}$. It is calculated from the known tangent stiffness and the total external load, i.e.,
	\begin{equation}
		\label{eq eq for numeric2}
		\textbf{K}_\text{T}^i {}^0\textbf{q}_\text{T}^{i+1} = \textbf{Q}^M.
	\end{equation}
The predictor load increment $\Delta \: \iii{\lambda}{0}{}{i+1}{}$ follows from a constraint equation \cite{1981crisfield, 2009smith} as
\begin{equation}
		\label{eq eq for numeric3}
		\Delta \: \iii{\lambda}{0}{}{i+1}{} = \pm \frac{\Delta \: \iii{l}{}{}{i+1}{}}{\sqrt{  {}^0\tilde{\textbf{q}}_\text{T}^{i+1} \: \cdotp \: {}^0\tilde{\textbf{q}}_\text{T}^{i+1}+1}},
\end{equation}
where $\Delta \iii{l}{}{}{i+1}{} $ is the arc-length value that is estimated for the first, and calculated for all the other increments as a function of the desired number of iterations per increment. For convenience, the displacement vectors in Eq.~\eqref{eq eq for numeric3} are scaled by the maximum displacement component of the corresponding linear solution from Eq.~\eqref{eq eq for numeric2}, and designated with tilde. The sign of the predictor solution is calculated from the condition that the projection of the predictor tangential displacement onto the previously converged displacement increment $\Delta \textbf{q}^i$ must be positive, ensuring continuation along the equilibrium path, i.e.,
	\begin{equation}
		\label{eq eq for numeric6}
		\Delta \, {}^0\!\lambda^{i+1} ( {}^0\tilde{\textbf{q}}^{i+1}_\text{T} \cdot \Delta \tilde{\textbf{q}}^i + \Delta\lambda ^i ) > 0.
	\end{equation}
In general, the predictor solution does not satisfy equilibrium and the AL iterations act as a corrector. In the remainder of this subsection, all quantities, unless specified  otherwise, are taken at the current time step $i+1$ and the time index is omitted to simplify notation. For the first iteration, the increments of displacement and load are initialized from the predictor solutions:
	\begin{equation}
		\label{eq eq for numeric7}
			\Delta  \, {}^1\!\textbf{q}= \Delta \, {}^0\!\lambda \, {}^0\!\textbf{q}_\text{T} \quad \textrm{and} \quad \Delta \, {}^1\!\lambda= \Delta \, {}^0\!\lambda.
	\end{equation}
If the convergence criteria is not met at the $j^{th}$ iteration, new tangential, ${}^{j+1}\textbf{q}_\text{T}$, and residual, $\delta {}^{j+1}\textbf{q}$, displacements are calculated from:
	\begin{equation}
	{}^j \textbf{K}_\text{T}\; {}^{j+1} \textbf{q}_\text{T} = \textbf{Q}^M \;\; \textrm{and} \;\; {}^j \textbf{K}_\text{T} \;\delta \, {}^{j+1}\!\textbf{q} = {^j}\textbf{R}.
	\label{eq:linearizeda}
\end{equation}
Then, we update the configuration with
		\begin{equation}
		\label{eq eq for numeric11a}
		\begin{aligned}
			{}^{j+1} \textbf{q} &= {}^j\textbf{q} + \delta  \, {}^{j+1}\! \textbf{q} + \delta \, {}^{j+1}\! \lambda \, \textbf{q}_\text{T}, \\
			\iii{\lambda}{j+1}{}{}{} &= {}^{j}\lambda +  \delta \, {}^{j+1}\! \lambda=\lambda ^{i} + \Delta \, {}^{j}\!\lambda + \delta \, {}^{j+1}\! \lambda,
		\end{aligned}
	\end{equation}
where the iterative load factor, $\delta ^{j+1} \lambda^{}$, follows from the condition that a new potential equilibrium point is orthogonal to the tangential displacement in the AL solution space, i.e.,
	\begin{equation}
	\label{eq eq for numeric12a}
	\delta \, {}^{j+1}\! \lambda^{} = -\frac{\delta \, {}^{j}\!\tilde{\textbf{q}}^{}\cdot {}^0\tilde{\textbf{q}}^{}_\text{T}}{{}^j \tilde{\textbf{q}}^{}_\textbf{T}\cdot {}^0\tilde{\textbf{q}}^{}_\text{T}+1}.
\end{equation}
The procedure is repeated until the convergence criteria is fulfilled \cite{2009smith}.

\subsubsection{Dynamic analysis using the HHT-$\alpha$ method}

When the inertial contributions are significant, the full equilibrium equation \eqref{eq:equil} is considered.
%
%
Using the Hilber-Hughes-Taylor (HHT)-$\alpha$ method, and by adding a velocity-proportional damping term, $\textbf{C} \dot{\textbf{q}}^{}$, this equation is replaced by
\begin{equation}
	\textbf{M} \ddot{\textbf{q}}^{i+1} +\textbf{C} \dot{\textbf{q}}^{i+1} + (1+\alpha)(\textbf{F}^{i+1}-\textbf{Q}^{i+1}-\pmb{\Phi}^{i+1})-\alpha(\textbf{F}^{i}-\textbf{Q}^{i}-\pmb{\Phi}^{i})=\textbf{R}^{i+1}= \textbf{0}, 
	\label{eq:equilDa}
\end{equation}
where the parameter $\alpha$ introduces numerical damping into the system, and it is restricted to $-1/2 \le\alpha \le0$ \cite{1977hilber}; $\alpha=0$ corresponds to Newmark's method. Similar as for the static analysis, we aim to solve \eqref{eq:equilDa} by an iterative scheme, linearizing the nonlinear terms w.r.t.~the previously converged configuration at time $t_i$.

The velocity at the current configuration is represented via the velocity at the previous increment, the acceleration, and the time increment
\begin{equation}
	\label{eq: 11a}
	\dot{\textbf{q}}^{i+1} = \dot{\textbf{q}}^{i} + \Delta t \ddot{\textbf{q}}_\gamma,
\end{equation}
where $\ddot{\textbf{q}}_\gamma$ is a linear combination of accelerations $\ddot{\textbf{q}}^{i}$ and $\ddot{\textbf{q}}^{i+1}$ controlled by parameter $\gamma$,

\begin{equation}
	\label{eq: 11}
	\ddot{\textbf{q}}_\gamma = \left(1-\gamma\right) \ddot{\textbf{q}}^i + \gamma \ddot{\textbf{q}}^{i+1}, \quad \gamma=\frac{ 1}{2}-\alpha.
\end{equation}
The displacement is represented similarly, considering that the acceleration varies within the time increment. Assume that the new displacement is a function of displacement and velocity at the start of the increment, and some value of acceleration
\begin{equation}
	\label{eq: vel1ss}
	\textbf{q}^{i+1}= \textbf{q}^{i} + \Delta t \dot{\textbf{q}}^i + \frac{1}{2} \Delta t^2 \ddot{\textbf{q}}_\beta,
\end{equation}
where we use the linear combination of accelerations
\begin{equation}
	\label{eq: acc1aa}
	\ddot{\textbf{q}}_\beta = \left(1-2\beta\right) \ddot{\textbf{q}}^i + 2 \beta \ddot{\textbf{q}}^{i+1}, \quad \beta = \frac{1}{4} (1-\alpha)^2.
\end{equation}
These equations allow us to update the velocity and the acceleration at the new iteration as a function of the displacement increment
\begin{equation}
	\label{eq: updateaa}
	\begin{aligned}
		{}^{j+1}\textbf{q}^{i+1} &= {}^j\textbf{q}^{i+1} + \Delta {}^{j+1}\textbf{q}^{i+1}\\
		{}^{j+1}\ddot{\textbf{q}}^{i+1} &= \frac{1}{\Delta t ^2 \beta}\left({}^{j+1}\textbf{q}^{i+1}-\textbf{q}^{i}\right) - \frac{1}{\Delta t \beta} \dot{\textbf{q}}^i - \frac{1-2\beta}{2\beta} \ddot{\textbf{q}}^i, \\
		{}^{j+1}\dot{\textbf{q}}^{i+1} &= \dot{\textbf{q}}^{i} + \Delta t [\left(1-\gamma\right) \ddot{\textbf{q}}^{i} + \gamma \; {}^{j+1}\ddot{\textbf{q}}^{i+1}].
	\end{aligned}
\end{equation}

After linearization, Eq.~\eqref{eq:equilDa} at the $j^{th}$ iteration becomes
\begin{equation}
	\textbf{M} \;{}^j\ddot{\textbf{q}}^{i+1} +\textbf{C} \;{}^j\dot{\textbf{q}}^{i+1} + (1+\alpha) \;{}^j\textbf{K}_\text{T}^{i+1} \;\Delta {}^{j+1}\textbf{q}^{i+1}={}^j\textbf{R}^{i+1}.
	\label{eq:equilDass}
\end{equation}
and by using Eq.~\eqref{eq: updateaa}, it can be written as
\begin{equation}
		{}^j \textbf{K}_\text{T,eq}^{i+1} \; \Delta {}^{j+1} \textbf{q}^{i+1} = {}^j\textbf{R}^{i+1},
		\label{eq:linearizedds}
\end{equation}
where
\begin{equation}
	{}^j \textbf{K}_\text{T,eq}^{i+1} = (1+\alpha) \;{}^j\textbf{K}_\text{T}^{i+1} +\frac{1}{\Delta t^2 \beta} \textbf{M} + \frac{\gamma}{\Delta t \beta} \textbf{C}.
	\label{eq:linearizedd}
\end{equation}
We continue iterating until the convergence is reached.

In our implementation, Rayleigh damping, which simply combines mass- and stiffness-proportional damping, i.e.~$\textbf{C}=\beta_k \textbf{K}+\beta_m \textbf{M}$, is adopted. Regarding the automatic time stepping, the half-increment residual approach is implemented \cite{1979hibbitt,2009smith}. 

\subsection{Numerical integration of the section-section potential}
\label{sec:cut}
		
\begin{figure}[h!]
	\centering
	\begin{tikzpicture}
		\begin{axis}[
			xlabel = {Log$_{10}$($\bar n_\text{GP}$)},
			ylabel = {Log$_{10}$(relative error)},
			ylabel near ticks,
			legend pos=south west,
			legend columns=2, 
			legend cell align=left,
			legend style={font=\tiny},
			width=0.47\textwidth,
			height=0.5\textwidth,
			ymin = -15, ymax =0,
			minor y tick num = 1,
			minor x tick num = 1,
			xticklabel style={/pgf/number format/fixed, /pgf/number format/precision=2},
			yticklabel style={/pgf/number format/fixed, /pgf/number format/precision=2},
			grid=both,clip=false];
			\addplot[black] table [col sep=comma]{data/dataIntegrationTestGap0.03R.csv};
			\addplot[black,dashed] table [col sep=comma]{data/dataIntegrationTestGap0.04R.csv};
			\addplot[magenta] table [col sep=comma]{data/dataIntegrationTestGap0.05R.csv};		
			\addplot[magenta,dashed] table [col sep=comma]{data/dataIntegrationTestGap0.08R.csv};		
			\addplot[green] table [col sep=comma]{data/dataIntegrationTestGap0.15R.csv};		
			\addplot[green,dashed] table [col sep=comma]{data/dataIntegrationTestGap0.3R.csv};		
			\addplot[blue] table [col sep=comma]{data/dataIntegrationTestGap0.4R.csv};		
			\addplot[blue,dashed] table [col sep=comma]{data/dataIntegrationTestGap1.R.csv};		
			\node [text width=1em,anchor=north west] at (rel axis cs: -0.28,1.1){a)};
			\legend{$\bar q_2=0.03$,$\bar q_2=0.04$
				,$\bar q_2=0.05$
				,$\bar q_2=0.08$
				,$\bar q_2=0.15$
				,$\bar q_2=0.3$,$\bar q_2=0.4$
				,$\bar q_2=1$
				}
		\end{axis}
	\end{tikzpicture}
	\begin{tikzpicture}
		\begin{axis}[
			xlabel = {Cutoff distance $\bar c = c/R$},
			ylabel = {Log$_{10}$(relative error)},
			ylabel near ticks,
			legend pos=north east,
			legend columns=2, 
			legend cell align=left,
			legend style={font=\tiny},
			width=0.47\textwidth,
			height=0.5\textwidth,
			xmin = 2, xmax =10,
			minor y tick num = 1,
			minor x tick num = 1,
			restrict x to domain=2:10,
			xticklabel style={/pgf/number format/fixed, /pgf/number format/precision=2},
			yticklabel style={/pgf/number format/fixed, /pgf/number format/precision=2},clip=false,
			grid=both];
			\addplot[black] table [col sep=comma]{data/dataIntegrationTestCut0.03R.csv};
			\addplot[black,dashed] table [col sep=comma]{data/dataIntegrationTestCut0.04R.csv};
			\addplot[magenta] table [col sep=comma]{data/dataIntegrationTestCut0.05R.csv};		
			\addplot[magenta,dashed] table [col sep=comma]{data/dataIntegrationTestCut0.08R.csv};		
			\addplot[green] table [col sep=comma]{data/dataIntegrationTestCut0.15R.csv};		
			\addplot[green,dashed] table [col sep=comma]{data/dataIntegrationTestCut0.3R.csv};		
			\addplot[blue] table [col sep=comma]{data/dataIntegrationTestCut0.4R.csv};	
			\addplot[blue,dashed] table [col sep=comma]{data/dataIntegrationTestCut1.R.csv};		
			\node [text width=1em,anchor=north west] at (rel axis cs: -0.28,1.1){b)};
	\end{axis}	\end{tikzpicture}
	\caption{Two sources of integration error for the disk-cylinder LJ force using the $\operatorname{D-D_{app}}$ law for eight different gaps. The cylinder's length is 2, $R_x=R_y=0.02$, $k_6=-10^{-7}$, and $k_{12}=5 \times 10^{-25}$. a) Relative error of numerical integration w.r.t.~the highly accurate numerical solution vs.~number of integration points per unit length. b) Relative error of numerical integration using cutoff w.r.t.~integration along the whole length vs.~fixed cutoff distance $c$  ($\bar n_\text{GP}=2000$). }
	\label{fig:LJpoten3ss}
\end{figure}
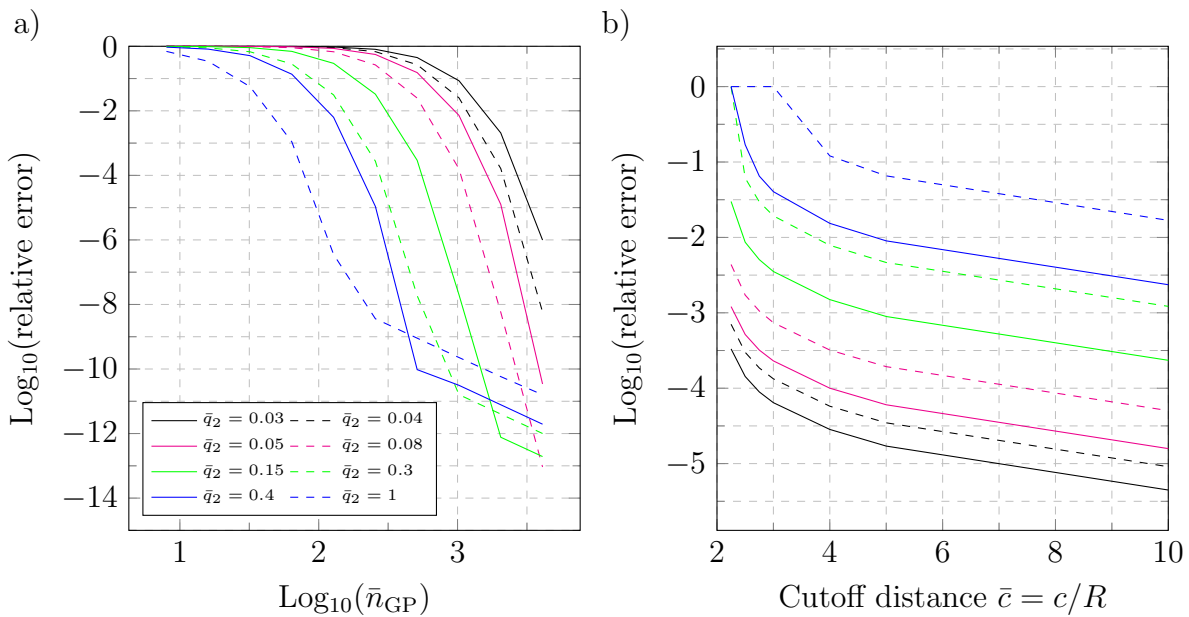

To define the fiber-fiber interaction, the section-section interaction potential needs to be numerically integrated along the axes of both beams. Due to high gradients of the disk-disk LJ interaction potential, this integration is challenging. In our previous work, we have concluded that mid-point rule is optimal for short-range interactions \cite{2024borkovićb}. The number of required quadrature points per unit length, $\bar n_\text{GP}$, depends on the minimum gap $q_\text{min}$. Let us consider an integration of LJ force between a disk and a cylinder of length 2. The relative error for different values of $\bar n_\text{GP}$ is plotted in Fig.~\ref{fig:LJpoten3ss}a for eight gap values.
A highly accurate solution requires large number of integration points which strongly affects the efficiency. For the adopted LJ parameters, the maximum adhesion occurs for $\bar q_{2} \approx  0.05$, while equilibrium occurs for $\bar q_\text{2,eq} \approx 0.042$, see Fig.~\ref{fig:LJpoten3s}a. From the obtained results in Fig.~\ref{fig:LJpoten3ss}a, it follows that the integration with $\bar n_\text{GP}=10^3$ gives a relative error of $1 \%$ for the gap at maximum adhesion.

Since the interaction forces are reciprocal to distance, only the closest point-pairs contribute significantly to the interaction in the regime of small separations. Thus, it is reasonable to define a circle around an integration point, and to neglect the interaction with points lying outside of it to improve efficiency. The radius of this circle is called \emph{cutoff distance} and is designated by $c$. The relative error of integrating the LJ disk-cylinder force using the cutoff distance is plotted in Fig.~\ref{fig:LJpoten3ss}b for eight gap values and $\bar n_\text{GP}=2000$. A strong influence of the normalized cutoff distance $\bar c=c/R$ on the integration accuracy is evident. For very small separations $\bar q_2 <0.08$, a cutoff slightly larger than $\bar c=2$ returns a relative error below $0.4\%$. Since the cutoff affects the computational time almost linearly, finding the appropriate $c$ is essential for an efficient and accurate analysis \cite{2024borkovićb}, and we will deal with it extensively in Section \ref{sec:num}.

Regarding moderate separations, the influence of more, if not all, points should be considered, and we address it with two approaches. The first approach defines two meshes of integration points: a dense one for the regime of small separations and a sparse one for moderate separations. Then, we use a large cutoff for moderate separations and a small cutoff for small separations. We switch between these two regimes when the minimum gap $q_\text{2,min}$ crosses a threshold value $q_\text{2,thr}$. The second approach is to employ a cutoff function $c_f$ with a minimal gap $q_\text{2,min}$ as an argument. There are many options to define a cutoff function, and we choose a piecewise linear one.
The function $c_f$ is adopted as constant for small separations $q_\text{2,min}\leq q_\text{2,eq}$, and linear for $q_\text{2,min} > q_\text{2,eq}$ with slope $s$, i.e.,
\begin{equation}
	\label{eq:1}
c_f(q_\text{2,min})=	\begin{cases} 
		w q_\text{2,eq} & q_\text{2,min}\leq q_\text{2,eq}  \\
		w q_\text{2,eq}+s (q_\text{2,min}-q_\text{2,eq}) & q_\text{2,eq}< q_\text{2,min} 
	\end{cases}.
\end{equation}
We will compare these two approaches in Subsection \ref{sec:collision}.

	\section{Numerical analysis}
	\label{sec:num}

The following analysis aims to demonstrate that the section-section approach can model very complex and fundamental, yet rarely considered, cases of fibers' interaction. We are not pursuing highly refined and fully converged solutions in all examples. Our focus is rather on discussing the modeled phenomena while comparing different computational models.

We exclusively consider quartic elements with $C^1$ interelement continuity.

\subsection{Snap-to-contact of two cantilever fibers}
\label{sec:num1}
	
\begin{figure}[h!]
		\begin{center}
			\includegraphics[width=0.4\textwidth]{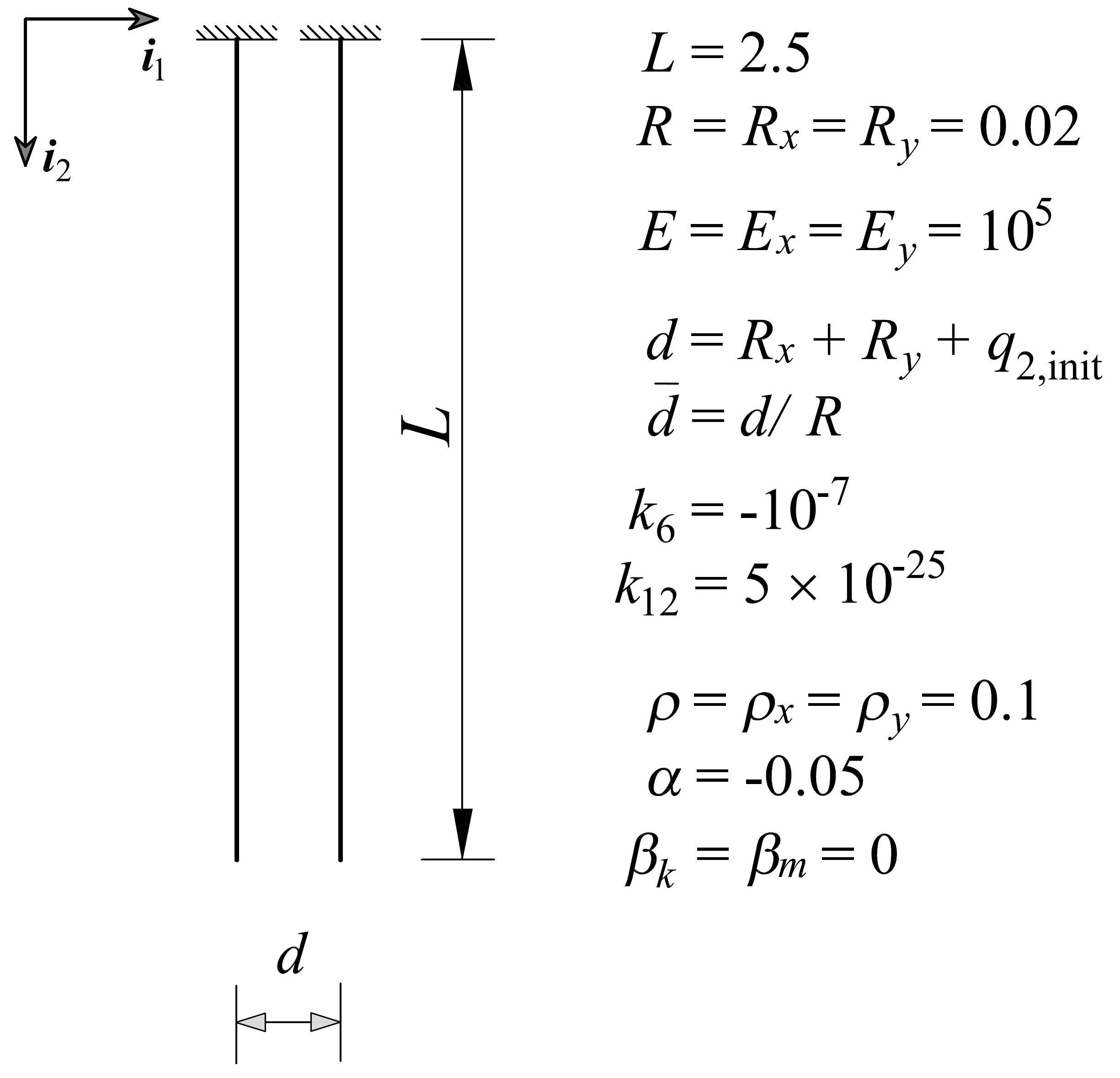}
			\caption{Snap-to-contact of two cantilever fibers: Problem setup and parameters.}
			\label{fig:intro0}
		\end{center}
	\end{figure}

Let us consider two fibers that are initially parallel and separated such that the LJ force is attractive, Fig.~\ref{fig:intro0}. 
This setup leads to a snap-to-contact (STC) between fibers, a phenomenon that is of highly nonlinear and dynamic nature. However, we can also model this problem using quasi-static analysis. Therefore, we consider three approaches for this example: (i) statics, using the NR method; (ii) statics, using the AL method; and (iii) dynamics, using the HHT-$\alpha$ method.

\subsubsection{Initial considerations}

Due to the high nonlinearity of the considered boundary value problem, it is necessary to apply the LJ force incrementally via load proportionality factor or quasi-time parameter, $\lambda=t\in\left[0,1\right]$. We aim to find the final equilibrium configuration for $t=1$. Our initial effort to solve the problem using the NR algorithm failed, so we focus on the AL method.

\begin{figure}[h!]
		\centering
		\begin{tikzpicture}
			\begin{axis}[
				xlabel = {t},
				ylabel = {Reaction force},
				ylabel near ticks,
				legend pos=north west,
				legend cell align=left,
				legend style={font=\scriptsize},
				width=0.45\textwidth,
				xmin = 0, xmax =0.004,
				ymin = 0, ymax =0.00012,
				minor y tick num = 1,
				minor x tick num = 1,
				xticklabel style={/pgf/number format/fixed, /pgf/number format/precision=2},
				yticklabel style={/pgf/number format/fixed, /pgf/number format/precision=2},
				clip=false,grid=both];
				\node [text width=1em,anchor=north west] at (rel axis cs: -0.25,1.1){a)};
				\addplot[red,only marks, mark =+,mark size = 0.9pt,restrict x to domain=0:0.004] table [col sep=comma] {data/dataNumExampleReactionDense105.csv};
				\addplot[blue,only marks, mark size = 0.9pt,restrict x to domain=0:0.004] table [col sep=comma] {data/dataNumExampleReactionSparse105.csv};
				\addplot[
				mark size=3pt,mark=o] 
				coordinates {
					(0.00264538, 0.00000935275)
					
					(	0.00036447, 0.000012988
					)
					
					(
					0.00266986, 0.0000445661
					)
					
					(
					0.00251447, 0.0000640126
					)
				}; 
				\node [above] at (axis cs:  0.00264538, 0.00000935275) {$P_1$};
				\node [above] at (axis cs:  0.00036447, 0.000012988) {$P_2$};
				\node [right] at (axis cs:  0.00266986, 0.0000445661) {$P_3$};
				\node [left] at (axis cs:  0.00251447, 0.0000640126) {$P_4$};
				\legend{restricted arc-length,unrestricted arc-length}
			\end{axis}
		\end{tikzpicture}
		\begin{tikzpicture}
			\begin{axis}[
				xlabel = {t},
				ylabel = {Reaction force},
				ylabel near ticks,
				legend pos=south west,
				legend cell align=left,
				legend style={font=\tiny},
				width=0.45\textwidth,
				xmin = 0, xmax =1,
				ymin = 0, ymax =0.007,
				minor y tick num = 1,
				minor x tick num = 1,
				xticklabel style={/pgf/number format/fixed, /pgf/number format/precision=2},
				yticklabel style={/pgf/number format/fixed, /pgf/number format/precision=2},
				clip=false,grid=both];
				\node [text width=1em,anchor=north west] at (rel axis cs: -0.195,1.1){b)};
				\addplot[red,only marks, mark=+,mark size = 0.9pt,restrict x to domain=0:1] table [col sep=comma] {data/dataNumExampleReactionDense105.csv};
				\addplot[blue,only marks, mark size = 0.9pt,restrict x to domain=0:1] table [col sep=comma] {data/dataNumExampleReactionSparse105.csv};
			\end{axis}
		\end{tikzpicture}
		\caption{Snap-to-contact of two cantilever fibers: Horizontal component of reaction force vs.~quasi-time for restricted and unrestricted values of arc-length ($\bar c = 3$, $\bar d=2.5$, $n_\text{el}=80$, $n_\text{GP}=100$, ISSIP). a) $t \in [0,4 \cdot 10^{-3}]$, b)  $t \in [0,1]$. The gaps in the a) are due to the fact that although the arc-length $\Delta l$ is restricted, the load increment $\Delta \lambda$ (or quasi-time increment $\Delta t$) is unrestricted. The near-singular tangent stiffness at the load limit points $P_i$, leads to a large value of the load increment.}
		\label{fig:LJpoten4xaaNEW}
\end{figure}
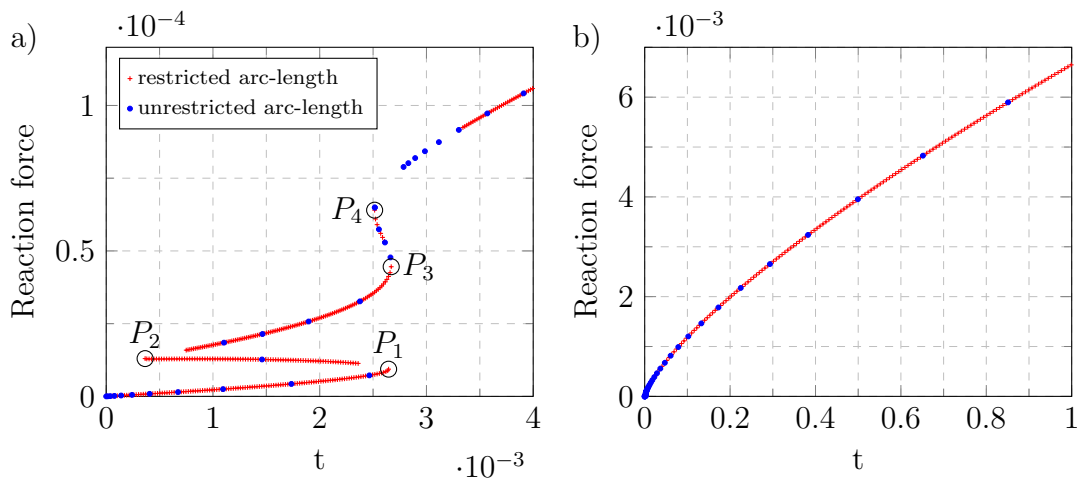

Since the example is computationally demanding, the first step in our analysis is to scrutinize the influence of the arc-length, $\Delta l$, and the resulting quasi-time step, $\Delta t$, on the accuracy. Let us consider the horizontal component of the reaction force, which is a representative macroscopic quantity of this setup. Equilibrium paths of the reaction force, using ISSIP and $\bar d = d/R=2.5$, are obtained for both restricted and unrestricted arc-lengths and displayed in Fig.~\ref{fig:LJpoten4xaaNEW}. 
 First, the fibers' response is highly nonlinear for $t \in [0,4 \cdot 10^{-3}]$ and consists of multiple load limit points, which provides rationale for the failure of the NR algorithm. Second, the results with both approaches are fully aligned, meaning that there is no need to restrict the arc-length value to find the final equilibrium position. However, even for the simulation with a restricted arc-length, quasi-time steps increase after the load limit points $P_i$, see gaps in data points in Fig.~\ref{fig:LJpoten4xaaNEW}a. This can be addressed by further restricting the arc-length or the quasi-time increment itself, but it is not pursued here.

From these initial results, we observe four load limit points, two upper ($P_1$ and $P_3$) and two lower ones ($P_2$ and $P_4$), Fig.~\ref{fig:LJpoten4xaaNEW}a. Interestingly, the upper load limit points occur for similar values, i.e., $t\approx 0.0026$. This suggests that, for a slightly smaller value of $t= 0.0026$, the considered quasi-static boundary value problem has at least five solutions. 

\begin{figure}[h!]
	\centering
	\begin{tikzpicture}
		\begin{axis}[
			xlabel = {$\ve{i}_1$ coordinate},
			ylabel = {$\ve{i}_2$ coordinate},
			ylabel near ticks,
			legend pos=north west,
			legend cell align=left,
			legend columns=2,
			legend style={font=\scriptsize,at={(axis cs:0.0050,0.1)},anchor=south east},
			width=0.5\textwidth,
			minor y tick num = 2,
			minor x tick num = 1,
			yticklabel=\pgfmathparse{abs(\tick)}\pgfmathprintnumber{\pgfmathresult},
			clip=true,clip mode=individual,grid=both];
			\node [text width=1em,anchor=north west] at (rel axis cs: -0.22,1.1){a)};
			\addplot[green,thick] table [col sep=comma] {data/dataCantileverSnapToCOntactDenseShapeTe0d0000.csv};
			\addplot[cyan,thick] table [col sep=comma] {data/dataCantileverSnapToCOntactDenseShapeTe0d00264.csv};
			\addplot[teal,thick] table [col sep=comma] {data/dataCantileverSnapToCOntactDenseShapeTe0d00036.csv};
			\addplot[violet,thick] table [col sep=comma] {data/dataCantileverSnapToCOntactDenseShapeTe0d002669.csv};
			\addplot[red,thick] table [col sep=comma] {data/dataCantileverSnapToCOntactDenseShapeTe0d002514.csv};
			\addplot[blue,thick] table [col sep=comma] {data/dataCantileverSnapToCOntactDenseShapeTe0d00399.csv};
			\addplot[purple,thick] table [col sep=comma] {data/dataCantileverSnapToCOntactDenseShapeTe0d01126.csv};
			\addplot[black,thick] table [col sep=comma] {data/dataCantileverSnapToCOntactDenseShapeTe1d0065.csv};
			
			\legend{t=0, t$\approx$0.00264 ($P_1$), t$\approx$0.00036 ($P_2$), t$\approx$0.00267 ($P_3$), t$\approx$0.00251 ($P_4$), t$\approx$0.00399, t=0.01126,t$\approx$1}
		\end{axis}
	\end{tikzpicture}
	\begin{tikzpicture}
		\begin{axis}[
			xlabel = {t},
			ylabel = {Displacement},
			ylabel near ticks,
			legend pos=south east,
			legend cell align=left,
			legend style={font=\scriptsize},
			width=0.5\textwidth,
			minor y tick num = 1,
			minor x tick num = 1,
			xticklabel style={/pgf/number format/fixed, /pgf/number format/precision=2},
			yticklabel style={/pgf/number format/fixed, /pgf/number format/precision=2},
			clip=false,grid=both];
			\node [text width=1em,anchor=north west] at (rel axis cs: -0.195,1.1){b)};
			\addplot[black,thick] table [col sep=comma] {data/dataCantileverSnapToCOntactDenseDispPathXI1d0.csv};
			\addplot[red,thick] table [col sep=comma] {data/dataCantileverSnapToCOntactDenseDispPathXI0d8.csv};
			\addplot[blue,thick] table [col sep=comma] {data/dataCantileverSnapToCOntactDenseDispPathXI0d6.csv};
			\addplot[green,thick] table [col sep=comma] {data/dataCantileverSnapToCOntactDenseDispPathXI0d4.csv};

			\legend{s=2.5,s=2.0,s=1.5,s=1.0}
		\end{axis}
	\end{tikzpicture}
	\caption{Snap-to-contact of two cantilever fibers: a) Configurations of the left fiber's axis for eight different quasi-time values. To allow visibility, coordinates on the horizontal axis are scaled w.r.t.~coordinates on the vertical axis by a factor of $f_s\approx650$. b) Horizontal component of displacement vs.~quasi-time for four different points on a left fiber for $t \in [0,4 \cdot 10^{-3}]$. ($\bar c = 3$, $\bar d=2.5$, $n_\text{el}=80$, $n_\text{GP}=100$, ISSIP).}
	\label{fig:LJpoten4xaaNEWaa}
\end{figure}
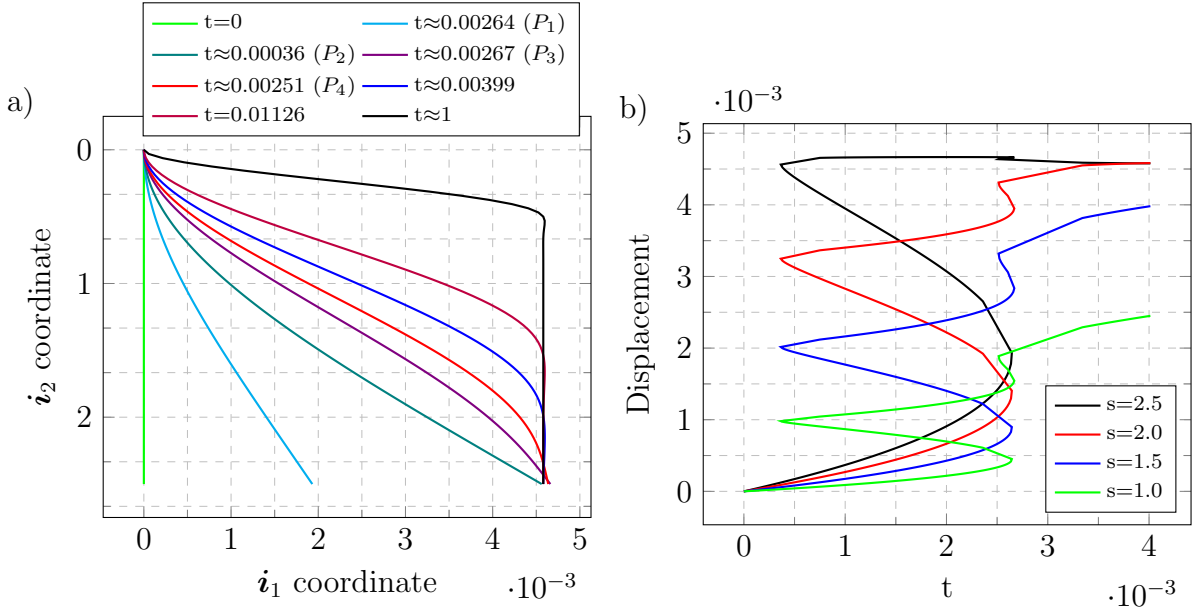

\begin{figure}[h!]
		\begin{center}
			\includegraphics[width=\textwidth]{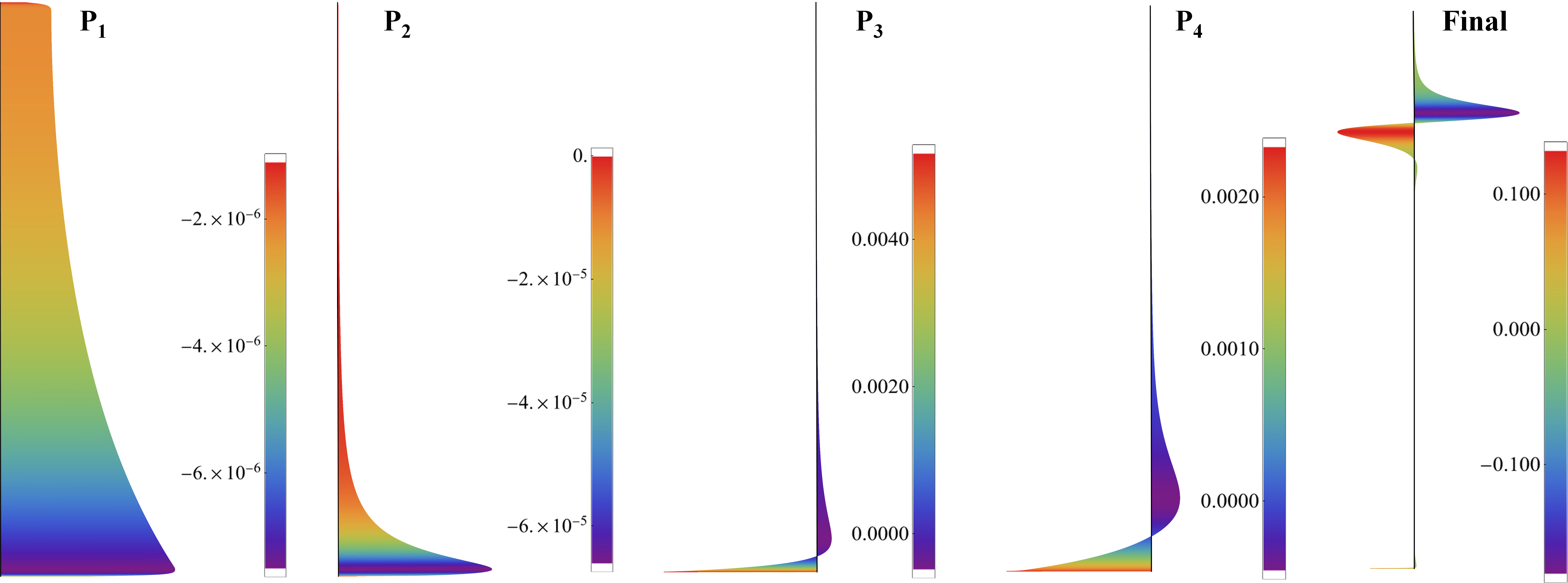}
			\caption{Snap-to-contact of two cantilever fibers: Normal component of the interaction force, $f_2$, for $\bar d =2.5$ is plotted on the left fiber for five characteristics configurations: at four load limit points, and at the end (t=1).}
			\label{fig:intro0x}
		\end{center}
\end{figure}

To further scrutinize the computational sensitivity of this problem, configurations of the left fiber's axis for eight different quasi-time steps are plotted in Fig.~\ref{fig:LJpoten4xaaNEWaa}a, while the displacement vs.~time plots of four points on the left fiber are displayed in Fig.~\ref{fig:LJpoten4xaaNEWaa}b. Furthermore, the normal component of the interaction force $f_2$ is plotted on the left fiber in Fig.~\ref{fig:intro0x} for five configurations: at the four load limit points and at the end ($t=1$).

At $P_1$, the distribution of the attractive force along the fiber's length has a small gradient and the fiber's tips are not yet in contact. This configuration is unstable since a small increase of $\Delta t$ causes a large change of the configuration. Therefore, the arc-length method finds the next equilibrium point for a negative $\Delta t$. The first contact between fiber's tips occurs at $P_2$, after which the system can again sustain positive $\Delta t$ and equilibrium branch $P_2-P_3$ becomes stable. Along this branch of the equilibrium path, the repulsive force develops at the tip and peaks at $P_3$. At this point, the gradient of the interaction force is very steep, the tangent stiffness is near singular, and the system is again unstable. By decreasing the quasi-time, the algorithm continues along equilibrium branch $P_3-P_4$. The gradient of the interaction force along the fiber's length decreases and, at $P_4$, the system can again sustain positive $\Delta t$. The stable equilibrium branch of the considered setup occurs afterward and it is characterized by an adhesion between parts of fibers where $f_2=0$. This is clearly illustrated in the final configuration where the major parts of fibers are straight and the interaction force between them is zero. The transition from $P_4$ to the next equilibrium point is characterized by a rotation and the adhesion of the fiber's tip area. These are depicted with the red and blue curves in Fig.~\ref{fig:LJpoten4xaaNEWaa}a.


\subsubsection{Calibration of the numerical model}

Before scrutinizing the final configuration, let us discuss the calibration of the numerical model by considering the mesh density, the number of integration points, and the cutoff distance. Again, we consider the horizontal component of the reaction force using the ISSIP law.
	
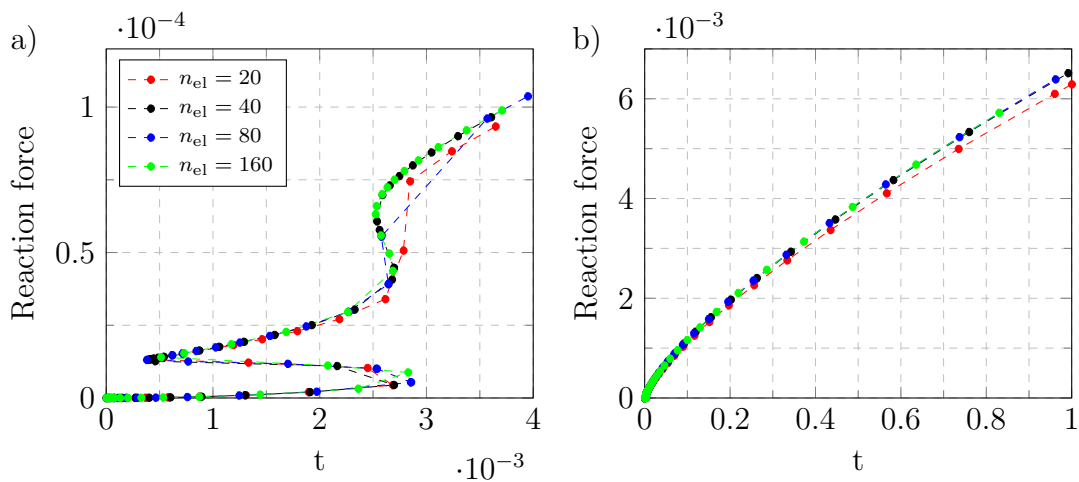
\begin{figure}[h!]
		\centering
		\begin{tikzpicture}
			\begin{axis}[
				xlabel = {t},
				ylabel = {Reaction force},
				ylabel near ticks,
				legend pos=north west,
				legend cell align=left,
				legend style={font=\scriptsize},
				width=0.45\textwidth,
				xmin = 0, xmax =0.004,
				ymin = 0, ymax =0.00012,
				minor y tick num = 1,
				minor x tick num = 1,
				xticklabel style={/pgf/number format/fixed, /pgf/number format/precision=2},
				yticklabel style={/pgf/number format/fixed, /pgf/number format/precision=2},
				clip=false,grid=both];
				\node [text width=1em,anchor=north west] at (rel axis cs: -0.25,1.1){a)};
					\addplot[red,dashed,line width=0.25pt,mark=*,mark size=1.4pt,
				restrict x to domain=0:0.004] table [col sep=comma] {data/dataNumExampleReaction20Elem105.csv};
				
				\addplot[black,dashed,line width=0.25pt,mark=*,mark size=1.4pt,
				restrict x to domain=0:0.004] table [col sep=comma] {data/dataNumExampleReaction40Elem105.csv};
			
				\addplot[blue,dashed,line width=0.25pt,mark=*,mark size=1.4pt,
				restrict x to domain=0:0.004] table [col sep=comma] {data/dataNumExampleReaction80Elem105.csv};
				\addplot[green,dashed,line width=0.25pt,mark=*,mark size=1.4pt,
				restrict x to domain=0:0.004] table [col sep=comma] {data/dataNumExampleReaction160Elem105.csv};
				\legend{$n_\text{el}=20$,$n_\text{el}=40$,$n_\text{el}=80$,$n_\text{el}=160$}
			\end{axis}
		\end{tikzpicture}
		\begin{tikzpicture}
			\begin{axis}[
				xlabel = {t},
				ylabel = {Reaction force},
				ylabel near ticks,
				legend pos=south west,
				legend cell align=left,
				legend style={font=\scriptsize},
				width=0.45\textwidth,
				xmin = 0, xmax =1,
				ymin = 0, ymax =0.007,
				minor y tick num = 1,
				minor x tick num = 1,
				xticklabel style={/pgf/number format/fixed, /pgf/number format/precision=2},
				yticklabel style={/pgf/number format/fixed, /pgf/number format/precision=2},
				clip=false,grid=both];
				\node [text width=1em,anchor=north west] at (rel axis cs: -0.195,1.1){b)};

			\addplot[red,dashed,line width=0.25pt,mark=*,mark size=1.4pt,
		restrict x to domain=0:1] table [col sep=comma] {data/dataNumExampleReaction20Elem105.csv};

				\addplot[black,dashed,line width=0.25pt,mark=*,mark size=1.4pt,
			restrict x to domain=0:1] table [col sep=comma] {data/dataNumExampleReaction40Elem105.csv};

				\addplot[blue,dashed,line width=0.25pt,mark=*,mark size=1.4pt,
				restrict x to domain=0:1] table [col sep=comma] {data/dataNumExampleReaction80Elem105.csv};

						\addplot[green,dashed,line width=0.25pt,mark=*,mark size=1.4pt,
				restrict x to domain=0:1] table [col sep=comma] {data/dataNumExampleReaction160Elem105.csv};

			\end{axis}
		\end{tikzpicture}
		\caption{Snap-to-contact of two cantilever fibers: Horizontal component of the reaction force vs.~quasi-time for different element numbers ($\bar c = 2.5$, $\bar d=2.5$, $n_\text{GP}=100$). a) $t \in [0,4 \cdot 10^{-3}]$, b)  $t \in [0,1]$.}
		\label{fig:LJpoten4xaaNEW1}
\end{figure}

The convergence w.r.t.~the number of elements is shown in Fig.~\ref{fig:LJpoten4xaaNEW1}.
Regrading the considered reaction force, the mesh with 40 elements provides converged values.

\begin{figure}[h!]
		\centering
		\begin{tikzpicture}
			\begin{axis}[
				xlabel = {t},
				ylabel = {Reaction force},
				ylabel near ticks,
				legend pos=north west,
				legend cell align=left,
				legend style={font=\scriptsize},
				width=0.45\textwidth,
				xmin = 0, xmax =0.004,
				ymin = 0, ymax =0.00012,
				minor y tick num = 1,
				minor x tick num = 1,
				xticklabel style={/pgf/number format/fixed, /pgf/number format/precision=2},
				yticklabel style={/pgf/number format/fixed, /pgf/number format/precision=2},
				clip=false,grid=both];
				\node [text width=1em,anchor=north west] at (rel axis cs: -0.25,1.1){a)};
					\addplot[red,dashed,line width=0.25pt,mark=*,mark size=1.4pt,
				restrict x to domain=0:0.004] table [col sep=comma] {data/dataNumExampleReaction80Elem105c005.csv};
					
				\addplot[black,dashed,line width=0.25pt,mark=*,mark size=1.4pt,
				restrict x to domain=0:0.004] table [col sep=comma] {data/dataNumExampleReaction80Elem105c0055.csv};

			\addplot[blue,dashed,line width=0.25pt,mark=*,mark size=1.4pt,
			restrict x to domain=0:0.004] table [col sep=comma] {data/dataNumExampleReaction80Elem105c006.csv};
				
				\addplot[green,dashed,line width=0.25pt,mark=*,mark size=1.4pt,
				restrict x to domain=0:0.004] table [col sep=comma] {data/dataNumExampleReaction80Elem105c007.csv};
			
				\legend{$\bar c=2.50$,$\bar c=2.75$,$\bar c=3.00$,$\bar c=3.50$}
			\end{axis}
		\end{tikzpicture}
		\begin{tikzpicture}
			\begin{axis}[
				xlabel = {t},
				ylabel = {Reaction force},
				ylabel near ticks,
				legend pos=south west,
				legend cell align=left,
				legend style={font=\tiny},
				width=0.45\textwidth,
				xmin = 0, xmax =1,
				ymin = 0, ymax =0.007,
				minor y tick num = 1,
				minor x tick num = 1,
				xticklabel style={/pgf/number format/fixed, /pgf/number format/precision=2},
				yticklabel style={/pgf/number format/fixed, /pgf/number format/precision=2},
				clip=false,grid=both];
\node [text width=1em,anchor=north west] at (rel axis cs: -0.195,1.1){b)};
						
					\addplot[red,dashed,line width=0.25pt,mark=*,mark size=1.4pt,
				restrict x to domain=0:1] table [col sep=comma] {data/dataNumExampleReaction80Elem105c005.csv};
				
				\addplot[black,dashed,line width=0.25pt,mark=*,mark size=1.4pt,
				restrict x to domain=0:1] table [col sep=comma] {data/dataNumExampleReaction80Elem105c0055.csv};
	
				\addplot[blue,dashed,line width=0.25pt,mark=*,mark size=1.4pt,
				restrict x to domain=0:1] table [col sep=comma] {data/dataNumExampleReaction80Elem105c006.csv};
				
				\addplot[green,dashed,line width=0.25pt,mark=*,mark size=1.4pt,
				restrict x to domain=0:1] table [col sep=comma] {data/dataNumExampleReaction80Elem105c007.csv};
			\end{axis}
		\end{tikzpicture}
		\caption{Snap-to-contact of two cantilever fibers: Horizontal component of the reaction force vs.~quasi-time for different values of cutoff distance $c$ ($n_\text{el}=80$, $\bar d=2.5$, $n_\text{GP}=100$, ISSIP): a) $t \in [0,4 \cdot 10^{-3}]$, b)  $t \in [0,1]$.}
		\label{fig:LJpoten4xaaNEW2}
	\end{figure}
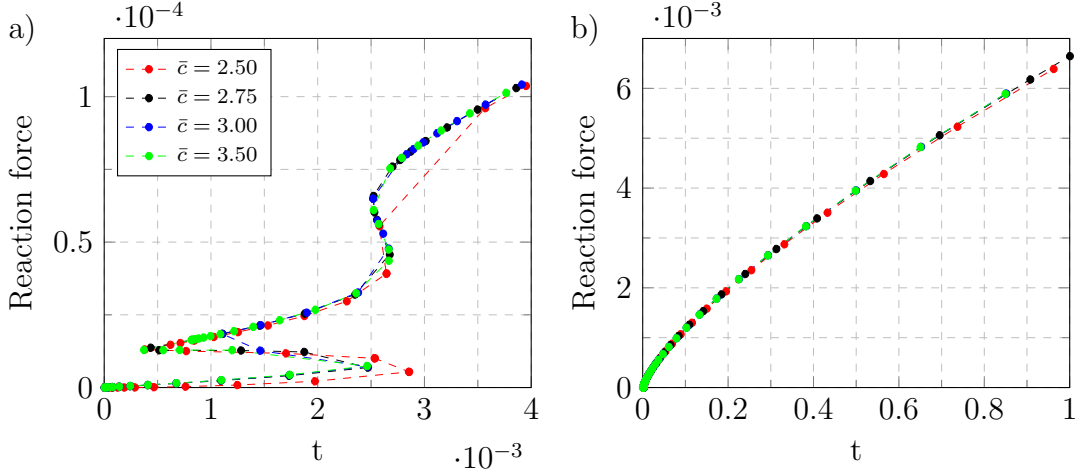

The influence of the fixed cutoff distance $c$ is considered in Fig.~\ref{fig:LJpoten4xaaNEW2}.
	For the reaction force, a normalized cutoff distance of $\bar c = 2.75$ returns practically the same result as $\bar c = 3.5$. Since some parts of the beams are always within the regime of moderate separations, the cutoff distance $\bar c = 2.5$ returns a small reduction in accuracy.

	Finally, we scrutinize the influence of the number of integration points per element, $n_\text{GP}$, using the mesh of 40 elements. 
	To be precise, each element is divided into $n_\text{GP}$ segments, where 1-point Gauss integration (mid-point rule) is applied, see Subsection \ref{sec:cut}.
	%
	\begin{figure}[h!]
		\centering
		\begin{tikzpicture}
			\begin{axis}[
				xlabel = {t},
				ylabel = {Reaction force},
				ylabel near ticks,
				legend pos=north west,
				legend cell align=left,
				legend style={font=\scriptsize},
				width=0.45\textwidth,
				xmin = 0, xmax =0.004,
				ymin = 0, ymax =0.00012,
				minor y tick num = 1,
				minor x tick num = 1,
				xticklabel style={/pgf/number format/fixed, /pgf/number format/precision=2},
				yticklabel style={/pgf/number format/fixed, /pgf/number format/precision=2},
				clip=false,grid=both];
\node [text width=1em,anchor=north west] at (rel axis cs: -0.25,1.1){a)};
					\addplot[red,dashed,line width=0.25pt,mark=*,mark size=1.4pt,
				restrict x to domain=0:0.004] table [col sep=comma] {data/dataNumExampleReaction40Elem50GP105.csv};
				\addplot[blue,dashed,line width=0.25pt,mark=*,mark size=1.4pt,
				restrict x to domain=0:0.004] table [col sep=comma] {data/dataNumExampleReaction40Elem105.csv};
			
				\addplot[green,dashed,line width=0.25pt,mark=*,mark size=1.4pt,
				restrict x to domain=0:0.004] table [col sep=comma] {data/dataNumExampleReaction40Elem200GP105.csv};
				\legend{$n_\text{GP}=50$,$n_\text{GP}=100$,$n_\text{GP}=200$}
			\end{axis}
		\end{tikzpicture}
		\begin{tikzpicture}
			\begin{axis}[
				xlabel = {t},
				ylabel = {Reaction force},
				ylabel near ticks,
				legend pos=south west,
				legend cell align=left,
				legend style={font=\scriptsize},
				width=0.45\textwidth,
				xmin = 0, xmax =1,
				ymin = 0, ymax =0.007,
				minor y tick num = 1,
				minor x tick num = 1,
				xticklabel style={/pgf/number format/fixed, /pgf/number format/precision=2},
				yticklabel style={/pgf/number format/fixed, /pgf/number format/precision=2},
				clip=false,grid=both];
				\node [text width=1em,anchor=north west] at (rel axis cs: -0.195,1.1){b)};
					\addplot[red,dashed,line width=0.25pt,mark=*,mark size=1.4pt,
				restrict x to domain=0:1.2] table [col sep=comma] {data/dataNumExampleReaction40Elem50GP105.csv};
				\addplot[blue,dashed,line width=0.25pt,mark=*,mark size=1.4pt,
				restrict x to domain=0:1.26] table [col sep=comma] {data/dataNumExampleReaction40Elem105.csv};

				\addplot[green,dashed,line width=0.25pt,mark=*,mark size=1.4pt,
				restrict x to domain=0:1.26] table [col sep=comma] {data/dataNumExampleReaction40Elem200GP105.csv};
			\end{axis}
		\end{tikzpicture}
		\caption{Snap-to-contact of two cantilever fibers: Horizontal component of the reaction force vs.~quasi-time for different number of integration points per element ($n_\text{el}=40$,  $\bar c = 2.5$, $\bar d=2.5$, ISSIP). a) $t \in [0,4 \cdot 10^{-3}]$, b)  $t \in [0,1]$.}
		\label{fig:LJpoten4xaaNEW3}
	\end{figure}
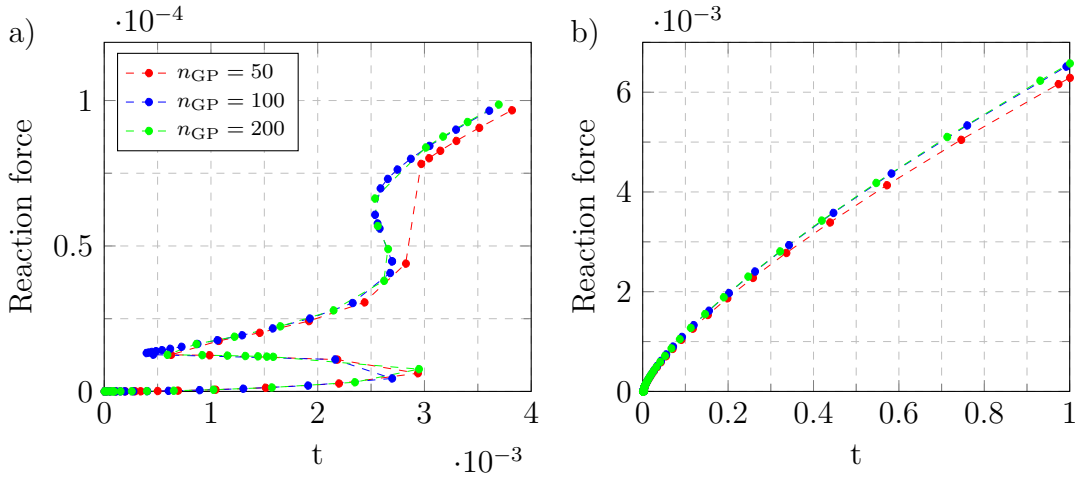
The difference between $n_\text{GP}=100$ and $n_\text{GP}=200$ models is negligible. In this example, $n_\text{GP}=100$ equals 1600 integration points per unit length, $\bar n_\text{GP}=1600$, which is in-line with the observations in Subsection \ref{sec:cut}. Therefore, the model with $n_\text{el}=40$, $\bar c = 2.75$ and $n_\text{GP}=100$ is adopted in the following.

\subsubsection{Parametric analysis}
\label{subsecParAn}
	
With the adopted numerical model, we compare our two section-section laws and vary the inter-support distance in this subsection.

\begin{figure}[h!]
		\centering
		\begin{tikzpicture}
			\begin{axis}[
				xlabel = {t},
				ylabel = {Reaction force},
				ylabel near ticks,
				legend pos=north west,
				legend cell align=left,
				legend style={font=\scriptsize},
				width=0.45\textwidth,
				xmin = 0, xmax =0.004,
				ymin = 0, ymax =0.00012,
				minor y tick num = 1,
				minor x tick num = 1,
				xticklabel style={/pgf/number format/fixed, /pgf/number format/precision=2},
				yticklabel style={/pgf/number format/fixed, /pgf/number format/precision=2},
				clip=false,grid=both];
				\node [text width=1em,anchor=north west] at (rel axis cs: -0.25,1.1){a)};
				\addplot[green,dashed,line width=0.25pt,mark=*,mark size=1.4pt,
				restrict x to domain=0:0.004] table [col sep=comma] {data/dataNumExampleReaction40Elem105c0055IP66.csv};
				\addplot[blue,dashed,line width=0.25pt,mark=*,mark size=1.4pt,
				restrict x to domain=0:0.004] table [col sep=comma] {data/dataNumExampleReaction40Elem105c0055IP669.csv};
				\legend{ISSIP, $\operatorname{D-D_{app}}$}
			\end{axis}
		\end{tikzpicture}
		\begin{tikzpicture}
			\begin{axis}[
				xlabel = {t},
				ylabel = {Reaction force},
				ylabel near ticks,
				legend pos=south west,
				legend cell align=left,
				legend style={font=\tiny},
				width=0.45\textwidth,
				xmin = 0, xmax =1,
				ymin = 0, ymax =0.007,
				minor y tick num = 1,
				minor x tick num = 1,
				xticklabel style={/pgf/number format/fixed, /pgf/number format/precision=2},
				yticklabel style={/pgf/number format/fixed, /pgf/number format/precision=2},
				clip=false,grid=both];
				\node [text width=1em,anchor=north west] at (rel axis cs: -0.195,1.1){b)};
				\addplot[green,dashed,line width=0.25pt,mark=*,mark size=1.4pt,
				restrict x to domain=0:1.26] table [col sep=comma] {data/dataNumExampleReaction40Elem105c0055IP66.csv};
				\addplot[blue,dashed,line width=0.25pt,mark=*,mark size=1.4pt,
				restrict x to domain=0:1.26] table [col sep=comma] {data/dataNumExampleReaction40Elem105c0055IP669.csv};
			\end{axis}
		\end{tikzpicture}
		\caption{Snap-to-contact of two cantilever fibers: Horizontal component of the reaction force vs.~quasi-time for two IP laws ($n_\text{el}=40$, $\bar c = 2.75$ $\bar d=2.5$, $n_\text{GP}=100$). a) $t \in [0,4 \cdot 10^{-3}]$, b)  $t \in [0,1]$.}
		\label{fig:LJpoten4xaaNEW4}
\end{figure}
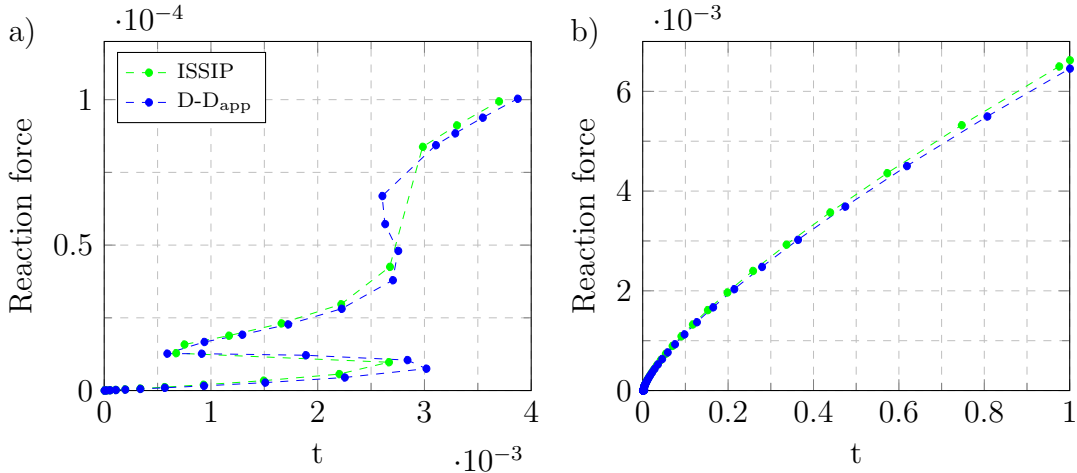

Equilibrium paths of the horizontal reaction force using the ISSIP and $\operatorname{D-D_{app}}$ laws are displayed in Fig.~\ref{fig:LJpoten4xaaNEW4}.
Although similar, these laws return different results.
	As outlined in Subsection \ref{sec:sec-sec}, the $\operatorname{D-D_{app}}$ law provides correct scaling for moderate and large separations, and it is more accurate than the ISSIP law. Therefore, we will employ it to further examine this example.

The previous results were calculated with a fixed cutoff distance $c$. Since we want to model cases when the beams are initially further apart, we adopt a more efficient definition of the cutoff. We combine both approaches considered in Subsection \ref{sec:cut}. The parameters of the cutoff function $c_f$ in \eqqref{eq:1} are $w=20$ and $s=2$. For the short-range regime, this function returns a cutoff distance of $\bar c_f=2.85$ which is a bit larger than the fixed value adopted previously. When the fibers are further apart, the cutoff distance slowly increases as a linear function of $q_\text{2,min}$, according to \eqqref{eq:1}. Since we are here only interested in the final configuration, we do not purse highly accurate forces in the regime of moderate separations. Furthermore, we use two values of $n_\text{GP}$, depending on the minimal gap, as discussed in Subsection \ref{sec:cut}. For moderate separations, we employ a relatively small number of integration points, $n_\text{GP,large}=10$. When a threshold value is crossed (here $\bar q_\text{2,thr}=0.4$), we consider that the fibers are in the short-range regime, and use $n_\text{GP,short}=100$. 


	 
\begin{figure}[h!]
		\centering
		\begin{tikzpicture}
			\begin{axis}[
				xlabel = {t},
				ylabel = {Reaction force},
				ylabel near ticks,
				legend pos=north west,
				legend cell align=left,
				legend style={font=\tiny},
				width=0.45\textwidth,
				xmin = 0, xmax =0.00031,
				ymin = 0, ymax =0.000013,
				minor y tick num = 1,
				minor x tick num = 1,
				restrict x to domain=0:0.00031,
				xticklabel style={/pgf/number format/fixed, /pgf/number format/precision=2},
				yticklabel style={/pgf/number format/fixed, /pgf/number format/precision=2},
				grid=both,clip=false];
				\node [text width=1em,anchor=north west] at (rel axis cs: -0.25,1.1){a)};
				\addplot[black,dashed,line width=0.25pt,mark=*,mark size=1.4pt] table [col sep=comma] {data/dataNumExampleReaction40Elem100gp105c20ip66901inD0045RestrictedDLOnlyStart.csv};
				\addplot[only marks,mark size=3pt,mark=o]
				coordinates{
					( 0.000242687, 0.00000447502053738)
				(0.0000902558983, 0.0000065955373569)
					};
			\end{axis}		
		\end{tikzpicture}
		\begin{tikzpicture}
			\begin{axis}[
				xlabel = {t},
				ylabel = {Reaction force},
				ylabel near ticks,
				legend pos=south west,
				legend cell align=left,
				legend style={font=\tiny},
				width=0.45\textwidth,
				xmin = 0, xmax =1,
				ymin = 0, ymax =0.01,
				minor y tick num = 1,
				minor x tick num = 1,		
				restrict y to domain=0:0.01,
				xticklabel style={/pgf/number format/fixed, /pgf/number format/precision=2},
				yticklabel style={/pgf/number format/fixed, /pgf/number format/precision=2},
				grid=both,clip=false];			
			\node [text width=1em,anchor=north west] at (rel axis cs: -0.25,1.1){b)};		
				\addplot[black,dashed,line width=0.25pt,mark=*,mark size=1.4pt,
				restrict x to domain=0:1] table [col sep=comma] {data/dataNumExampleReaction40Elem100gp105c20ip66901inD0045.csv};			
			\end{axis}
		\end{tikzpicture}
		\caption{Snap-to-contact of two cantilever fibers: Horizontal component of the reaction force vs.~quasi-time for $\bar d=2.25$. The load limit points are marked in a). ($\operatorname{D-D_{app}}$, $n_\text{el}=40$, $c_f$, $n_\text{GP,m/s}=10/100$): a) $t \in [0,3 \cdot 10^{-4}]$, b)  $t \in [0,1]$.}
		\label{fig:LJpoten4xaaNEW42xxd}
\end{figure}
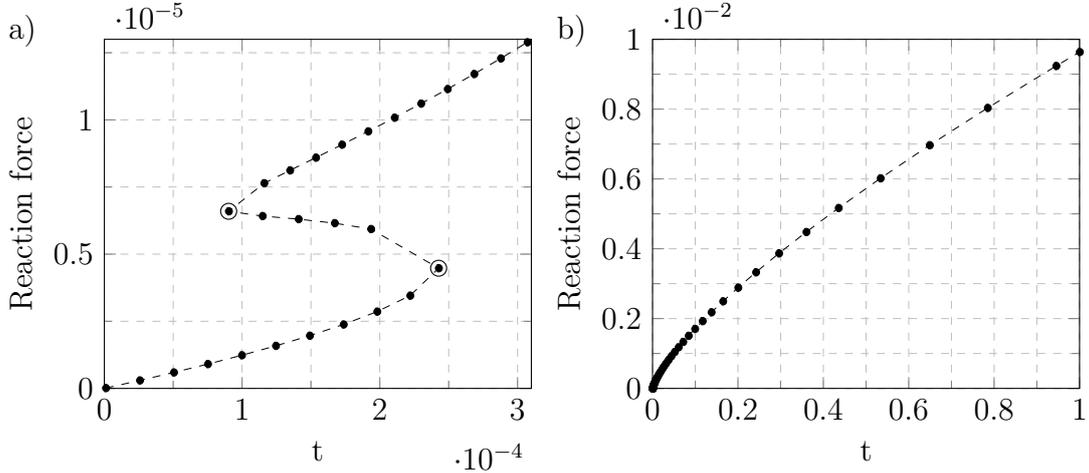

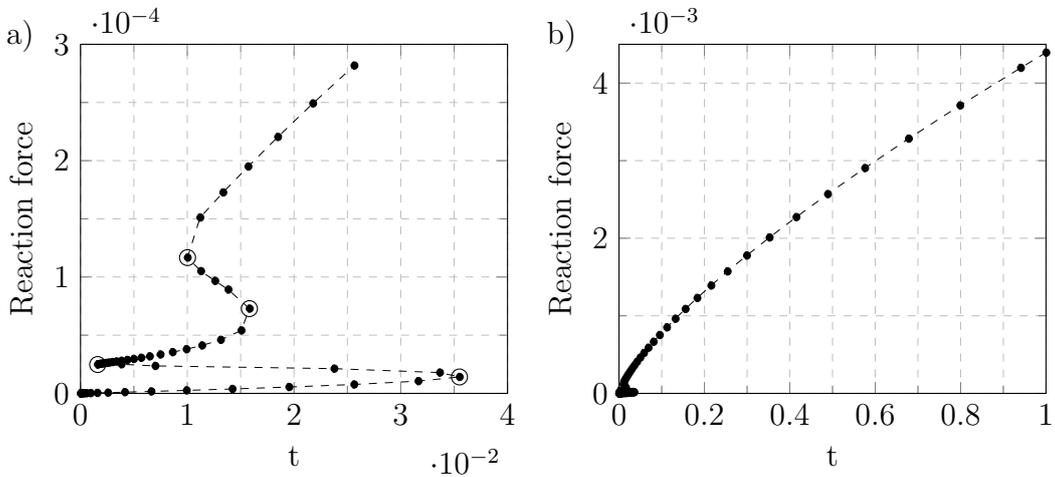
\begin{figure}[h!]
		\centering
		\begin{tikzpicture}
			\begin{axis}[
				xlabel = {t},
				ylabel = {Reaction force},
				ylabel near ticks,
				legend pos=north west,
				legend cell align=left,
				legend style={font=\tiny},
				width=0.45\textwidth,
				xmin = 0, xmax =0.04,
				ymin = 0, ymax =0.0003,
				minor y tick num = 1,
				minor x tick num = 1,
				restrict y to domain=0:0.0003,
				xticklabel style={/pgf/number format/fixed, /pgf/number format/precision=2},
				yticklabel style={/pgf/number format/fixed, /pgf/number format/precision=2},grid=both,clip=false];
				\node [text width=1em,anchor=north west] at (rel axis cs: -0.2,1.1){a)};
				\addplot[black,dashed,line width=0.25pt,mark=*,mark size=1.4pt,restrict x to domain=0:0.04] table [col sep=comma] {data/dataNumExampleReaction40Elem100gp105c20ip66901inD006.csv};		
				\addplot[only marks,mark size=3pt,mark=o]
				coordinates{
					( 0.035498213,	0.0000140998 )
					(0.001606211,	0.0000247645)
					(0.015822393,	0.0000728396)
					(0.010022865,	0.000116689)
				};			
			\end{axis}
		\end{tikzpicture}
		\begin{tikzpicture}
			\begin{axis}[
				xlabel = {t},
				ylabel = {Reaction force},
				ylabel near ticks,
				legend pos=south west,
				legend cell align=left,
				legend style={font=\tiny},
				width=0.45\textwidth,
				xmin = 0, xmax =1,
				ymin = 0, ymax =0.0045,
				minor y tick num = 1,
				minor x tick num = 1,		
				restrict y to domain=0:0.0045,
				xticklabel style={/pgf/number format/fixed, /pgf/number format/precision=2},
				yticklabel style={/pgf/number format/fixed, /pgf/number format/precision=2},
				grid=both,clip=false];			
				\node [text width=1em,anchor=north west] at (rel axis cs: -0.195,1.1){b)};		
				\addplot[black,dashed,line width=0.25pt,mark=*,mark size=1.4pt,
				restrict x to domain=0:1] table [col sep=comma] {data/dataNumExampleReaction40Elem100gp105c20ip66901inD006.csv};
			\end{axis}
		\end{tikzpicture}
		\caption{Snap-to-contact of two cantilever fibers: Horizontal component of the reaction force vs.~quasi-time for $\bar d=3$. The load limit points are marked in a).  ($\operatorname{D-D_{app}}$, $n_\text{el}=40$, $c_f$, $n_\text{GP,m/s}=10/100$): a) $t \in [0,4 \cdot 10^{-2}]$, b)  $t \in [0,1]$.}
		\label{fig:LJpoten4xaaNEW42xx}
	\end{figure}

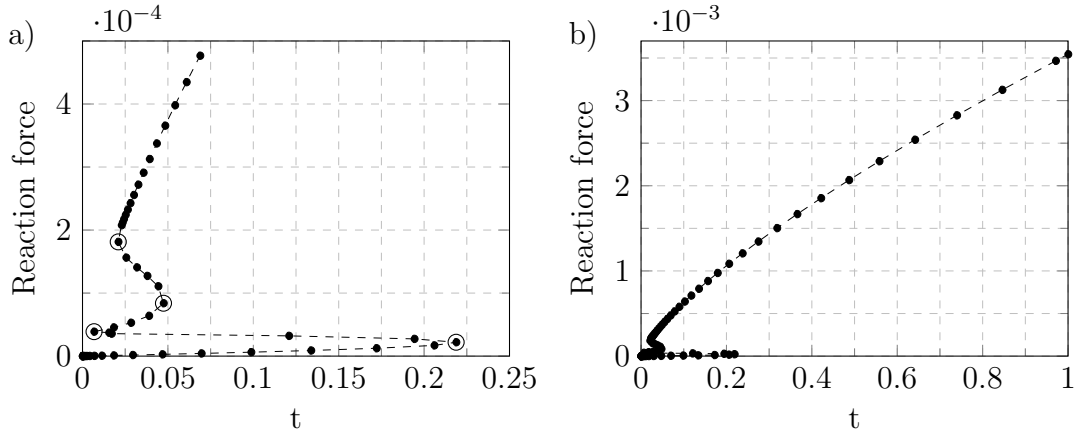
\begin{figure}[h!]
		\centering
		\begin{tikzpicture}
			\begin{axis}[
				xlabel = {t},
				ylabel = {Reaction force},
				ylabel near ticks,
				legend pos=north west,
				legend cell align=left,
				legend style={font=\tiny},
				width=0.45\textwidth,
				height=0.36\textwidth,
				xmin = 0, xmax =0.25,
				ymin = 0, ymax =0.0005,
				minor y tick num = 1,
				minor x tick num = 1,
				restrict y to domain=0:0.0005,
				xticklabel style={/pgf/number format/fixed, /pgf/number format/precision=2},
				yticklabel style={/pgf/number format/fixed, /pgf/number format/precision=2},grid=both,clip=false];
				\node [text width=1em,anchor=north west] at (rel axis cs: -0.2,1.1){a)};
				\addplot[black,dashed,line width=0.25pt,mark=*,mark size=1.4pt,restrict x to domain=0:0.25] table [col sep=comma] {data/dataNumExampleReaction40Elem100gp105c20ip66901inD007.csv};
								\addplot[only marks,mark size=3pt,mark=o]
				coordinates{
					( 0.218747107,	0.0000220131
					 )
					(0.006760034,	0.0000387678
					)
					(0.047408464,	0.0000838984
					)
					(0.020928452,	0.000181232
					)
				};					
			\end{axis}
		\end{tikzpicture}
		\begin{tikzpicture}
			\begin{axis}[
				xlabel = {t},
				ylabel = {Reaction force},
				ylabel near ticks,
				legend pos=south west,
				legend cell align=left,
				legend style={font=\tiny},
				width=0.45\textwidth,
				height=0.36\textwidth,
				xmin = 0, xmax =1,
				ymin = 0, ymax =0.0037,
				minor y tick num = 1,
				minor x tick num = 1,
				restrict y to domain=0:0.0038,		
				xticklabel style={/pgf/number format/fixed, /pgf/number format/precision=2},
				yticklabel style={/pgf/number format/fixed, /pgf/number format/precision=2},grid=both,clip=false];			
				\node [text width=1em,anchor=north west] at (rel axis cs: -0.195,1.1){b)};			
				\addplot[black,dashed,line width=0.25pt,mark=*,mark size=1.4pt,
				restrict x to domain=0:1.26] table [col sep=comma] {data/dataNumExampleReaction40Elem100gp105c20ip66901inD007.csv};			
			\end{axis}
		\end{tikzpicture}
		\caption{Snap-to-contact of two cantilever fibers: Horizontal component of the reaction force vs.~quasi-time for $\bar d=3.5$. The load limit points are marked on a).  ($\operatorname{D-D_{app}}$, $n_\text{el}=40$, $c_f$, $n_\text{GP,m/s}=10/100$): a) $t \in [0,0.25]$, b)  $t \in [0,1]$.}
		\label{fig:LJpoten4xaaNEW42x}
	\end{figure}

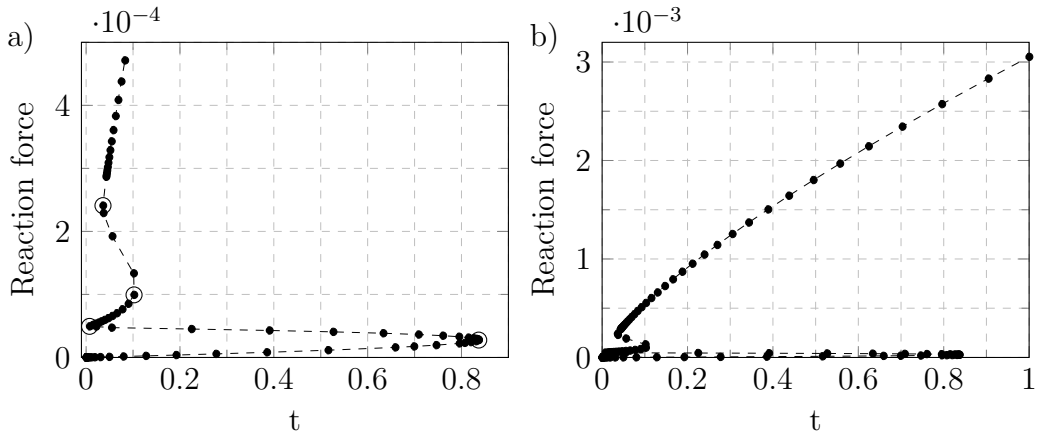
\begin{figure}[h!]
		\centering
		\begin{tikzpicture}
			\begin{axis}[
				xlabel = {t},
				ylabel = {Reaction force},
				ylabel near ticks,
				legend pos=north west,
				legend cell align=left,
				legend style={font=\tiny},
				width=0.45\textwidth,
				height=0.36\textwidth,
				xmin = -0.01, xmax =0.9,
				ymin = 0, ymax =0.0005,
				minor y tick num = 1,
				minor x tick num = 1,
				restrict y to domain=0:0.0005,
				xticklabel style={/pgf/number format/fixed, /pgf/number format/precision=2},
				yticklabel style={/pgf/number format/fixed, /pgf/number format/precision=2},grid=both,clip=false];
				\node [text width=1em,anchor=north west] at (rel axis cs: -0.2,1.1){a)};
				\addplot[black,dashed,line width=0.25pt,mark=*,mark size=1.4pt,restrict x to domain=0:0.9] table [col sep=comma] {data/dataNumExampleReaction40Elem100gp105c20ip66901inD008.csv};
				\addplot[only marks,mark size=3pt,mark=o]
				coordinates{
					( 0.836736561,	0.0000275981
					)
					(0.00735722,	0.0000493337
				)
					(0.102147403,	0.0000993092
					)
					(0.035933807,	0.000240974
					)
				};				
			\end{axis}
		\end{tikzpicture}
		\begin{tikzpicture}
			\begin{axis}[
				xlabel = {t},
				ylabel = {Reaction force},
				ylabel near ticks,
				legend pos=south west,
				legend cell align=left,
				legend style={font=\tiny},
				width=0.45\textwidth,
				height=0.36\textwidth,
				xmin = 0, xmax =1,
				ymin = 0, ymax =0.0032,
				minor y tick num = 1,
				minor x tick num = 1,
				restrict y to domain=0:0.0032,		
				xticklabel style={/pgf/number format/fixed, /pgf/number format/precision=2},
				yticklabel style={/pgf/number format/fixed, /pgf/number format/precision=2},grid=both,clip=false];			
				\node [text width=1em,anchor=north west] at (rel axis cs: -0.195,1.1){b)};		
				\addplot[black,dashed,line width=0.25pt,mark=*,mark size=1.4pt,
				restrict x to domain=0:1.26] table [col sep=comma] {data/dataNumExampleReaction40Elem100gp105c20ip66901inD008.csv};			
			\end{axis}
		\end{tikzpicture}
		\caption{Snap-to-contact of two cantilever fibers: Horizontal component of the reaction force vs.~quasi-time for $\bar d=4$. The load limit points are marked on a). ($\operatorname{D-D_{app}}$, $n_\text{el}=40$, $c_f$, $n_\text{GP,m/s}=10/100$): a) $t \in [0,0.9]$, b)  $t \in [0,1]$.}
		\label{fig:LJpoten4xaaNEW42xxx}
\end{figure}

We vary the inter-support distance $d$ and observe the equilibrium path of the horizontal reaction force 
in Figs.~\ref{fig:LJpoten4xaaNEW42xxd}, \ref{fig:LJpoten4xaaNEW42xx}, \ref{fig:LJpoten4xaaNEW42x}, and \ref{fig:LJpoten4xaaNEW42xxx}. 
Qualitatively, the fibers behave similarly for all the considered inter-support distances. There are always four load limit points on the equilibrium path, except for the case with $\bar d=2.25$, where only two exists. This decrease of the number of load limit points with a decrease of $d$ suggests that the limit points fully vanish for smaller $d$. To test this hypothesis, we have run an additional simulation with $\bar d=2.1$. Indeed, the NR algorithm can trace the complete equilibrium path since there are no load limit points in this case. These results are omitted for brevity. 

From the results in Figs.~\ref{fig:LJpoten4xaaNEW}, \ref{fig:LJpoten4xaaNEW42xxd}, \ref{fig:LJpoten4xaaNEW42xx}, \ref{fig:LJpoten4xaaNEW42x}, and \ref{fig:LJpoten4xaaNEW42xxx}, we observe that the first load limit point, $P_1$, approaches $t=1$ as $d$ increases. Also, the difference between the two upper limit points increases with $d$. However, the most significant difference between the cases with different $d$ lies in the value of the reaction force: Its value at $P_1$ increases while its value at $t=1$ decreases with the increase of $d$. If $P_1$ is reached for $t\ge1$, the fibers do not snap to contact for the adopted LJ potential. In our simulations, this occurs for approximately $\bar d>4$. Nevertheless, the final adhered configuration at $t=t_\text{max}=1$ for $\bar d>4$ can be calculated using quasi-statics in the following way: We first allow $t>t_\text{max}$ to find and turn around $P_1$, and then stop the calculation when the algorithm reaches the load level $t=t_\text{max}$ again. For example, consider the case plotted in Fig.~\ref{fig:LJpoten4xaaNEW42xxx}: If the quasi-time parameter is limited to $t_\text{max}=0.8$, finding the adhered configuration would require implementation of the described procedure.

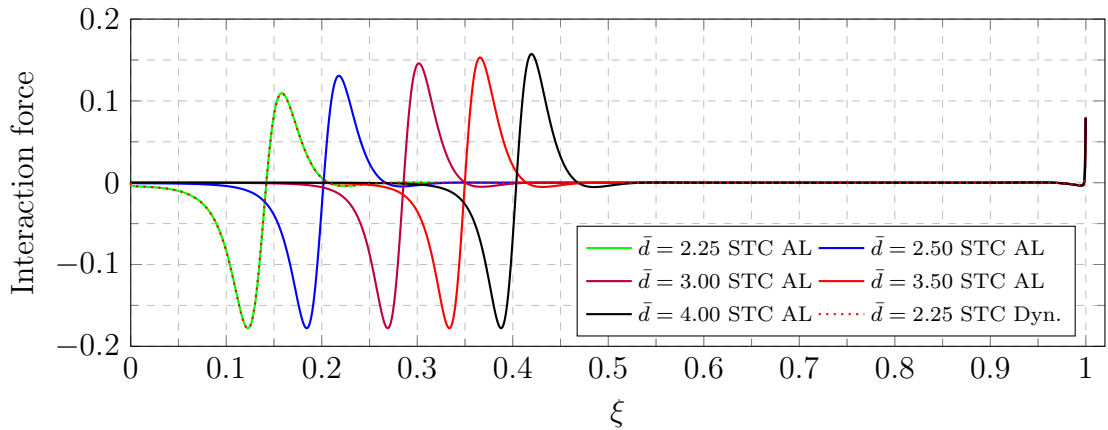
\begin{figure}[h!]
	\centering
	\begin{tikzpicture}
		\begin{axis}[
			xlabel = {$\xi$},
			ylabel = {Interaction force},
			ylabel near ticks,
			legend pos=south east,
			legend columns=2, 
			legend cell align=left,
			legend style={font=\scriptsize},
			width=0.9\textwidth,
			height=0.37\textwidth,
			xmin = 0, xmax =1.02,
			ymin = -0.2, ymax =0.2,
			minor y tick num = 1,
			minor x tick num = 1,
			xticklabel style={/pgf/number format/fixed, /pgf/number format/precision=2},
			yticklabel style={/pgf/number format/fixed, /pgf/number format/precision=2},
			clip=false,grid=both];
			\addplot[green,thick] 
			table [col sep=comma] {data/dataNumExampleReaction40Elem100gp105c20ip66901inD0045IFdistt1.csv};	
			\addplot[blue,thick] 
			table [col sep=comma] {data/dataNumExampleReaction40Elem100gp105c20ip66901inD005IFdistt1.csv};	
			\addplot[purple,thick] 
			table [col sep=comma] {data/dataNumExampleReaction40Elem100gp105c20ip66901inD006IFdistt1.csv};	
			\addplot[red,thick] 
			table [col sep=comma] {data/dataNumExampleReaction40Elem100gp105c20ip66901inD007IFdistt1.csv};	
			\addplot[black,thick] table [col sep=comma] {data/dataNumExampleReaction40Elem100gp105c20ip66901inD008IFdistt1.csv};	
			
%
			\addplot[red,dotted,thick] 
			table [col sep=comma]
			{data/dataNumExample40Elem100gp105c20ip66901inD0045IFdistDynamics.csv};	
%

				\legend{$\bar d = 2.25$ STC AL,$\bar d = 2.50$ STC AL,$\bar d = 3.00$ STC AL,$\bar d = 3.50$ STC AL,$\bar d = 4.00$ STC AL,$\bar d = 2.25$ STC Dyn.}
		\end{axis}
	\end{tikzpicture}
	\caption{Snap-to-contact of two cantilever fibers: Distribution of the interaction force at $t=1$ for five different values of inter-support distance $d$. We consider STC static case using the AL method. For $\bar d=2.25$, the solution obtained by using dynamic analysis is plotted as well. ($\operatorname{D-D_{app}}$, $n_\text{el}=40$, $c_f$, $n_\text{GP,m/s}=10/100$)}
	\label{fig:LJpoten4xaaNEW42xas1}
\end{figure}

Next, we plot the distribution of the interaction force $\ve{f}_2$ for $t=1$ and all five considered values of $d$ in Fig.~\ref{fig:LJpoten4xaaNEW42xas1}.
We observe that the maximum attraction force is similar for all simulations, while the maximum repulsion force increases with $d$. Also, there is a strong end-effect that is characterized by the peak repulsive force at the fiber's free end.
	
\clearpage
	
	\subsubsection{Comparison with other approaches}
	
To verify the results obtained with the static STC simulations, we have also run a peeling simulation between these two fibers. A peeling simulation is done as in \cite{2024borkovićb}, using the NR algorithm. In essence, we place fibers close to the equilibrium distance ($\bar d_\text{init} = 2.0425$) and separate the fibers by increasing the inter-support distance. 

\begin{figure}[h!]
	\centering
	\begin{tikzpicture}
		\begin{axis}[
			xlabel = {$\bar d$},
			ylabel = {Reaction force},
			ylabel near ticks,
			legend pos=north east,
			legend cell align=left,
			legend style={font=\scriptsize},
			width=0.6\textwidth,
			height=0.45\textwidth,
			ymin = 0, ymax = 0.02,
			minor y tick num = 1,
			minor x tick num = 1,		
			xticklabel style={/pgf/number format/fixed, /pgf/number format/precision=2},
			yticklabel style={/pgf/number format/fixed, /pgf/number format/precision=2},
			grid=both];			
			\addplot[black,line width=0.5pt] table [col sep=comma] {data/dataNumExamplePeelOffReaction40Elem100gp105c20ip66901vsdbar.csv};	
			
			\addplot[only marks,mark size=3pt,mark=square] coordinates{(2.25,0.0096350966870744
				)};
			
			\addplot[only marks, mark size=3pt,mark=o] coordinates{(2.5,0.00645919449564628
				)};
			\addplot[only marks, mark size=3pt,mark=triangle] coordinates{(3,0.00439308110145351
				)};
			\addplot[only marks, mark size=3pt,mark=asterisk] coordinates{(3.5,0.00354378044502742
				)};
			\addplot[only marks, mark size=3pt,mark=diamond] coordinates{(4,0.003052314562775
				)};
				\addplot[black, only marks,mark size=1.5pt] coordinates{(2.25,0.0096350966870744
					)};
			\legend{Peeling NR,$\bar d = 2.25$ STC AL,$\bar d = 2.50$ STC AL,$\bar d = 3.00$ STC AL,$\bar d = 3.50$ STC AL,$\bar d = 4.00$ STC AL,$\bar d = 2.25$ STC Dyn.}		
		\end{axis}
	\end{tikzpicture}
	\caption{Snap-to-contact of two cantilever fibers: Horizontal component of the reaction force vs.~$\bar d$ obtained with the static peeling simulation, static STC simulation, and dynamics. ($\operatorname{D-D_{app}}$, $n_\text{el}=40$, $c_f$, $n_\text{GP,m/s}=10/100$) }
	\label{fig:LJpoten4xaaNEW42xaa}
\end{figure}
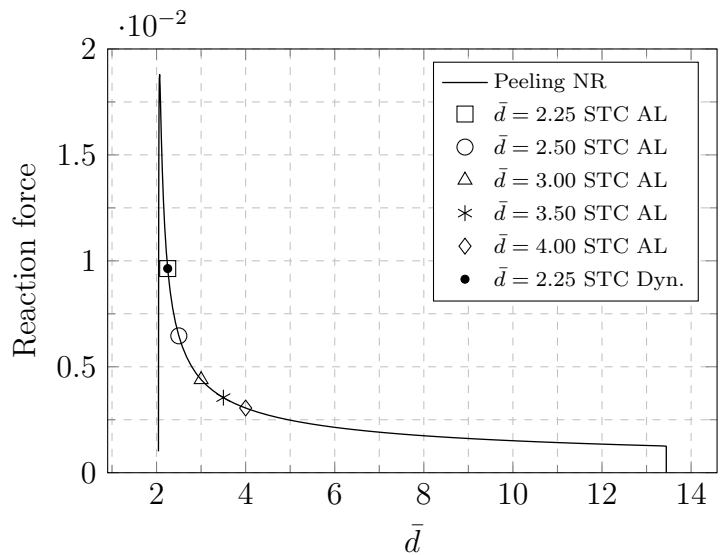

The value of the horizontal reaction force, obtained with the peeling simulation is plotted in Fig.~\ref{fig:LJpoten4xaaNEW42xaa} as a function of the inter-support distance $d$. 
A typical force peak at the initiation of the peeling is observed. It is followed by the decrease of the force as the peeling develops while pull-off occurs for $\bar d \approx 13.45$. For a comparison, the values of the reaction force that are obtained for specific values of $d$ using the arc-length STC analysis in Subsection \ref{subsecParAn} are plotted as well. The results are in full agreement, which provides confidence that our approach is correct. 

\begin{figure}[h!]
	\centering
	\begin{tikzpicture}
		\begin{axis}[
			xlabel = {$\xi$},
			ylabel = {Interaction force},
			ylabel near ticks,
			legend pos=north west,
			legend columns=4, 
			legend cell align=left,
			legend style={font=\scriptsize,at={(axis cs:0.1,0.2)},anchor=south west},
			width=0.9\textwidth,
			height=0.4\textwidth,
			xmin = 0, xmax =1.02,
			ymin = -0.2, ymax =0.25,
			minor y tick num = 1,
			minor x tick num = 1,
			xticklabel style={/pgf/number format/fixed, /pgf/number format/precision=2},
			yticklabel style={/pgf/number format/fixed, /pgf/number format/precision=2},
			clip=false,grid=both];

			\addplot[green,thick] table [col sep=comma] {data/dataNumExamplePeel40Elem100gp105c20ip66901inD0045IFdistt1.csv};	
			\addplot[blue,thick] table [col sep=comma] {data/dataNumExamplePeel40Elem100gp105c20ip66901inD005IFdistt1.csv};	
			\addplot[purple,thick] table [col sep=comma] {data/dataNumExamplePeel40Elem100gp105c20ip66901inD006IFdistt1.csv};	
			\addplot[red,thick] table [col sep=comma] {data/dataNumExamplePeel40Elem100gp105c20ip66901inD007IFdistt1.csv};	
			\addplot[black,thick] table [col sep=comma] {data/dataNumExamplePeel40Elem100gp105c20ip66901inD008IFdistt1.csv};	
			
			\addplot[brown,dashed,thick] table [col sep=comma] {data/dataNumExamplePeel40Elem100gp105c20ip66901inD01IFdistt1.csv};	
			\addplot[cyan, dashed,thick] table [col sep=comma] {data/dataNumExamplePeel40Elem100gp105c20ip66901inD016IFdistt1.csv};	
			\addplot[magenta,dashed,thick] table [col sep=comma] {data/dataNumExamplePeel40Elem100gp105c20ip66901inD0269IFdistt1.csv};	
			
			\legend{$\bar d \approx 2.25$ , $\bar d \approx 2.50$ ,$\bar d \approx 3.00$ ,$\bar d \approx 3.55$ ,$\bar d \approx 4.10$ , $\bar d \approx 5.00$,$\bar d \approx 8.00$, $\bar d \approx 13.45$ }
		\end{axis}
	\end{tikzpicture}
	\caption{Snap-to-contact of two cantilever fibers: Distribution of the interaction force at $t=1$ for eight different values of inter-support distance $d$. We consider static peeling simulations using the NR method. ($\operatorname{D-D_{app}}$, $n_\text{el}=40$, $c_f$, $n_\text{GP,m/s}=10/100$)}
	\label{fig:LJpoten4xaaNEW42xas1x}
\end{figure}
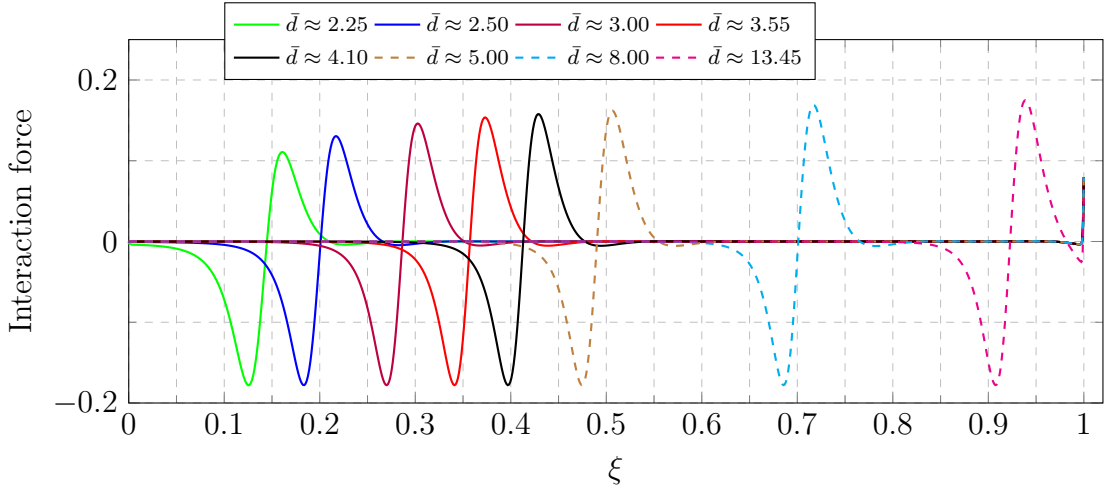

The resulting distributions of the interaction force for eight different inter-support distances are are plotted in Fig.~\ref{fig:LJpoten4xaaNEW42xas1x}. 
These results are practically the same as those obtained with STC analysis in Subsection \ref{subsecParAn} for $\bar d \le 4$, see Fig.~\ref{fig:LJpoten4xaaNEW42xas1}. An increase in repulsive force with an increase in inter-support distance continues for $\bar d > 4$.


To asses the accuracy of the quasi-static approach, we have run an implicit dynamic simulation for $\bar d = 2.25$. The dynamic simulation is done without gradually increasing the LJ potential. Due to inertial effects, the simulation results with the collision, oscillation, and adhesion of fibers. The dynamic behavior will be considered in more detail in the following examples. Here, we focus on the final steady-state configuration when the kinetic energy completely dissipates, which can be considered as a static case. The distribution of the interaction force at the final configuration is plotted in Fig.~\ref{fig:LJpoten4xaaNEW42xas1}, while the value of the horizontal reaction force is plotted in Fig.~\ref{fig:LJpoten4xaaNEW42xas1}. The results obtained with the static analysis are fully aligned with those from the dynamic analysis, which suggest that the influence of inertia on the final configuration is negligible in this setting.

\subsection{Dynamic snap-to-contact of two free fibers}

\begin{figure}[h!]
		\begin{center}
			\includegraphics[width=0.4\textwidth]{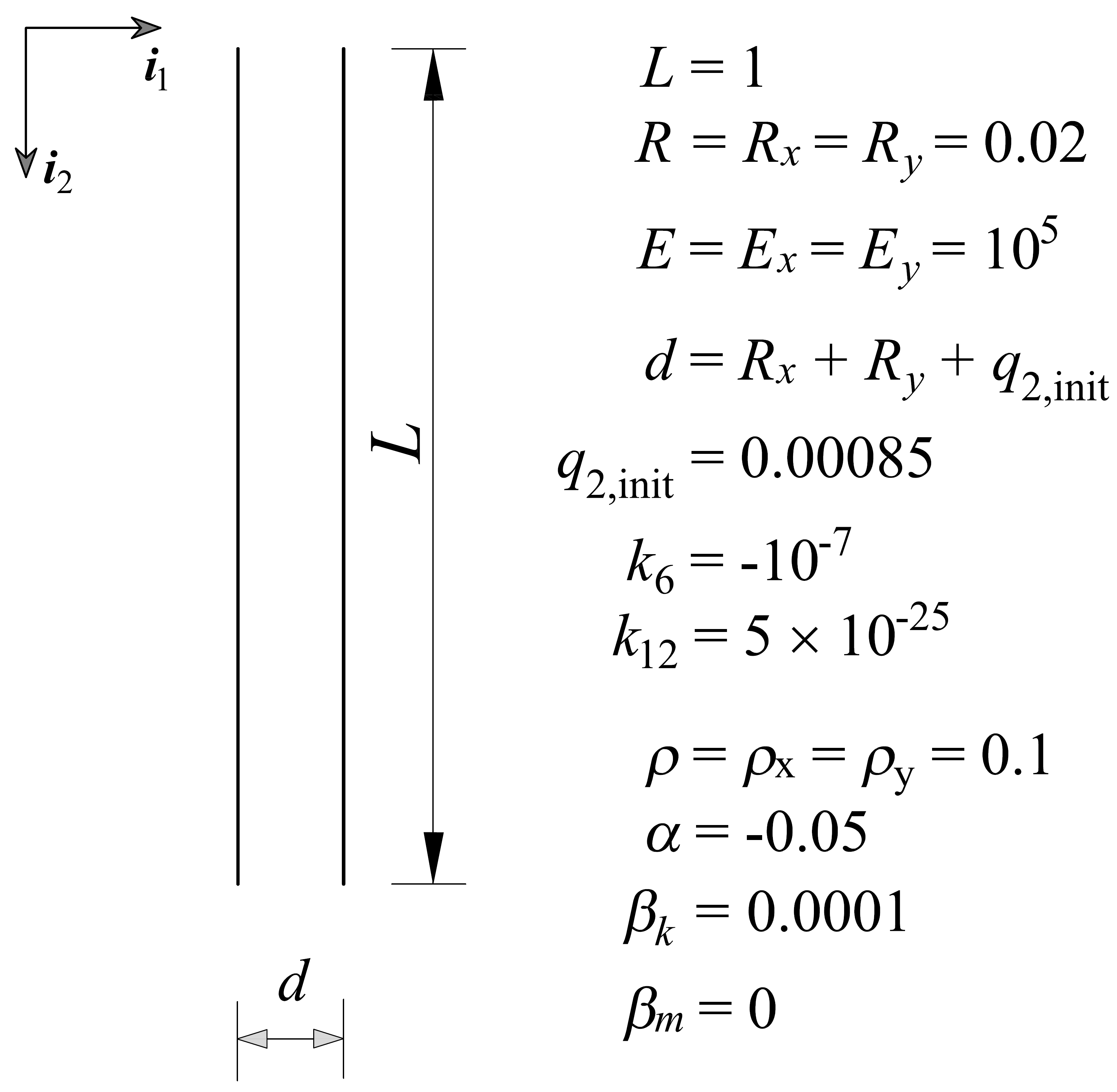}
			\caption{Dynamic snap-to-contact of two free fibers: Problem setup and parameters}
			\label{fig:intro1}
		\end{center}
\end{figure}

Next we investigate dynamic snap-to-contact of two free deformable fibers interacting via the LJ potential. Two identical, free, straight, and parallel fibers that are relatively close to each other are shown in Fig.~\ref{fig:intro1}.
	%
	%
%
%
If the fibers are relatively far apart, a strong collision can occur, which significantly complicates calculation and increases the computational time. Since we are mainly interested in the final equilibrium configuration, the initial distance between the fibers is near the disk-infinite cylinder static equilibrium distance. For the adopted values of physical constants, the disk-infinite cylinder equilibrium gap is $q_\text{2,eq}=0.00083913$, see Eq.~\eqref{eq:2}, and we set the initial gap to $q_{2,\text{init}}=0.00085$. The example exclusively falls into the small separation regime; therefore, we employ the ISSIP law.
	
Stiffness proportional damping with $\beta_k=0.0001$ is applied, and it gives approximately 1 $\%$ of relative damping in the first eigenmode. Furthermore, we use the recommended numerical damping for the HHT-$\alpha$ method, i.e.,~$\alpha=-0.05$.
	
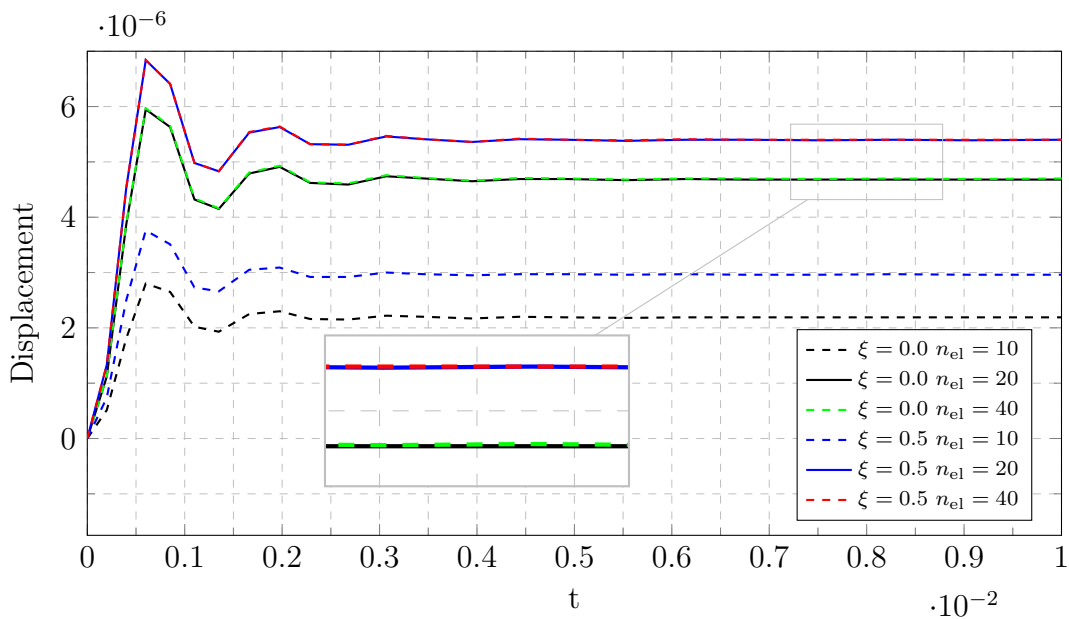
\begin{figure}[h]
		\centering
		\begin{tikzpicture}[spy using outlines=	{rectangle, magnification=2.0, connect spies}]
			\begin{axis}[
				xlabel = {t},
				ylabel = {Displacement},
				ylabel near ticks,
				legend pos=south east,
				legend cell align=left,
				legend style={font=\scriptsize},
				width=0.9\textwidth,
				height=0.5\textwidth,
				ymin = -0.00000175, ymax = 0.000007,
				xmin = 0, xmax =0.01,
				minor y tick num = 1,
				minor x tick num = 1,		
				xticklabel style={/pgf/number format/fixed, /pgf/number format/precision=2},
				yticklabel style={/pgf/number format/fixed, /pgf/number format/precision=2},grid=both];	
				\coordinate (spypoint) at (axis cs:0.008,0.000005);
				\coordinate (magnifyglass) at (axis cs:0.004,0.0000005);		
				\addplot[black,thick,dashed] table [col sep=comma] {data/dataStraightToContactXDispXi0E1041GP160C0045Kprig00001Mprig0maxDTunrestr.csv};		
				\addplot[black,thick] table [col sep=comma] {data/dataStraightToContactXDispXi0E2041GP160C0045Kprig00001Mprig0maxDTunrestr.csv};
				\addplot[green,dashed,thick] table [col sep=comma] {data/dataStraightToContactXDispXi0E4041GP160C0045Kprig00001Mprig0maxDTunrestr.csv};
			\addplot[blue,thick,dashed] table [col sep=comma] {data/dataStraightToContactXDispXi05E1041GP160C0045Kprig00001Mprig0maxDTunrestr.csv};
				\addplot[blue,thick] table [col sep=comma] {data/dataStraightToContactXDispXi05E2041GP160C0045Kprig00001Mprig0maxDTunrestr.csv};			
			\addplot[red,dashed,thick] table [col sep=comma] {data/dataStraightToContactXDispXi05E4041GP160C0045Kprig00001Mprig0maxDTunrestr.csv};	
				\legend{$\xi=0.0$ $n_\text{el}=10$,$\xi=0.0$ $n_\text{el}=20$,$\xi=0.0$  $n_\text{el}=40$,$\xi=0.5$ $n_\text{el}=10$,$\xi=0.5$ $n_\text{el}=20$,$\xi=0.5$ $n_\text{el}=40$}		
			\end{axis}
			\spy [lightgray, height=2cm,width=4cm] on (spypoint) in node[fill=white] at (magnifyglass);
		\end{tikzpicture}
		\caption{Dynamic snap-to-contact of two free fibers: Horizontal displacement component of the left fiber at two characteristic points vs.~$t$ for different numbers of elements $n_\text{el}$. (ISSIP, $\bar c=2.25$, $n_\text{GP}=160$, $\Delta t_\text{max}=\infty$) }
		\label{fig:LJpoten4xaaNEW42xaavtta}
	\end{figure}

First, let us calibrate the numerical model by considering the spatial discretization, the number of integration points, the cutoff distance, and the maximum time increment. We observe the horizontal displacement component of two characteristic points at the start ($\xi=0$) and at the middle ($\xi=0.5$) of the left fiber. The comparison of results for three meshes is displayed in Fig.~\ref{fig:LJpoten4xaaNEW42xaavtta}.
The beams do not deform significantly in this setup, and the mesh with 20 elements provides satisfactory accuracy. With this analysis, we have also tested the required number of integration points, since it is fixed to $n_\text{GP}=160$ for the results in Fig.~\ref{fig:LJpoten4xaaNEW42xaavtta}.

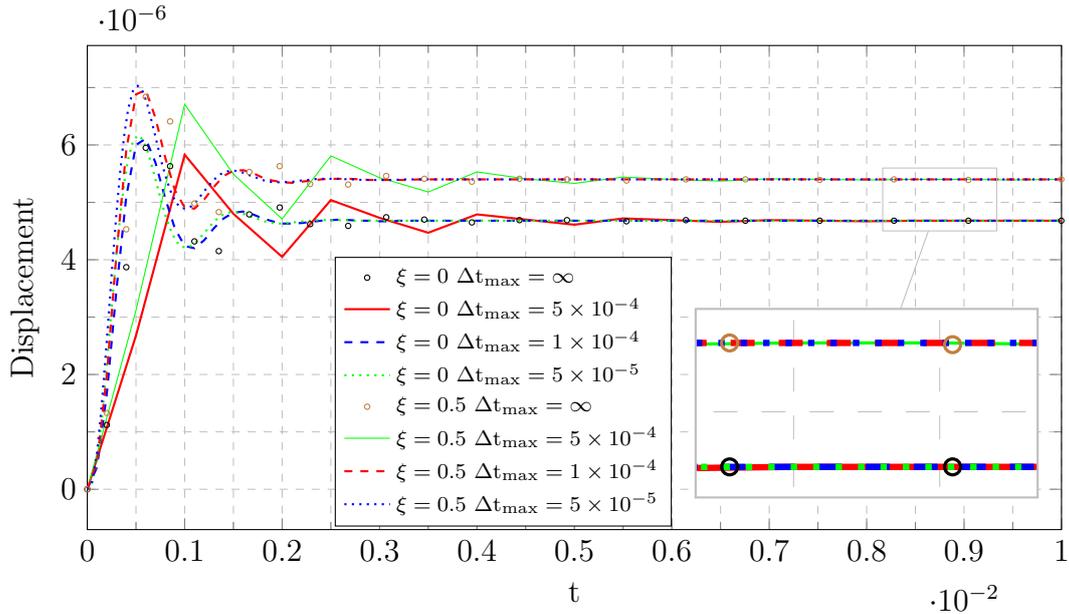
\begin{figure}[h]
	\centering
	\begin{tikzpicture}[spy using outlines=	{rectangle, magnification=3, connect spies}]
		\begin{axis}[
			xlabel = {t},
			ylabel = {Displacement},
			ylabel near ticks,
			legend pos=south east,
			legend cell align=left,
			legend style={font=\scriptsize,at={(axis cs:0.006,-0.00000065)},anchor=south east},
			width=0.9\textwidth,
			height=0.5\textwidth,
			xmin = 0.0, xmax =0.01,
			minor y tick num = 1,
			minor x tick num = 1,		
			xticklabel style={/pgf/number format/fixed, /pgf/number format/precision=2},
			yticklabel style={/pgf/number format/fixed, /pgf/number format/precision=2},grid=both];			
			\coordinate (spypoint) at (axis cs:0.00875,0.00000505);
			\coordinate (magnifyglass) at (axis cs:0.008,0.0000015);
			\addplot[black,only marks, mark size=1pt,mark=o] table [col sep=comma] {data/dataStraightToContactXDispXi0E2041GP160C0045Kprig00001Mprig0maxDTunrestr.csv};
			\addplot[red,thick] table [col sep=comma] {data/dataStraightToContactXDispXi0E2041GP160C0045Kprig00001Mprig0maxDT00005.csv};
			\addplot[blue,dashed,thick] table [col sep=comma] {data/dataStraightToContactXDispXi0E2041GP160C0045Kprig00001Mprig0maxDT00001.csv};
			\addplot[green,dotted,thick] table [col sep=comma] {data/dataStraightToContactXDispXi0E2041GP160C0045Kprig00001Mprig0maxDT000005.csv};\addplot[brown,only marks, mark size=1pt,mark=o] table [col sep=comma] {data/dataStraightToContactXDispXi05E2041GP160C0045Kprig00001Mprig0maxDTunrestr.csv};
			\addplot[green] table [col sep=comma] {data/dataStraightToContactXDispXi05E2041GP160C0045Kprig00001Mprig0maxDT00005.csv};
			\addplot[red,dashed,thick] table [col sep=comma] {data/dataStraightToContactXDispXi05E2041GP160C0045Kprig00001Mprig0maxDT00001.csv};
				\addplot[blue,dotted,thick] table [col sep=comma] {data/dataStraightToContactXDispXi05E2041GP160C0045Kprig00001Mprig0maxDT000005.csv};
			\legend{$\xi=0$ $\operatorname{\Delta t_\text{max}}=\infty$,
				$\xi=0$  $\operatorname{\Delta t_\text{max}}=5\times10^{-4}$,
				$\xi=0$  $\operatorname{\Delta t_\text{max}}=1\times10^{-4}$,
				$\xi=0$  $\operatorname{\Delta t_\text{max}}=5\times10^{-5}$,
				$\xi=0.5$ $\operatorname{\Delta t_\text{max}}=\infty$,
				$\xi=0.5$  $\operatorname{\Delta t_\text{max}}=5\times10^{-4}$,
			$\xi=0.5$  $\operatorname{\Delta t_\text{max}}=1\times10^{-4}$,
			$\xi=0.5$  $\operatorname{\Delta t_\text{max}}=5\times10^{-5}$}		
		\end{axis}
		\spy [lightgray, height=2.5cm,width=4.5cm] on (spypoint) in node[fill=white] at (magnifyglass);
	\end{tikzpicture}
	\caption{Dynamic snap-to-contact of two free fibers: Horizontal displacement component of the left fiber at two characteristic points vs.~$t$ for different $\Delta t_\text{max}$. (ISSIP, $n_\text{el}=20$, $\bar c=2.25$, $n_\text{GP}=160$)  }
	\label{fig:LJpoten4xaaNEW42xaavtt}
\end{figure}

Next, we consider the influence of the maximum time increment, $\Delta t_\text{max}$. The results for four different $\Delta t_\text{max}$ are displayed in Fig.~\ref{fig:LJpoten4xaaNEW42xaavtt}, where $\Delta t_\text{max}=\infty$ refers to the case of an unrestricted time increment.
$\Delta t_\text{max}$ strongly affects the transient part of the response. Large values of $\Delta t_\text{max}$ increase the duration of the transient oscillations. However, the final equilibrium configuration, that can be considered as the steady-state or the static equilibrium, is invariant w.r.t.~$\Delta t_\text{max}$.

Regarding the influence of the fixed cutoff distance $c$, we have run simulations for three values, $\bar c\in[2.25,2.5,2.75]$. Differences between the results exist, but they are not significant, since the complete response belongs to the regime of small separations. The cutoff distance $\bar c=2.5$ is adopted for further calculations and the results are omitted for brevity.

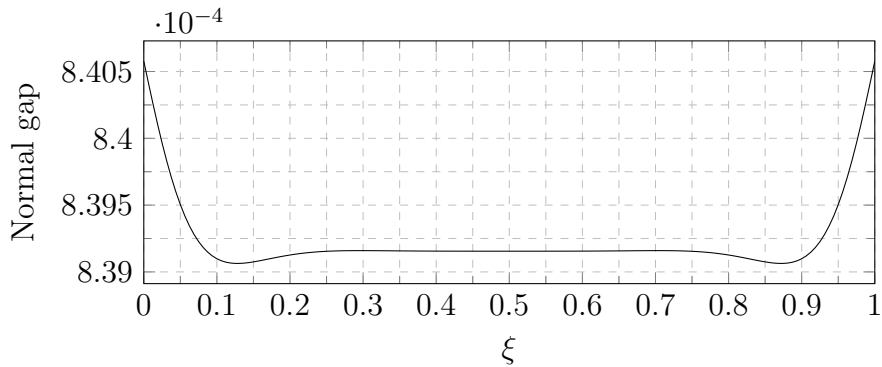
\begin{figure}[h!]
	\centering
	\begin{tikzpicture}
		\begin{axis}[
			xlabel = {$\xi$},
			ylabel = {Normal gap},
			ylabel near ticks,
			legend pos=north east,
			legend cell align=left,
			legend style={font=\tiny},
			width=0.7\textwidth,
			height=0.3\textwidth,
			ytick={0.0008390,0.0008395,0.0008400,0.0008405},	
			xmin = 0.0, xmax =1,
			minor y tick num = 1,
			minor x tick num = 1,	
			xticklabel style={/pgf/number format/fixed, /pgf/number format/precision=2},
			yticklabel style={/pgf/number format/fixed, /pgf/number format/precision=7},grid=both];			
			\addplot[black] table [col sep=comma] {data/dataStraightToContactGAPE4041GP160C005Kprig00001Mprig0maxDTunrestr.csv};

		\end{axis}
	\end{tikzpicture}
	\caption{Dynamic snap-to-contact of two free fibers: Normal gap between the two fibers at static equilibrium vs.~parametric coordinate $\xi$. (ISSIP, $n_\text{el}=40$, $\bar c=2.5$, $n_\text{GP}=160$, $\Delta t_\text{max}$=$\infty$) }
	\label{fig:LJpoten4xaaNEW42xaavttsa}
\end{figure}

After calibrating the numerical model, we focus on the equilibrium configuration between two symmetric deformable fibers that interact via the LJ potential. The normal gap between the fibers at the equilibrium configuration is plotted in Fig.~\ref{fig:LJpoten4xaaNEW42xaavttsa} as a function of the parametric coordinate $\xi$. 
Since the fibers are symmetric w.r.t.~vertical axis at each configuration, the normal gap is simply obtained by subtracting the horizontal displacement components of both fibers at a set of $\xi$ coordinates. This result shows that the final equilibrium configuration of these fibers is straight along the middle parts of fibers. Outside of these areas, fibers bend due to end-effects. At first, it is counter-intuitive that normal gaps at the fibers' ends are greater than those in the middle. This behavior is due to end-effects: end sections interact with a smaller part of the other fiber compared to the middle sections. 

As aforementioned, the analytical solution for the disk-infinite cylinder equilibrium distance is $q_\text{2,eq}=0.00083913$. The results obtained with numerical simulations are summarized in Table~\ref{tab:1}.
	\begin{table}[h!]
		\begin{center}
			\caption{Dynamic snap-to-contact of two free fibers: Comparison of the normal gap at equilibrium for $\xi=0.5$. The analytical solution is $q_\text{2,eq}=0.00083913$. (ISSIP, $n_\text{GP}=160$) }
			\label{tab:1}	
			\begin{tabular}{c c c }
				$\bar c$ & $n_\text{el}=20$	& $n_\text{el}=40$ \\
				\hline
				$2.25$ & $0.00083921$	 & $0.00083920$     \\
				$2.50$ & $0.00083916$	 & $0.00083915$        \\
				$2.75$ & $0.00083915$	 & $0.00083914$        \\
				\hline		
			\end{tabular}
		\end{center}
	\end{table}
The results converge towards the analytical solution, w.r.t.~to the mesh and the cutoff distance. The solution with $n_\text{el}=20$ and $\bar c = 2.5$ is accurate up to the fourth significant digit.

\begin{figure}[h!]
		\begin{center}
			\includegraphics[width=1\textwidth]{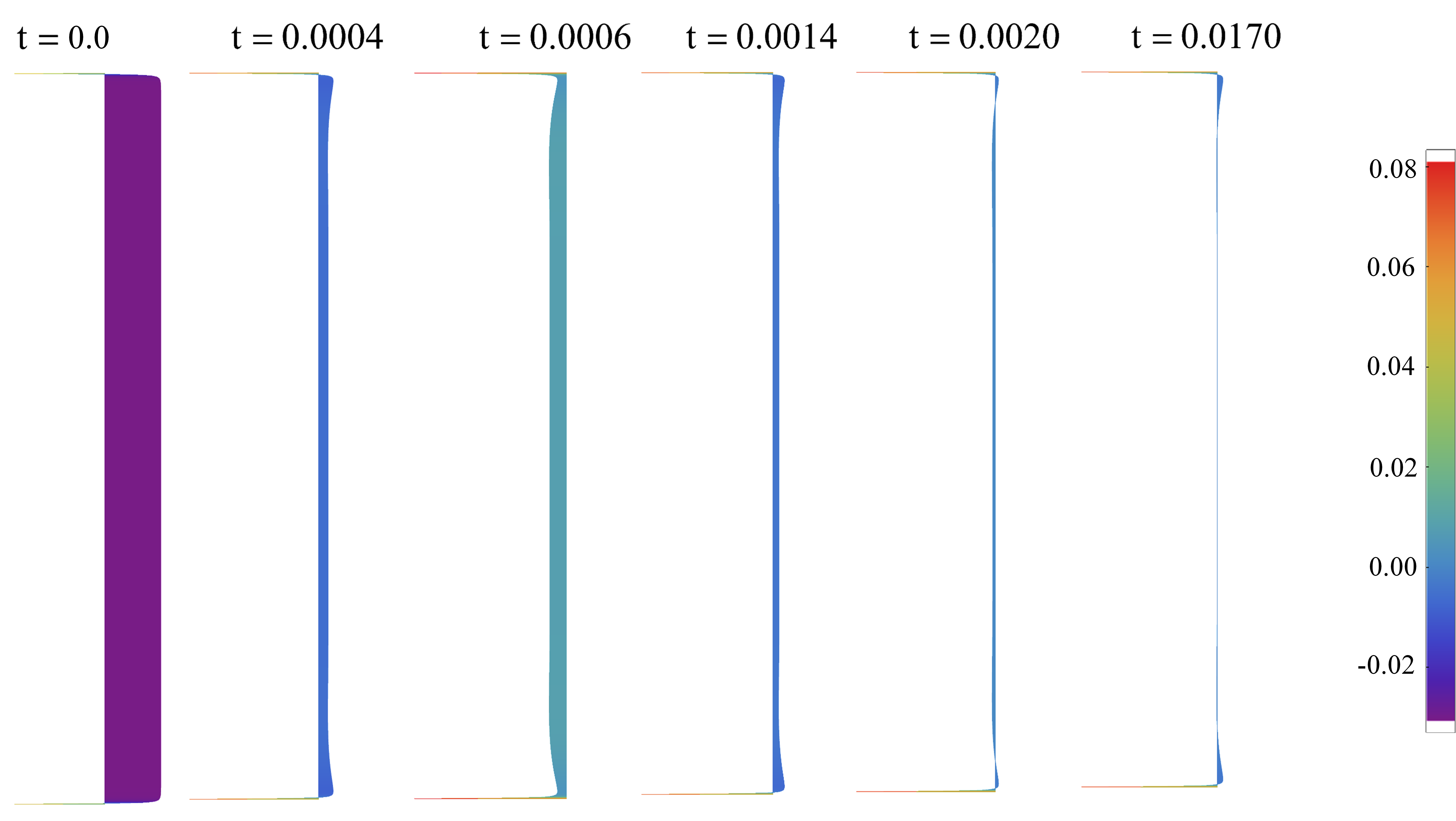}
			\caption{Dynamic snap-to-contact of two free fibers: Normal component of the interaction force is plotted on the left fiber for six different time instances.}
			\label{fig:intro1s}
		\end{center}
	\end{figure}

Finally, the normal component of the interaction force is plotted on the left fiber at six different time instances in Fig.~\ref{fig:intro1s}.
At the beginning, we observe almost constant attraction along the fiber's length, except at the end sections, where a repulsion occurs. Afterward, the interaction force in the middle parts damps to zero, and fibers are held together by end effects where the attraction and the repulsion balance each other. Compared with Fig.~\ref{fig:LJpoten4xaaNEW42xaavttsa}, the gradient of the interaction force is much steeper than that of the normal gap, which is due to the high exponents of the LJ interaction law.
	
Finding the equilibrium configuration between deformable fibers is not a trivial task. For the symmetric case, we showed that the equilibrium distance along the inner parts of fibers corresponds to the disk-infinite cylinder equilibrium distance. However, due to end effects, it is expected that such a configuration is difficult to obtain in physical simulations. Even in our numerical simulations, large impact forces can easily disrupt the symmetry of the problem, as in the following example.
	
\subsection{Strong collision of two interacting fibers}
\label{sec:collision}

\begin{figure}[h!]
	\begin{center}
		\includegraphics[width=0.5\textwidth]{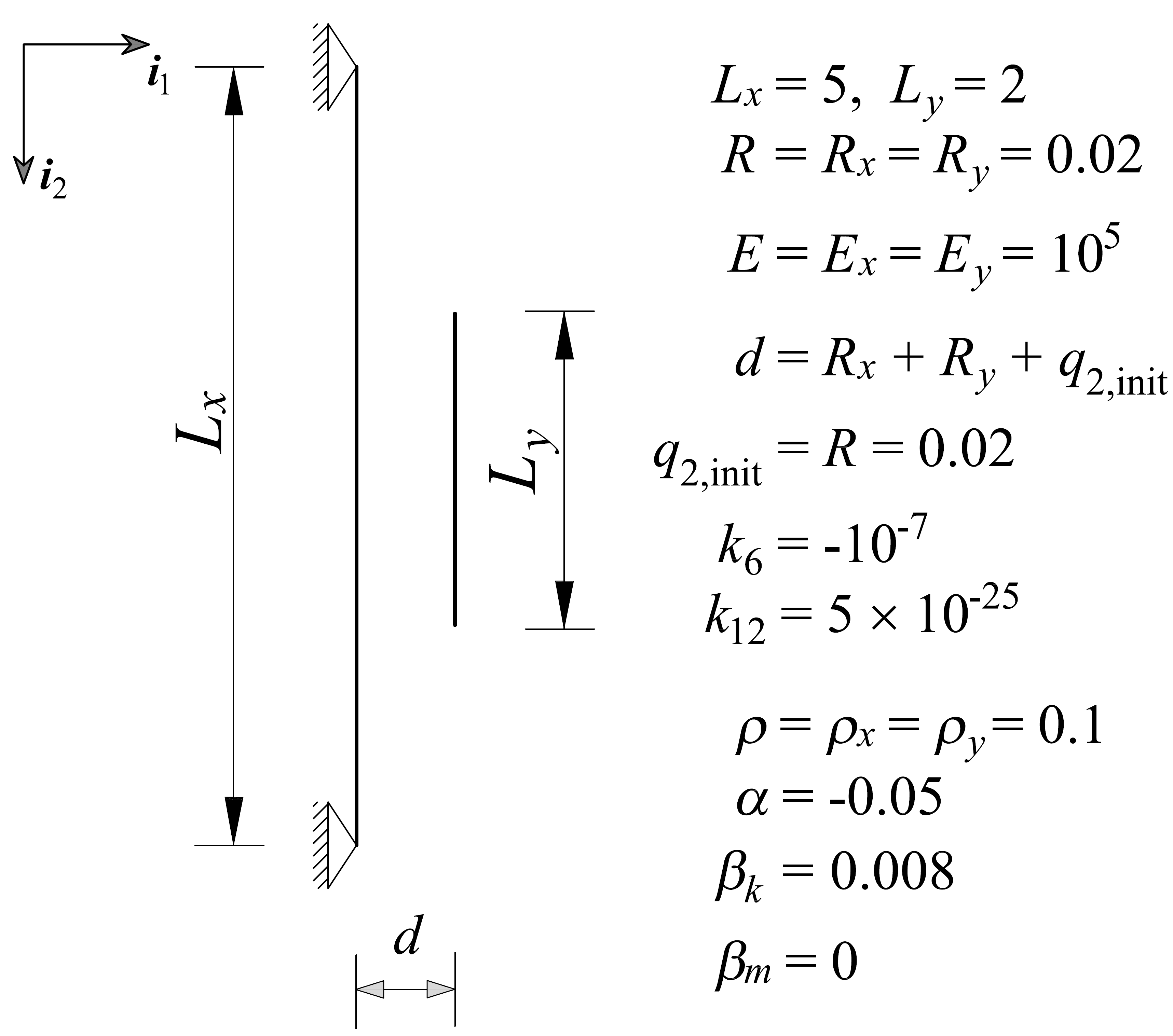}
		\caption{Strong collision of two interacting fibers: Problem setup and parameters.}
		\label{fig:collisionSetup}
	\end{center}
\end{figure}

In this example, we consider two fibers that are relatively far apart, $q_{2,\text{init}}=0.02=R$, and in the regime of moderate separations. The left fiber has restrained end translations, while the right fiber is free, see Fig.~\ref{fig:collisionSetup}. 
They interact via the LJ potential, which leads to a strong collision and an adhesion of the right fiber to the left one. Then, this coupled system continues with the classic transient free damped vibration response. The aim of this example is to compare the ISSIP and $\operatorname{D-D_{app}}$ laws in the case of moderate separations. Our considerations in Section \ref{sec:sec-sec} suggest that the $\operatorname{D-D_{app}}$ is more accurate in this case. 

We adopt meshes of 50 and 20 elements, for the right and the left fiber, respectively, $n_\text{el,x/y}=50/20$. Regarding integration, we employ two distributions of Gauss points, 16 for moderate and 80 for small separations, $n_\text{GP,m/s}=16/80$ with the threshold of $\bar q_\text{2,thr}=0.4$. For moderate separations, we do not employ any cutoff, but integrate over all quadrature points. As a comparison, we will also consider the results obtained with the cutoff function $c_f$ at the end of this subsection. Furthermore, the time increment is left unrestricted.

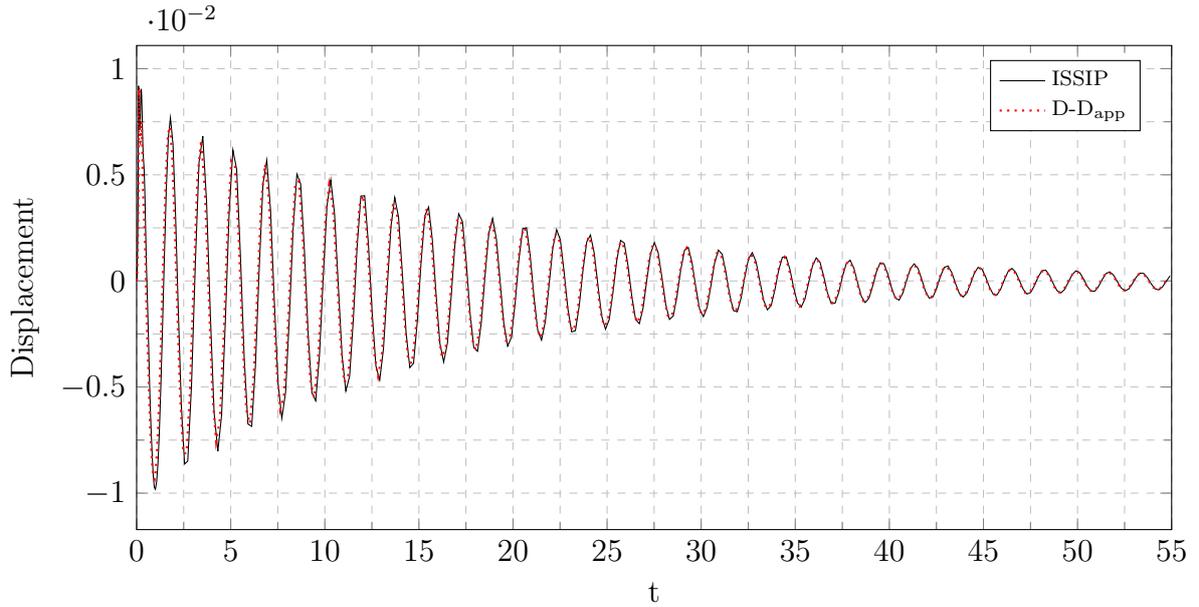
\begin{figure}[h!]
	\centering
	\begin{tikzpicture}
		\begin{axis}[
			xlabel = {t},
			ylabel = {Displacement},
			ylabel near ticks,
			legend pos=north east,
			legend cell align=left,
			legend style={font=\scriptsize},
			width=0.95\textwidth,
			height=0.5\textwidth,
			xmin = 0.0, xmax = 55,
			minor y tick num = 1,
			minor x tick num = 1,		
			xticklabel style={/pgf/number format/fixed, /pgf/number format/precision=2},
			yticklabel style={/pgf/number format/fixed, /pgf/number format/precision=2},grid=both];			
			\addplot[black,
			restrict x to domain=0:55
			] 	table [col sep=comma] {data/dataDispStraightColisionLSipip6601new1680c0055tr008restarted.csv};	
			\addplot[red,dotted,thick,
			restrict x to domain=0:55
			] 	table [col sep=comma] {data/dataDispStraightColisionLSipip66901new1680c0055tr008restarted.csv};	
			\legend{ISSIP,$\operatorname{D-D_{app}}$}	
		\end{axis}
	\end{tikzpicture}
	\caption{Strong collision of two interacting fibers: Horizontal displacement component at the center of left fiber. Comparison of two interaction laws for $t\in[0,55]$.  }
	\label{fig:squeezeReactosn123}
\end{figure}

\begin{figure}[h!]
		\centering
		\begin{tikzpicture}[spy using outlines=	{rectangle, magnification=4.0, connect spies}]
			\begin{axis}[
				xlabel = {t},
				ylabel = {Displacement},
				ylabel near ticks,
				legend pos=north west,
				legend cell align=left,
				legend style={font=\scriptsize},
				width=0.95\textwidth,
				height=0.45\textwidth,
				xtick distance=0.04,
				xmin = 0.0, xmax = 0.3,
				minor y tick num = 1,
				minor x tick num = 2,		
				xticklabel style={/pgf/number format/fixed, /pgf/number format/precision=2},
				yticklabel style={/pgf/number format/fixed, /pgf/number format/precision=2},grid=both];	
				\coordinate (spypoint) at (axis cs:0.115,0.009);
				\coordinate (magnifyglass) at (axis cs:0.2,0.003);		
				\addplot[black,
				restrict x to domain=0:1
				] 	table [col sep=comma] {data/dataDispStraightColisionLSipip6601new1680c0055tr008restarted.csv};	
				\addplot[red,
				restrict x to domain=0:1
				] 	table [col sep=comma] {data/dataDispStraightColisionLSipip66901new1680c0055tr008restarted.csv};	
				\legend{ISSIP, $\operatorname{D-D_{app}}$}	
			\end{axis}
			\spy [lightgray, height=3cm,width=7.5cm] on (spypoint)
			in node[fill=white] at (magnifyglass);
		\end{tikzpicture}
		\caption{Strong collision of two interacting fibers: Horizontal displacement component at the center of left fiber. Comparison of two interaction laws for $t\in[0,0.3]$.  }
		\label{fig:squeezeReactosn12}
	\end{figure}
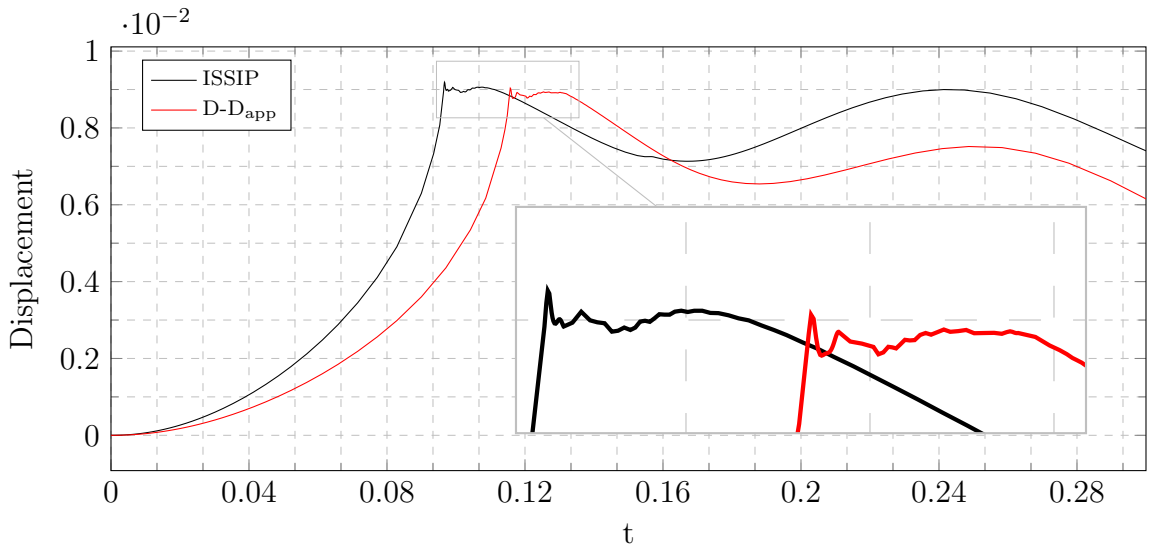

If we observe the horizontal displacement component at the center of the left fiber, apparently both laws return similar results, Fig.~\ref{fig:squeezeReactosn123}.
However, if we take a closer look at the instance of collision, the difference between laws becomes apparent, Fig.~\ref{fig:squeezeReactosn12}.
The ISSIP law does not consider proper scaling of the interaction potential at moderate and large separations, c.f.~Fig.~\ref{fig:LJpoten3}a. This implies that the ISSIP interaction force is stronger and brings fibers together earlier than the $\operatorname{D-D_{app}}$ one. 
	
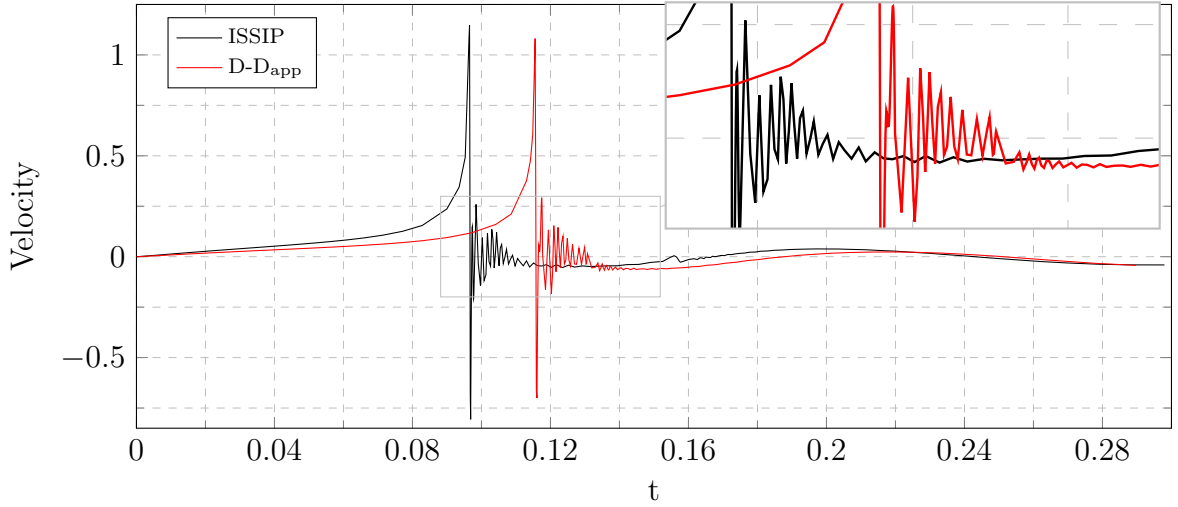
\begin{figure}[h!]
		\centering
		\begin{tikzpicture}[spy using outlines=	{rectangle, magnification=2.25, connect spies}]
			\begin{axis}[
				xlabel = {t},
				ylabel = {Velocity},
				ylabel near ticks,
				legend pos=north west,
				legend cell align=left,
				legend style={font=\scriptsize},
				width=0.95\textwidth,
				height=0.45\textwidth,
				xtick distance = 0.04,
				ymin = -0.85, ymax = 1.25,
				xmin = 0.0, xmax = 0.3,
				minor y tick num = 1,
				minor x tick num = 1,		
				xticklabel style={/pgf/number format/fixed, /pgf/number format/precision=2},
				yticklabel style={/pgf/number format/fixed, /pgf/number format/precision=2},grid=both];			
				\coordinate (spypoint) at (axis cs:0.12,0.05);
				\coordinate (magnifyglass) at (axis cs:0.225,0.7);	

				\addplot[black,
				restrict x to domain=0:0.3
				] 	table [col sep=comma] {data/dataVelStraightColisionLSipip6601new1680c0055tr008restarted.csv};
				
					\addplot[red,
				restrict x to domain=0:0.3
				] 	table [col sep=comma] {data/dataVelStraightColisionLSipip66901new1680c0055tr008restarted.csv};

				\legend{ISSIP,$\operatorname{D-D_{app}}$}	
			\end{axis}
			\spy [lightgray, height=3.0cm,width=6.5cm] on (spypoint)
			in node[fill=white] at (magnifyglass);
		\end{tikzpicture}
		\caption{Strong collision of two interacting fibers: Horizontal velocity component at the center of left fiber. Comparison of two interaction laws for $t\in[0,0.3]$.  }
		\label{fig:squeezeReactosn12a}
	\end{figure}

\begin{figure}[h!]
		\centering
		\begin{tikzpicture}[spy using outlines=	{rectangle, magnification=3.0, connect spies}]
			\begin{axis}[
				xlabel = {t},
				ylabel = {Interaction force},
				ylabel near ticks,
				legend pos=north east,
				legend cell align=left,
				legend style={font=\scriptsize},
				width=0.95\textwidth,
				height=0.45\textwidth,
				xtick distance = 0.02,
				xmin = 0.0, xmax = 0.15,
				minor y tick num = 1,
				minor x tick num = 1,		
				xticklabel style={/pgf/number format/fixed, /pgf/number format/precision=3},
				yticklabel style={/pgf/number format/fixed, /pgf/number format/precision=2},grid=both];		
				\coordinate (spypoint) at (axis cs:0.11,0.0);
				\coordinate (magnifyglass) at (axis cs:0.0475,1.5);				
				\addplot[black,
				restrict x to domain=0:1
				] 	table [col sep=comma] {data/dataIForceKsi05StraightColisionip6601gp1680c0055.csv};
				
				\addplot[red,
				restrict x to domain=0:1
				] 	table [col sep=comma] {data/dataIForceKsi05StraightColisionip66901gp1680c0055.csv};
				\legend{ISSIP,$\operatorname{D-D_{app}}$}	
			\end{axis}
			\spy [lightgray, height=3cm,width=8.5cm] on (spypoint)
			in node[fill=white] at (magnifyglass);
		\end{tikzpicture}
		\caption{Strong collision of two interacting fibers: Normal component of the interaction force at the center of left fiber. Comparison of two laws for $t\in[0,0.15]$. }
		\label{fig:squeezeReactosn12s}
	\end{figure}
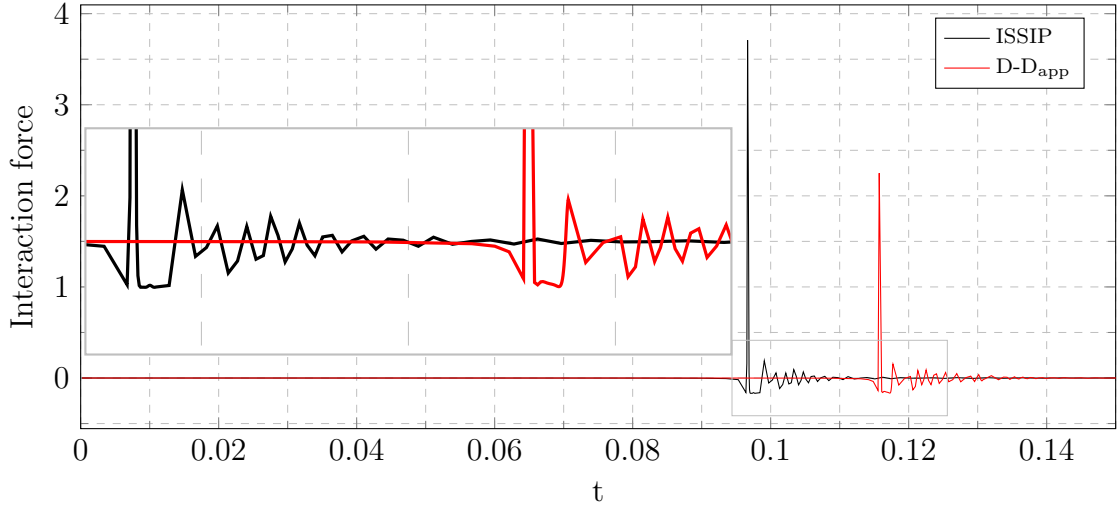

These differences are also apparent by observing velocities and interaction forces shown in Figs.~\ref{fig:squeezeReactosn12a} and \ref{fig:squeezeReactosn12s}.
The velocity before the collision is larger for the ISSIP model, as well as the interaction force at collision, in comparison with the $\operatorname{D-D_{app}}$ model. Therefore, the ISSIP model causes larger accelerations than the $\operatorname{D-D_{app}}$ model, which results in a stronger collision.

\begin{figure}[h!]
		\begin{center}
			\includegraphics[width=\textwidth]{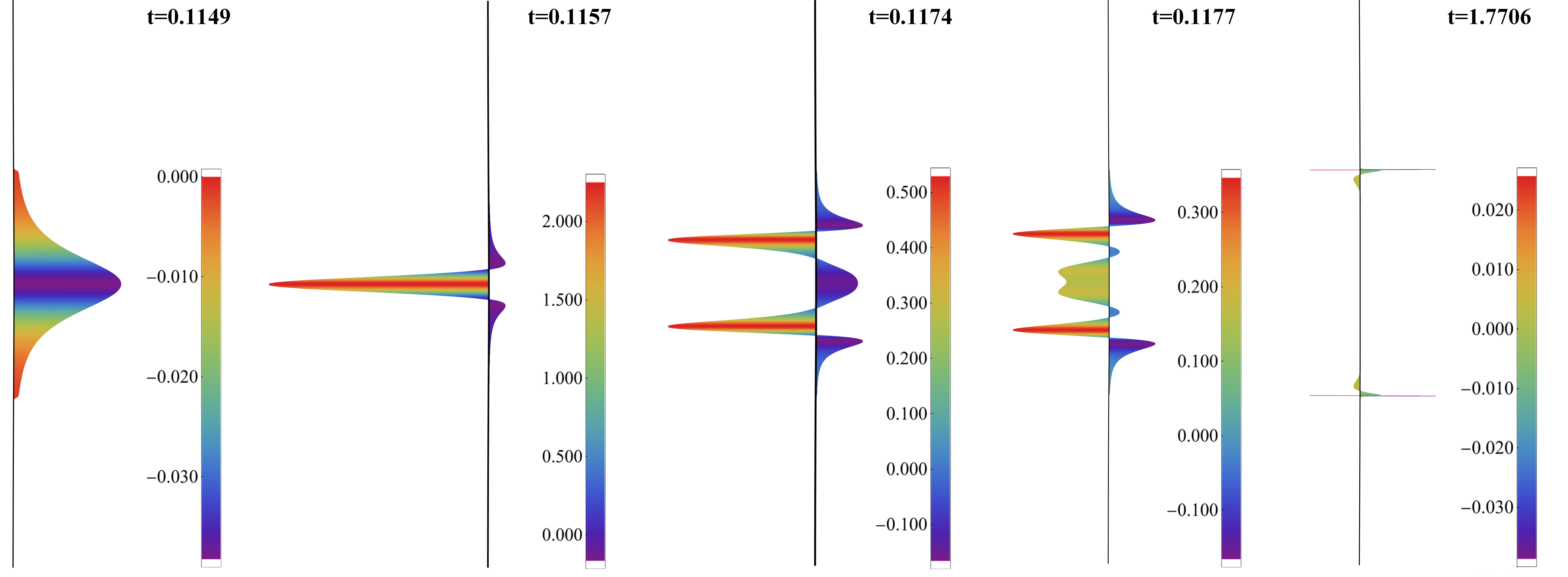}
			\caption{Strong collision of two interacting fibers: Normal component of the interaction force is plotted on the left fiber for five characteristics time instances.}
			\label{fig:intro0xs}
		\end{center}
\end{figure}

Furthermore, the distribution of the normal component of the interaction force on the left fiber is displayed in Fig.~\ref{fig:intro0xs} for five characteristic configurations. Before the collision occurs at $t\approx0.115$, only the smooth distribution of the attractive force is present. The strongest peak of repulsive force occurs at the fiber's center at the instance of collision. During the collision and right after it, the distribution of the interaction force changes abruptly. As the system damps accumulated energy, peaks of the interaction force only remain at the ends of fiber's contact area.

As aforementioned, due to the strong collision, symmetry is easily lost in this example. A small slip of the order of the distance between integration points occurs between fibers after the collision. It can be alleviated by increasing the density of integration points or by excluding the tangential component of the interaction force, $\ve{f}_1$. Alternatively this issue can be addressed by introducing the sliding friction. However, this is not done here. Additionally, a refinement of the maximum time step is necessary for a fully converged response, which is not pursued here either.
	
	

\begin{figure}[h!]
		\centering
		\begin{tikzpicture}[
			zoombubblerectangular/.style = {thin, draw=black, fill=white, rectangle, minimum width = 0.37\textwidth, minimum height = 0.2\textwidth,anchor=south west}
			]
			\def\zoomxmin{0.11}
			\def\zoomxmax{0.13}
			\def\zoomymin{0.007}
			\def\zoomymax{0.0095}
			
			\def\zoomxtargetmin{0.005}
			\def\zoomxtargetmax{0.07}
			\def\zoomytargetmin{0.0025}
			\def\zoomytargetmax{0.009}
			\newif\ifannotatezoomlines
			\annotatezoomlinestrue
			\newif\ifzoomcircularbubble
			\zoomcircularbubblefalse
			\begin{axis}[
				xlabel = {t},
				ylabel = {Displacement at $\xi=0.5$},
				ylabel near ticks,
				legend cell align=left,
				legend style={font=\scriptsize,at={(axis cs:0.14,-0.0007)},anchor=south east},
				width=0.9\textwidth,
				height=0.45\textwidth,
				xtick distance=0.04,
				xmin = 0.00, xmax = 0.14,
				minor y tick num = 1,
				minor x tick num = 2,		
				xticklabel style={/pgf/number format/fixed, /pgf/number format/precision=2},
				yticklabel style={/pgf/number format/fixed, /pgf/number format/precision=2},grid=both];	
				
				
				
				\addplot[black,
				restrict x to domain=0:1
				] 	table [col sep=comma] {data/dataDispStraightColisionLSip66901new8080cInf006tr0008.csv};
				
				\addplot[red,dashed,
				restrict x to domain=0:1
				] 	table [col sep=comma] {data/dataDispStraightColisionLSip66901new1680cInf0055tr0008.csv};

				\addplot[magenta,dotted,thick,
				restrict x to domain=0:1
				] 	table [col sep=comma] {data/dataDispStraightColisionLSipip66901new880c006tr008.csv};
		
				\addplot[green,dashed,thick,
				restrict x to domain=0:1
				] 	table [col sep=comma] {data/dataDispStraightColisionNewAutomC1533ip669.csv};	

				\addplot[blue,dotted,thick,
				restrict x to domain=0:1
				] 	table [col sep=comma] {data/dataDispStraightColisionNewAutomC1566ip669.csv};
				
				\coordinate (zoomtargetbl) at ({axis cs: \zoomxtargetmin, \zoomytargetmin});
				\coordinate (zoomtargettr) at ({axis cs: \zoomxtargetmax, \zoomytargetmax});
				\coordinate (zoomsourcebl) at ({axis cs: \zoomxmin, \zoomymin});
				\coordinate (zoomsourcetr) at ({axis cs: \zoomxmax, \zoomymax});
				\coordinate (zoomtargetlowerleft) at ({axis cs: \zoomxtargetmin, \zoomytargetmin});
					
				\legend{$n_\text{GP,m/s}=80/80$ $\bar c=3.00$,$n_\text{GP,m/s}=16/80$ $\bar c=2.75$,$n_\text{GP,m/s}=8/80$ $\bar c=3.00$, $n_\text{GP}=80$ $w=15$ $s=3$,$n_\text{GP}=80$ $w=15$ $s=6$}
			\end{axis}
				 \node[zoombubblerectangular] (bubbletarget) at (zoomtargetlowerleft) {};
%
				\ifannotatezoomlines
				\draw (zoomsourcebl) rectangle (zoomsourcetr);
				
				\begin{axis}
					[
					no markers,
					at = {(zoomtargetlowerleft)},
					anchor=south west,
					enlarge x limits = false,
					enlarge y limits = false,
					yshift = 4pt,
					scale only axis,
					width = 0.35\textwidth,
					height = 0.175\textwidth,
					axis lines = none,
					xmin = \zoomxmin,
					xmax = \zoomxmax,
					clip mode = individual
					]
				\addplot[black,
				restrict x to domain=0:1
				] 	table [col sep=comma] {data/dataDispStraightColisionLSip66901new8080cInf006tr0008.csv};
				
				\addplot[red,dashed,
				restrict x to domain=0:1
				] 	table [col sep=comma] {data/dataDispStraightColisionLSip66901new1680cInf0055tr0008.csv};

				\addplot[magenta,dotted,thick,
				restrict x to domain=0:1
				] 	table [col sep=comma] {data/dataDispStraightColisionLSipip66901new880c006tr008.csv};
				
				\addplot[green,dashed,thick,
				restrict x to domain=0:1
				] 	table [col sep=comma] {data/dataDispStraightColisionNewAutomC1533ip669.csv};	
				
				\addplot[blue,dotted,thick,
				restrict x to domain=0:1
				] 	table [col sep=comma] {data/dataDispStraightColisionNewAutomC1566ip669.csv};
				\end{axis}
		\end{tikzpicture}
		\caption{Strong collision of two interacting fibers: Horizontal displacement component at the center of left fiber. Comparison of cutoff functions approaches. ($\operatorname{D-D_{app}}$)}
		\label{fig:cutoffvar}	\end{figure}
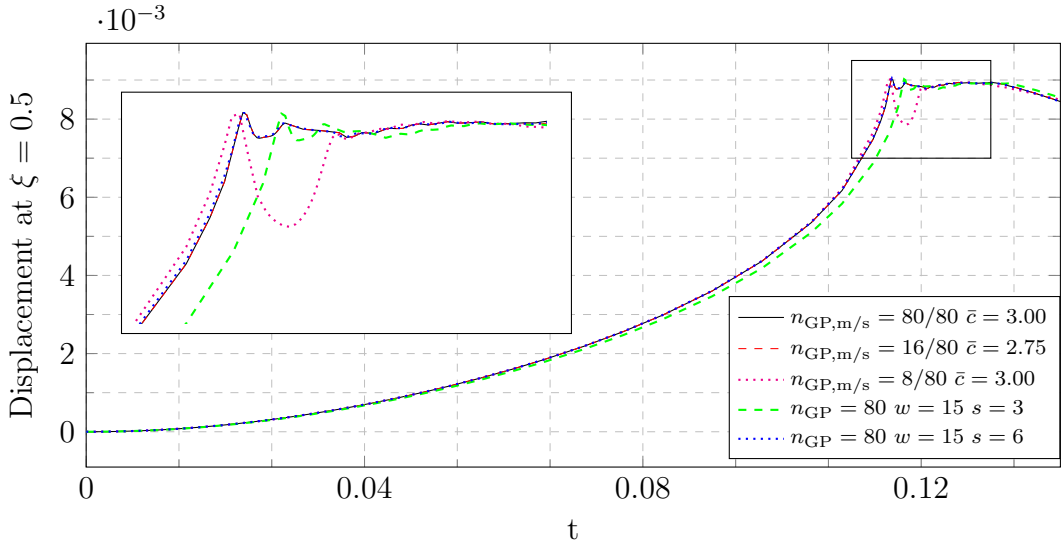

Finally, let us consider the influence of the cutoff function definition. We define a reference solution by using 80 integration points per element for both moderate and short separations, $n_\text{GP,m/s}=80/80$. Then we reduce the number of integration points for the moderate separations regime and vary the cutoff distance for the short-range regime. The results are shown in Fig.~\ref{fig:cutoffvar}.
The model with $n_\text{GP,m/s}=16/80$ returns the same result as the reference solution. The model with $n_\text{GP,m/s}=8/80$ results in a strong repulsion after the collision, which differs from the somewhat smaller repulsion of the reference model. Next, we show results using the cutoff function $c_f$, see Eq.~\eqref{eq:1}, and a constant number of integration points for all separations, $n_\text{GP}=80$. We observe that a relatively large slope parameter $s$ is needed for the converged solution. This suggests that a relatively large cutoff distance is required in the regime of moderate separations, which significantly reduces efficiency due to the dense distribution of integration points. Therefore, the approach that employs the threshold value between small and moderate separations is recommended compared to the cutoff function approach.

	
\subsection{Fiber bending by an adhering fiber}

\begin{figure}[h!]
		\begin{center}
			\includegraphics[width=0.5\textwidth]{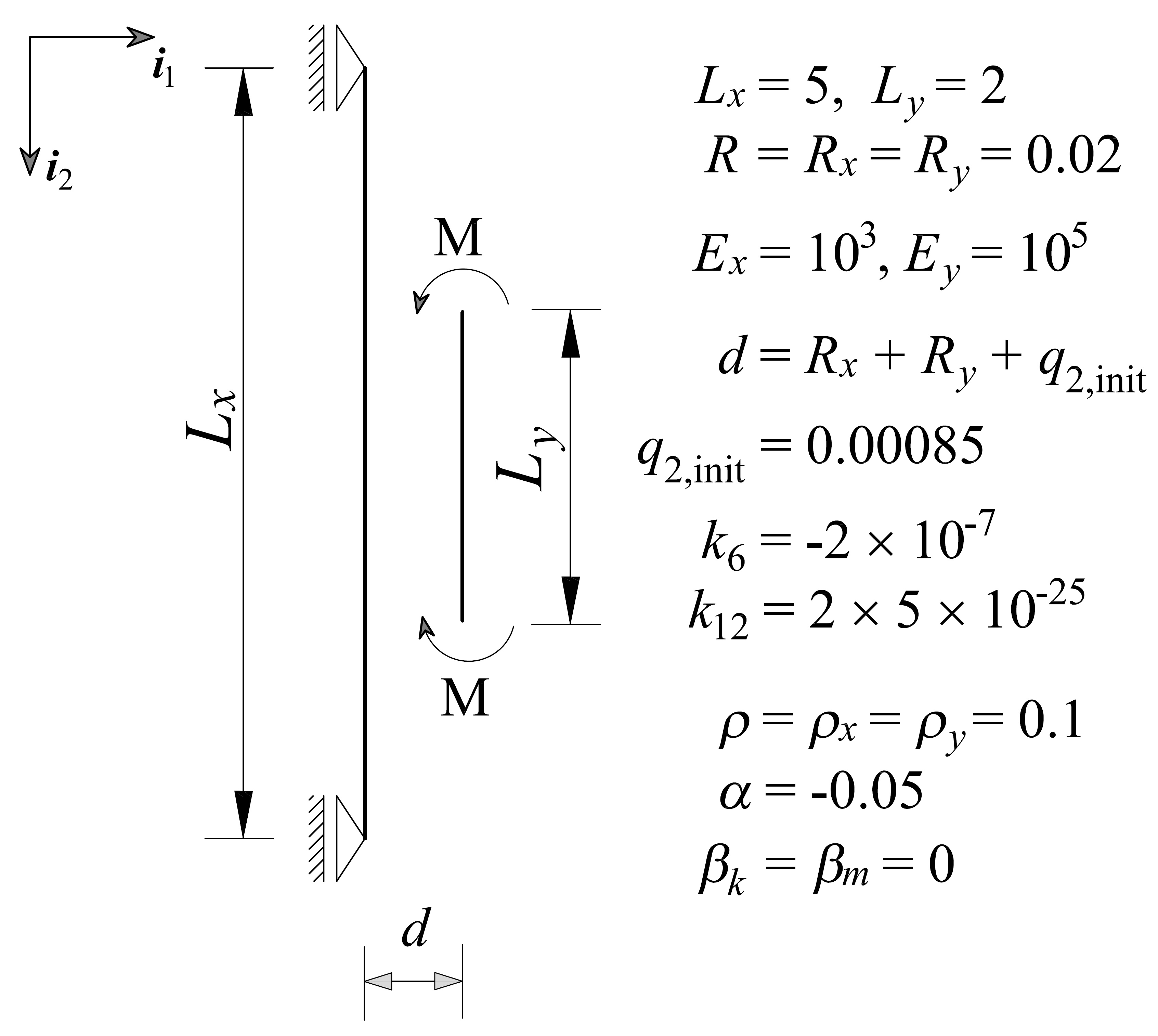}
			\caption{Fiber bending by an adhering fiber: Problem setup and parameters.}
			\label{fig:intro0s}
		\end{center}
	\end{figure}

In this example, we consider a similar setup as in Fig.~\ref{fig:intro0s}.
Two parallel fibers with different lengths and stiffness are interacting via the LJ potential. Initially, the fibers are close to equilibrium. The left fiber has horizontal supports at its ends, while the right fiber is free but loaded with two external end moments $M$. The moments are linear functions of time $M=\pi E_{y} I_{y} t$. Due to this setup, the fibers snap to contact, the right fiber bends due to the external moments and simultaneously imposes a curvature to the left fiber. This can be considered as a 2D approximation of the case when a protein attaches to a cell membrane and imposes a curvature \cite{2011qualmann}. In 4D printing, we can distinguish between active and passive parts/layers \cite{2024manikandan}, and this is one way to model passive fiber bending by an adhering active fiber.
	
The dynamic analysis with $\alpha=-0.05$ without stiffness or mass proportional damping is considered. Since the fibers are in the range of small separations, the ISSIP law is used with a fixed cutoff distance.
	
During the analysis, the problem of negative gaps for Case 4 in Subsection \ref{sec:beam-beam} occurs. As aforementioned, we address it by setting the minimum allowed value for the gap to $q_\text{2,lim}=10^{-8}$. 
	
\begin{figure}[h!]
		\centering
		
		\includegraphics[width=\textwidth]{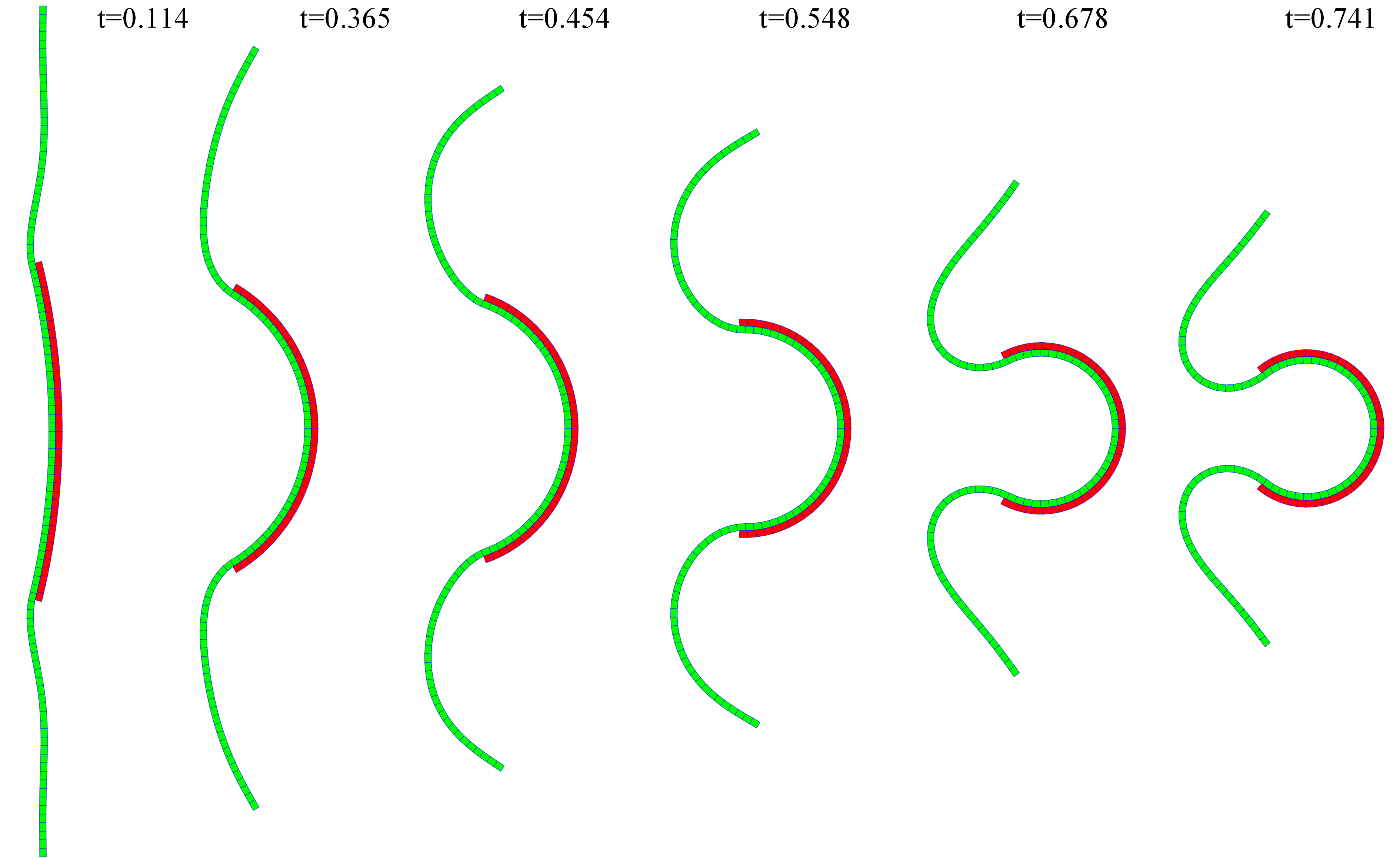}
		\caption{Fiber bending by an adhering fiber: Configurations at six different time instances. (ISSIP, $n_\text{el,x/y}=100/40$, $n_\text{GP}=160$, $\bar c = 2.25$)}
		\label{fig:squeezeIForceDistx2uaaa}
	\end{figure}

The deformed configurations at six instances are shown in Fig.~\ref{fig:squeezeIForceDistx2uaaa}.
As anticipated, the right fiber adheres to the left fiber and deforms it. The fibers have practically constant curvature along the contact length and the response is symmetric w.r.t.~the horizontal axis through $\xi=0.5$.

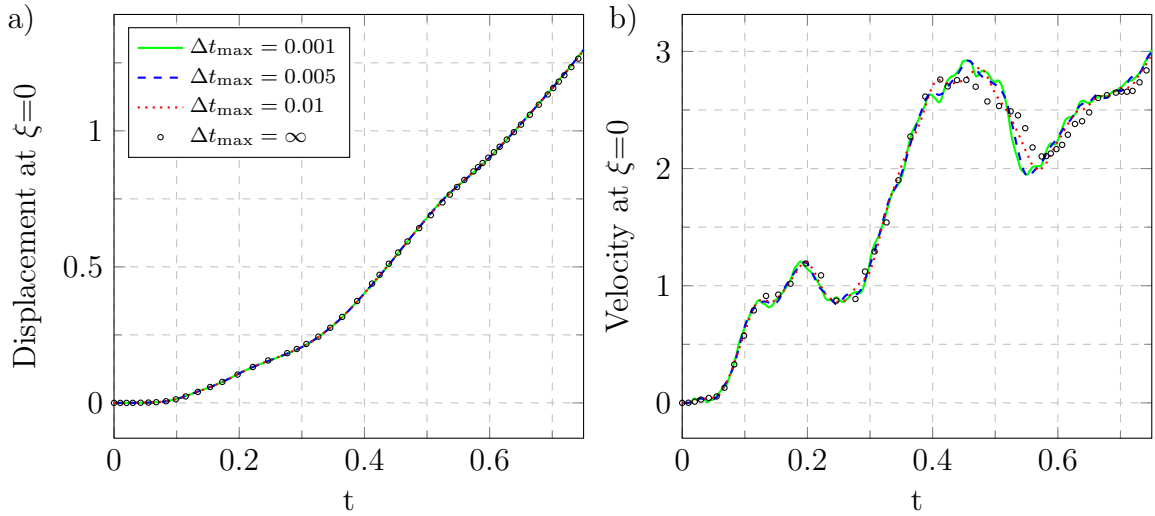
\begin{figure}[h!]
	\centering
		\begin{tikzpicture}
	\begin{axis}[
	xlabel = {t},
	ylabel = {Displacement at $\xi$=0},
	ylabel near ticks,
	legend pos=north west,
	legend cell align=left,
	legend style={font=\scriptsize},
	width=0.485\textwidth,
	height=0.45\textwidth,
	xmin = 0, xmax =0.75,
	minor y tick num = 1,
	minor x tick num = 1,	
	restrict x to domain=0:0.75,	
	xticklabel style={/pgf/number format/fixed, /pgf/number format/precision=2},
	yticklabel style={/pgf/number format/fixed, /pgf/number format/precision=2},grid=both,clip=false];			
	\node [text width=1em,anchor=north west] at (rel axis cs: -0.25,1.05){a)};		
	\addplot[green,thick] table [col sep=comma] {data/dataDispYxi0SqueezeE100D41G160c0045YE105103DoubleAtMaxDT1x10m3.csv};
	\addplot[blue,dashed,thick] 	table [col sep=comma] {data/dataDispYxi0SqueezeE100D41G160c0045YE105103DoubleAtMaxDT5x10m3.csv};
	\addplot[red,dotted,thick] 	table [col sep=comma] {data/dataDispYxi0SqueezeE100D41G160c0045YE105103DoubleAtMaxDT1x10m2.csv};	
	
	\addplot[black, only marks,mark size=1pt,mark=o] 	table [col sep=comma] {data/dataDispYxi0SqueezeE100D41G160c0045YE105103DoubleAtMaxDTinfinity.csv};	
	\legend{$\Delta t_\text{max}=0.001$,$\Delta t_\text{max}=0.005$,$\Delta t_\text{max}=0.01$, 
		$\Delta t_\text{max}=\infty$}		
	\end{axis}
	\end{tikzpicture}
	\begin{tikzpicture}
		\begin{axis}[
			xlabel = {t},
			ylabel = {Velocity at $\xi$=0},
			ylabel near ticks,
			legend columns=2, 
			legend pos=south east,
			legend cell align=left,
			width=0.485\textwidth,
			height=0.45\textwidth,
			xmin = 0, xmax =0.75,
			minor y tick num = 1,
			minor x tick num = 1,
			restrict x to domain=0:0.75,		
			xticklabel style={/pgf/number format/fixed, /pgf/number format/precision=2},
			yticklabel style={/pgf/number format/fixed, /pgf/number format/precision=2},grid=both,clip=false];			
			\node [text width=1em,anchor=north west] at (rel axis cs: -0.18,1.05){b)};	
			\addplot[green,thick] table [col sep=comma] {data/dataVelYxi0SqueezeE100D41G160c0045YE105103DoubleAtMaxDT1x10m3.csv};
			\addplot[blue,dashed,thick] 	table [col sep=comma] {data/dataVelYxi0SqueezeE100D41G160c0045YE105103DoubleAtMaxDT5x10m3.csv};	
			\addplot[red,dotted,thick] 	table [col sep=comma] {data/dataVelYxi0SqueezeE100D41G160c0045YE105103DoubleAtMaxDT1x10m2.csv};			
			\addplot[black, only marks,mark size=1pt,mark=o] 	table [col sep=comma] {data/dataVelYxi0SqueezeE100D41G160c0045YE105103DoubleAtMaxDTinfinity.csv};

		\end{axis}
	\end{tikzpicture}
	\caption{Fiber bending by an adhering fiber: Vertical component of displacement and velocity of left fiber at $\xi=0$: a) Displacement. b) Velocity. (ISSIP, $n_\mathrm{el,x/y}=100/40$, $n_\text{GP}=160$, $\bar c = 2.25$, $\Delta t_\text{max}=\infty$)  }
	\label{fig:squeezeReactonsaa}
\end{figure}

Next, the influence of the maximum time increment on the displacement and the velocity at the boundary of the left fiber is considered. The results of four simulations, one with the unrestricted and three with restricted values of $\Delta t_\text{max}$, are displayed in Fig.~\ref{fig:squeezeReactonsaa}. The size of the maximum time increment does not affect the observed displacement component, but it does affect the velocity. Since configuration-dependent interaction forces are of the main interest in this research, we use the unrestricted time increment in the following.

\begin{figure}[h!]
	\centering
	\begin{tikzpicture}[
		zoombubblerectangular/.style = {thin, draw=black, fill=white, rectangle, minimum width = 0.14\textwidth, minimum height = 0.075\textwidth,anchor=south west}
		]
		\def\zoomxmin{0.302}
		\def\zoomxmax{0.3045}
		\def\zoomymin{-0.2}
		\def\zoomymax{0.15}
		
		\def\zoomxtargetmin{0.3065}
		\def\zoomxtargetmax{0.317}
		\def\zoomytargetmin{0.3}
		\def\zoomytargetmax{1}
		\newif\ifannotatezoomlines
		\annotatezoomlinestrue
		\newif\ifzoomcircularbubble
		\zoomcircularbubblefalse
		\begin{axis}[
			xlabel = {$\xi$},
			ylabel = {Interaction force},
			ylabel near ticks,
			legend pos=north east,
			legend cell align=left,
			legend style={font=\scriptsize},
			width=0.325\textwidth,
			height=0.35\textwidth,
			xmin = 0.3, xmax =0.32,
			xtick distance=0.01,
			minor y tick num = 1,
			minor x tick num = 1,	
			restrict x to domain=0.3:0.32,	
			xticklabel style={/pgf/number format/fixed, /pgf/number format/precision=4},
			yticklabel style={/pgf/number format/fixed, /pgf/number format/precision=4},
			clip=false,grid=both];			
			\node [text width=1em,anchor=north west] at (rel axis cs: -0.28,1.1){a)};
			\addplot[blue
			] 	table [col sep=comma] {data/dataIForceDistribt0114SqueezeE100D41G160c0045YE105103DoubleAt.csv};		
			\addplot[red,densely dashed
			] 	table [col sep=comma] {data/dataIForceDistribt0114SqueezeE100D41G160c005YE105103DoubleAt.csv};	
			\addplot[green,dotted,each nth point=1
			] 	table [col sep=comma] {data/dataIForceDistribt0114SqueezeE100D41G160c0055YE105103DoubleAt.csv};	
			\legend{$\bar c = 2.25$, $\bar c = 2.50$, $\bar c = 2.75$}		
			\coordinate (zoomtargetbl) at ({axis cs: \zoomxtargetmin, \zoomytargetmin});
			\coordinate (zoomtargettr) at ({axis cs: \zoomxtargetmax, \zoomytargetmax});
			\coordinate (zoomsourcebl) at ({axis cs: \zoomxmin, \zoomymin});
			\coordinate (zoomsourcetr) at ({axis cs: \zoomxmax, \zoomymax});
			\coordinate (zoomtargetlowerleft) at ({axis cs: \zoomxtargetmin, \zoomytargetmin});
			
			\node[zoombubblerectangular] (bubbletarget) at (zoomtargetlowerleft) {};
			
			\ifannotatezoomlines
			\draw (zoomsourcebl) rectangle (zoomsourcetr);
			
		\end{axis}
		\begin{axis}
			[
			no markers,
			at = {(zoomtargetlowerleft)},
			anchor=south west,
			enlarge x limits = false,
			enlarge y limits = false,
			yshift = 4pt,
			scale only axis,
			width = 0.15\textwidth,
			height = 0.067\textwidth,
			axis lines = none,
			xmin = \zoomxmin,
			xmax = \zoomxmax,
			ymin = \zoomymin,
			ymax = \zoomymax,
			restrict x to domain=0.3:0.32
			];
			\addplot[blue,thick
			] 	table [col sep=comma] {data/dataIForceDistribt0114SqueezeE100D41G160c0045YE105103DoubleAt.csv};		
			\addplot[red,thick,densely dashed
			] 	table [col sep=comma] {data/dataIForceDistribt0114SqueezeE100D41G160c005YE105103DoubleAt.csv};	
			\addplot[green, thick,dotted,each nth point=1
			] 	table [col sep=comma] {data/dataIForceDistribt0114SqueezeE100D41G160c0055YE105103DoubleAt.csv};	
			
		\end{axis}
	\end{tikzpicture}
	\begin{tikzpicture}[
		zoombubblerectangular/.style = {thin, draw=black, fill=white, rectangle, minimum width = 0.14\textwidth, minimum height = 0.075\textwidth,anchor=south west}
		]
		\def\zoomxmin{0.302}
		\def\zoomxmax{0.3045}
		\def\zoomymin{-0.2}
		\def\zoomymax{0.15}
		
		\def\zoomxtargetmin{0.3065}
		\def\zoomxtargetmax{0.317}
		\def\zoomytargetmin{0.3}
		\def\zoomytargetmax{1}
		\newif\ifannotatezoomlines
		\annotatezoomlinestrue
		\newif\ifzoomcircularbubble
		\zoomcircularbubblefalse
		\begin{axis}[
			xlabel = {$\xi$},
			ylabel = {Interaction force},
			ylabel near ticks,
			legend pos=north east,
			legend cell align=left,
			legend style={font=\scriptsize},
			width=0.325\textwidth,
			height=0.35\textwidth,
			xmin = 0.3, xmax =0.32,
			xtick distance=0.01,
			minor y tick num = 1,
			minor x tick num = 1,	
			restrict x to domain=0.3:0.32,	
			xticklabel style={/pgf/number format/fixed, /pgf/number format/precision=4},
			yticklabel style={/pgf/number format/fixed, /pgf/number format/precision=4},
			clip=false,grid=both];			
			\coordinate (spypoint) at (axis cs:0.303,0);
			\coordinate (magnifyglass) at (axis cs:0.3125,2.5);		
			\node [text width=1em,anchor=north west] at (rel axis cs: -0.28,1.1){b)};		
			\addplot[blue
			] 	table [col sep=comma] {data/dataIForceDistribt0114SqueezeE100D41G160c0045YE105103DoubleAt.csv};	
			
			\addplot[red, densely dashed
			] 	table [col sep=comma] {data/dataIForceDistribt0114SqueezeE100D41G320c0045YE105103DoubleAt.csv};	
			\legend{$n_\text{GP}=160$,$n_\text{GP}=320$}	
			\coordinate (zoomtargetbl) at ({axis cs: \zoomxtargetmin, \zoomytargetmin});
			\coordinate (zoomtargettr) at ({axis cs: \zoomxtargetmax, \zoomytargetmax});
			\coordinate (zoomsourcebl) at ({axis cs: \zoomxmin, \zoomymin});
			\coordinate (zoomsourcetr) at ({axis cs: \zoomxmax, \zoomymax});
			\coordinate (zoomtargetlowerleft) at ({axis cs: \zoomxtargetmin, \zoomytargetmin});
			
			\node[zoombubblerectangular] (bubbletarget) at (zoomtargetlowerleft) {};
			
			\ifannotatezoomlines
			\draw (zoomsourcebl) rectangle (zoomsourcetr);	
		\end{axis}
		\begin{axis}
			[
			no markers,
			at = {(zoomtargetlowerleft)},
			anchor=south west,
			enlarge x limits = false,
			enlarge y limits = false,
			yshift = 4pt,
			scale only axis,
			width = 0.15\textwidth,
			height = 0.067\textwidth,
			axis lines = none,
			xmin = \zoomxmin,
			xmax = \zoomxmax,
			ymin = \zoomymin,
			ymax = \zoomymax,
			restrict x to domain=0.3:0.32
			];
			\addplot[blue,thick
			] 	table [col sep=comma] {data/dataIForceDistribt0114SqueezeE100D41G160c0045YE105103DoubleAt.csv};	
			
			\addplot[red, thick,densely dashed
			] 	table [col sep=comma] {data/dataIForceDistribt0114SqueezeE100D41G320c0045YE105103DoubleAt.csv};	
		\end{axis}
	\end{tikzpicture}
	\begin{tikzpicture}[
		zoombubblerectangular/.style = {thin, draw=black, fill=white, rectangle, minimum width = 0.14\textwidth, minimum height = 0.075\textwidth,anchor=south west}
		]
		\def\zoomxmin{0.302}
		\def\zoomxmax{0.3045}
		\def\zoomymin{-0.2}
		\def\zoomymax{0.15}
		
		\def\zoomxtargetmin{0.3065}
		\def\zoomxtargetmax{0.317}
		\def\zoomytargetmin{0.3}
		\def\zoomytargetmax{1}
		\newif\ifannotatezoomlines
		\annotatezoomlinestrue
		\newif\ifzoomcircularbubble
		\zoomcircularbubblefalse
		\begin{axis}[
			xlabel = {$\xi$},
			ylabel = {Interaction force},
			ylabel near ticks,
			legend pos=north east,
			legend cell align=left,
			legend style={font=\scriptsize,at={(axis cs:0.321,6)},anchor=south east},
			width=0.325\textwidth,
			height=0.35\textwidth,
			xmin = 0.3, xmax =0.32,
			xtick distance=0.01,
			minor y tick num = 1,
			minor x tick num = 1,	
			restrict x to domain=0.3:0.32,	
			xticklabel style={/pgf/number format/fixed, /pgf/number format/precision=4},
			yticklabel style={/pgf/number format/fixed, /pgf/number format/precision=4},
			clip=false,grid=both];			
			\coordinate (spypoint) at (axis cs:0.303,0);
			\coordinate (magnifyglass) at (axis cs:0.3125,2.5);	
			\node [text width=1em,anchor=north west] at (rel axis cs: -0.28,1.1){c)};			
			\addplot[blue]	table [col sep=comma] {data/dataIForceDistribt0114SqueezeE100D41G160c005YE105103DoubleAt.csv};	
			\addplot[red,densely dashed] 	table [col sep=comma] {data/dataIForceDistribt0114SqueezeE200D41G160c005YE105103DoubleAt.csv};	
			\legend{$n_\text{el,x/y}=100/40$,$n_\text{el,x/y}=200/80$}		
			\coordinate (zoomtargetbl) at ({axis cs: \zoomxtargetmin, \zoomytargetmin});
			\coordinate (zoomtargettr) at ({axis cs: \zoomxtargetmax, \zoomytargetmax});
			\coordinate (zoomsourcebl) at ({axis cs: \zoomxmin, \zoomymin});
			\coordinate (zoomsourcetr) at ({axis cs: \zoomxmax, \zoomymax});
			\coordinate (zoomtargetlowerleft) at ({axis cs: \zoomxtargetmin, \zoomytargetmin});
			
			\node[zoombubblerectangular] (bubbletarget) at (zoomtargetlowerleft) {};
			
			\ifannotatezoomlines
			\draw (zoomsourcebl) rectangle (zoomsourcetr);
		\end{axis}
		\begin{axis}
			[
			no markers,
			at = {(zoomtargetlowerleft)},
			anchor=south west,
			enlarge x limits = false,
			enlarge y limits = false,
			yshift = 4pt,
			scale only axis,
			width = 0.15\textwidth,
			height = 0.067\textwidth,
			axis lines = none,
			xmin = \zoomxmin,
			xmax = \zoomxmax,
			ymin = \zoomymin,
			ymax = \zoomymax,
			restrict x to domain=0.3:0.32
			];
			\addplot[blue,thick]	table [col sep=comma] {data/dataIForceDistribt0114SqueezeE100D41G160c005YE105103DoubleAt.csv};	
			\addplot[red,thick,densely dashed] 	table [col sep=comma] {data/dataIForceDistribt0114SqueezeE200D41G160c005YE105103DoubleAt.csv};	
		\end{axis}
	\end{tikzpicture}
	\caption{Fiber bending by an adhering fiber: Distribution of the normal component of the interaction force on the left fiber vs.~$\xi$ at $t=0.114$. a) For three values of the fixed cutoff distance $c$. (ISSIP, $n_\text{el,x/y}=100/40$, $n_\text{GP}=160$)  b) For different $n_\text{GP}$. ( $n_\text{el,x/y}=100/40$, $\bar c = 2.25$) c) For two meshes: $n_\text{el,x/y}=100/40$ with $n_\text{GP}=160$ and $n_\text{el,x/y}=200/80$ with $n_\text{GP}=80$. ($\bar c = 2.50$)  }
	\label{fig:squeezeIForceDistx}
\end{figure}
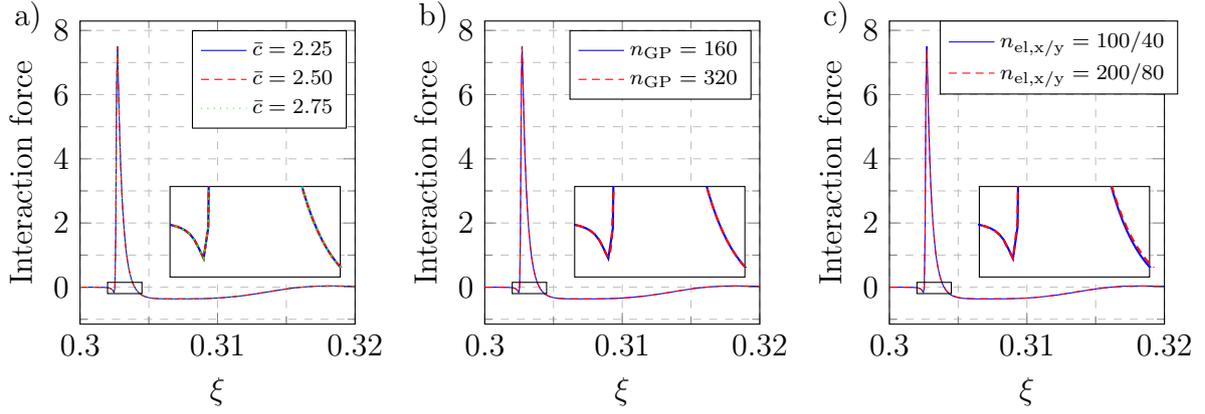

To further calibrate our numerical model, the influence of the cutoff distance, the number of elements, and the number of integration points per unit length are considered using the unrestricted $\Delta t_\text{max}$. The distribution of the normal component of the interaction force acting on the left fiber at instance $t=0.114$ is shown in Fig.~\ref{fig:squeezeIForceDistx}. Since the results are symmetric with respect to the fiber's center, we only consider the distribution near $\xi=0.3$. 
We observe a significant peak in the repulsive interaction force that occurs at the point where the right fiber's end presses the left fiber. From the results in Fig.~\ref{fig:squeezeIForceDistx}, we use the model with $n_\text{el,x/y}=100/40$, $n_\text{GP}=160$, and $\bar c = 2.25$ for obtaining the following results.

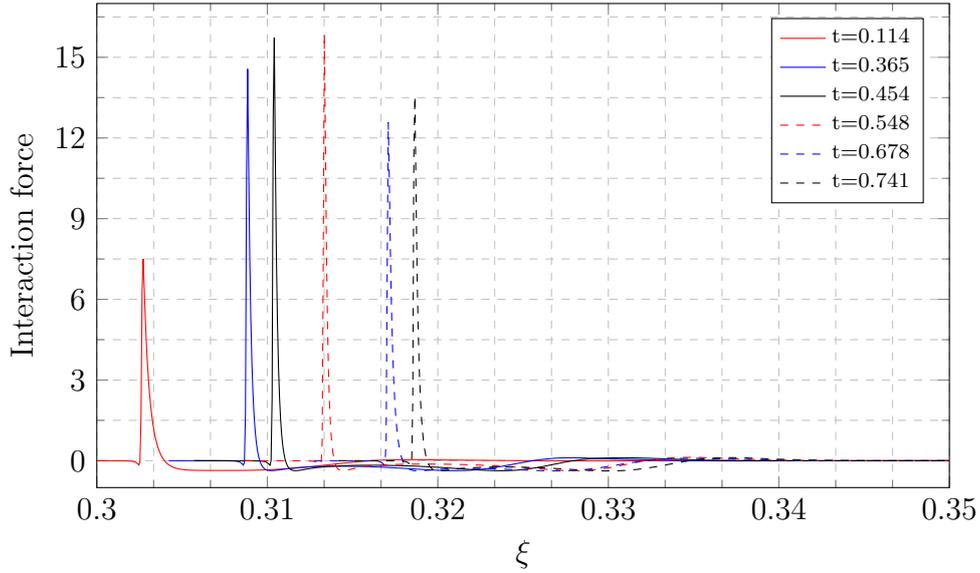
\begin{figure}[h!]
	\centering
	\begin{tikzpicture}
		\begin{axis}[
			xlabel = {$\xi$},
			ylabel = {Interaction force},
			ylabel near ticks,
			legend pos=north east,
			legend cell align=left,
			legend style={font=\scriptsize},
			width=0.8\textwidth,
			height=0.5\textwidth,
			ymin = -1, ymax = 17,
			xmin = 0.3, xmax =0.35,
			minor y tick num = 1,
			minor x tick num = 2,	
			xtick distance=0.01,
			ytick distance=3,
			restrict x to domain=0.3:0.35,	
			xticklabel style={/pgf/number format/fixed, /pgf/number format/precision=4},
			yticklabel style={/pgf/number format/fixed, /pgf/number format/precision=4},
			grid=both];			
			%
		\node [text width=1em,anchor=north west] at (rel axis cs: -0.3,1.1){\subcaption{\label{fig:d}}};
		\addplot[red, each nth point=1	] 	table [col sep=comma] {data/dataIForceDistribt0114SqueezeE100D41G160c0045YE105103DoubleAt.csv};	
		\addplot[blue, each nth point=1	] 	table [col sep=comma] {data/dataIForceDistribt0365SqueezeE100D41G160c0045YE105103DoubleAtTEST.csv};	
		\addplot[black, each nth point=1	] 	table [col sep=comma] {data/dataIForceDistribt0454SqueezeE100D41G160c0045YE105103DoubleAtTEST.csv};	
		\addplot[red,dashed, each nth point=1	] 	table [col sep=comma] {data/dataIForceDistribt0548SqueezeE100D41G160c0045YE105103DoubleAtTEST.csv};	
		\addplot[blue,dashed, each nth point=1	] 	table [col sep=comma] {data/dataIForceDistribt0678SqueezeE100D41G160c0045YE105103DoubleAtTEST.csv};	
		\addplot[black,dashed, each nth point=1	] 	table [col sep=comma] {data/dataIForceDistribt0741SqueezeE100D41G160c0045YE105103DoubleAtTEST.csv};	
		\legend{t=0.114, t=0.365, t=0.454, t=0.548,t=0.678,t=0.741}		
		\end{axis}
	\end{tikzpicture}
	\caption{Fiber bending by an adhering fiber: Distribution of the normal component of the interaction force on the left fiber vs.~$\xi$ for six different time instances. (ISSIP, $n_\text{el,x/y}=100/40$, $n_\text{GP}=160$, and $\bar c=2.25$)  }
	\label{fig:squeezeIForceDistx2uaaaa}
\end{figure}

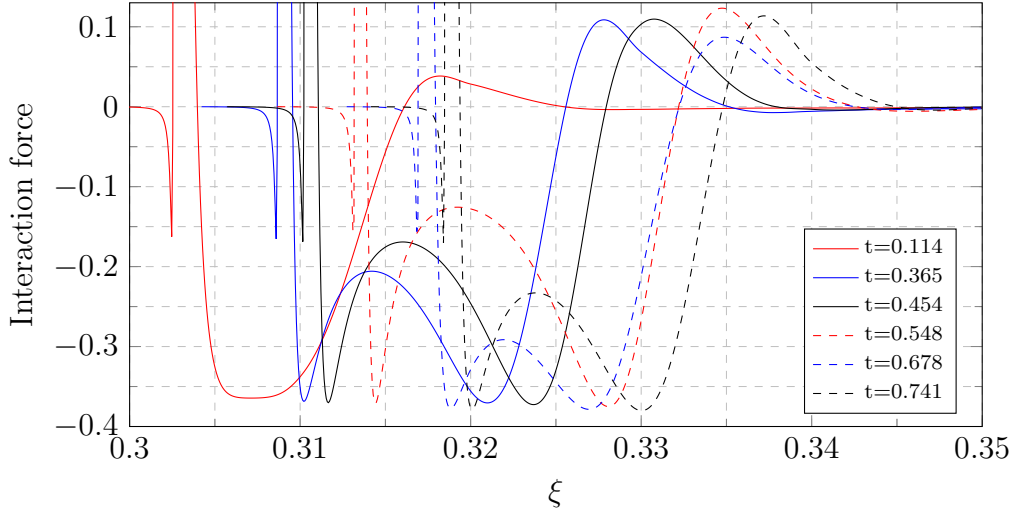
\begin{figure}[h!]
	\centering
	\begin{tikzpicture}
		
		\begin{axis}[
			xlabel = {$\xi$},
			ylabel = {Interaction force},
			ylabel near ticks,
			legend pos=south east,
			legend cell align=left,
			legend style={font=\scriptsize},
			width=0.8\textwidth,
			height=0.45\textwidth,
			ymin = -0.4, ymax = 0.13,
			xmin = 0.3, xmax =0.35,
			xtick={0.3,0.31,0.32,0.33,0.34,0.35},
			minor y tick num = 1,
			minor x tick num = 1,	
			restrict x to domain=0.0.3:0.35,	
			xticklabel style={/pgf/number format/fixed, /pgf/number format/precision=4},
			yticklabel style={/pgf/number format/fixed, /pgf/number format/precision=4},
			grid=both];			
			%
			\node [text width=1em,anchor=north west] at (rel axis cs: -0.3,1.1){\subcaption{\label{fig:d}}};
		
				\addplot[red, each nth point=1	] 	table [col sep=comma] {data/dataIForceDistribt0114SqueezeE100D41G160c0045YE105103DoubleAt.csv};	
				\addplot[blue, each nth point=1	] 	table [col sep=comma] {data/dataIForceDistribt0365SqueezeE100D41G160c0045YE105103DoubleAtTEST.csv};	
				\addplot[black, each nth point=1	] 	table [col sep=comma] {data/dataIForceDistribt0454SqueezeE100D41G160c0045YE105103DoubleAtTEST.csv};	
				\addplot[red,dashed, each nth point=1	] 	table [col sep=comma] {data/dataIForceDistribt0548SqueezeE100D41G160c0045YE105103DoubleAtTEST.csv};	
				\addplot[blue,dashed, each nth point=1	] 	table [col sep=comma] {data/dataIForceDistribt0678SqueezeE100D41G160c0045YE105103DoubleAtTEST.csv};	
				\addplot[black,dashed, each nth point=1	] 	table [col sep=comma] {data/dataIForceDistribt0741SqueezeE100D41G160c0045YE105103DoubleAtTEST.csv};	
			\legend{t=0.114, t=0.365, t=0.454, t=0.548,t=0.678,t=0.741}		
		\end{axis}
	\end{tikzpicture}
	\caption{Fiber bending by an adhering fiber: Distribution of the normal component of the interaction force on the left fiber vs.~$\xi$ for six different time instances. Peak repulsive part of the force is clipped for a better visibility. (ISSIP, $n_\text{el,x/y}=100/40$, $n_\text{GP}=160$, and $\bar c=2.25$)  }
	\label{fig:squeezeIForceDistx2uaa}
\end{figure}

The distribution of the normal component of the interaction force at six different time instances is shown in Fig.~\ref{fig:squeezeIForceDistx2uaaaa}.
The repulsive peak is present during the whole simulation time, but its value varies, as well as its position, since sliding occurs between the fibers. To get a better insight into the distribution of the interaction force, we have clipped these repulsive peaks and plotted the same distributions in Fig.~\ref{fig:squeezeIForceDistx2uaa}.
Besides the usual oscillation between repulsive and attractive forces, here we also observe oscillations of the attractive force. This occurs due to the strong competition of repulsion, attraction, and stiffness of the beams near the right fiber ends.

With this example, we show that the section-section approach can model complex contact problems between deformable fibers. Note that we model only the first part of the dynamic response here, and the system is far from reaching the steady-state. We would need to stop increasing external moments, introduce damping, and add self-contact to find the steady-state configuration. However, this is left for future research.

\section{Conclusions}

We have shown that the section-section interaction potential concept is a powerful tool for the analysis of various cases of interactions between deformable fibers. The problems involving snap-to-contact phenomena are highly nonlinear, dynamic, and require careful computational modeling.

We consider and compare two section-section interaction laws, designated with ISSIP and D-D$_\text{app}$. By using the Lennard-Jones interaction potential and implicit dynamics, we have obtained the following insights:
\begin{itemize}
	\item Snap-to-contact phenomena can be modeled by quasi-statics if the rigid-body motions are restrained. An arc-length solver is generally required to calculate the equilibrium path due to the anticipated instability of a system.
	\item The equilibrium configuration between two free, deformable, and symmetric interacting fibers is straight along the central part and curved near the ends. The equilibrium distance between the fibers along the straight parts is practically equal to the disk-infinite cylinder equilibrium distance.
	\item The section-section approach allows the modeling of a fiber's complex response due to the deformation of an adhering fiber.
	\item If fibers interact within the range of moderate separations, the $\operatorname{D-D_{app}}$ law is preferable over the ISSIP law. 
	\item Efficient integration of the section-section interaction potential requires a sensible definition of the cutoff distance.
\end{itemize}



The application range of these laws is broad, from biology and cell mechanics over the adhesion of wafers to crack propagation modeling.
Adding a tangential contact into the present formulation would be beneficial to model the behavior of sliding fibers. Furthermore, the efficiency of numerical integration can be improved by considering adaptive algorithms.

\section*{Acknowledgments}


This research was funded in part by the Austrian Science Fund (FWF) \href{https://doi.org/10.55776/P36019}{10.55776/P36019}. For the purpose of open access, the authors have applied a CC BY public copyright licence to any Author Accepted Manuscript version arising from this submission.

\printbibliography

@article{1972langbein,
  title = {Van der {{Waals}} attraction between cylinders, rods or fibers},
  author = {Langbein, D.},
  date = {1972-09-01},
  journaltitle = {Phys. Kondens. Mater.},
  volume = {15},
  number = {1},
  pages = {61--86},
  doi = {10.1007/BF02422580},
  langid = {english}
}

@article{1977hilber,
  title = {Improved numerical dissipation for time integration algorithms in structural dynamics},
  author = {Hilber, Hans M. and Hughes, Thomas J. R. and Taylor, Robert L.},
  date = {1977},
  journaltitle = {Earthq. Eng. Struct. Dyn.},
  volume = {5},
  number = {3},
  pages = {283--292},
  doi = {10.1002/eqe.4290050306},
  langid = {english}
}

@report{1979hibbitt,
  title = {Analysis of pipe whip},
  author = {Hibbitt, H.D. and Karlsson, B.I.},
  date = {1979},
  pages = {82},
  location = {United States}
}

@article{1981crisfield,
  title = {A fast incremental/iterative solution procedure that handles “snap-through”},
  author = {Crisfield, M. A.},
  date = {1981-06-01},
  journaltitle = {Computers \& Structures},
  volume = {13},
  number = {1},
  pages = {55--62},
  doi = {10.1016/0045-7949(81)90108-5},
  langid = {english}
}

@article{1997argento,
  title = {Surface formulation for molecular interactions of macroscopic bodies},
  author = {Argento, C. and Jagota, A. and Carter, W. C.},
  date = {1997-07-01},
  journaltitle = {J Mech Phys Solids},
  volume = {45},
  number = {7},
  pages = {1161--1183},
  doi = {10.1016/S0022-5096(96)00121-4},
  langid = {english}
}

@article{2000feng,
  title = {Contact behavior of spherical elastic particles: a computational study of particle adhesion and deformations},
  shorttitle = {Contact behavior of spherical elastic particles},
  author = {Feng, James Q.},
  date = {2000-10-25},
  journaltitle = {Colloids and Surfaces A: Physicochemical and Engineering Aspects},
  volume = {172},
  number = {1},
  pages = {175--198},
  doi = {10.1016/S0927-7757(00)00580-X}
}

@article{2002crisfield,
  title = {Adaptive hierarchical enrichment for delamination fracture using a decohesive zone model},
  author = {Crisfield, M. A. and Alfano, G.},
  date = {2002},
  journaltitle = {Int. J. Numer. Methods Eng.},
  volume = {54},
  number = {9},
  pages = {1369--1390},
  doi = {10.1002/nme.469},
  langid = {english}
}

@book{2005parsegian,
  title = {Van der {{Waals Forces}}: {{A Handbook}} for {{Biologists}}, {{Chemists}}, {{Engineers}}, and {{Physicists}}},
  shorttitle = {Van der {{Waals Forces}}},
  author = {Parsegian, V. Adrian},
  date = {2005},
  publisher = {Cambridge University Press},
  location = {Cambridge},
  doi = {10.1017/CBO9780511614606}
}

@thesis{2006sauer,
  type = {phdthesis},
  title = {An atomic interaction based continuum model for computational multiscale contact mechanics},
  author = {Sauer, R. A.},
  date = {2006},
  institution = {University of California},
  location = {Berkeley}
}

@book{2006wriggers,
  title = {Computational {{Contact Mechanics}}},
  author = {Wriggers, P.},
  date = {2006},
  publisher = {Springer-Verlag},
  location = {Berlin Heidelberg},
  doi = {10.1007/978-3-540-32609-0},
  langid = {english}
}

@article{2006wu,
  title = {Adhesive contact between a nano-scale rigid sphere and an elastic half-space},
  author = {Wu, Jiunn-Jong},
  date = {2006-01},
  journaltitle = {J. Phys. D: Appl. Phys.},
  volume = {39},
  number = {2},
  pages = {351},
  doi = {10.1088/0022-3727/39/2/017},
  langid = {english}
}

@article{2007sauer,
  title = {A contact mechanics model for quasi-continua},
  author = {Sauer, Roger A. and Li, S.},
  date = {2007},
  journaltitle = {Int. J. Numer. Methods Eng.},
  volume = {71},
  number = {8},
  pages = {931--962},
  doi = {10.1002/nme.1970},
  langid = {english}
}

@article{2008alavinasab,
  title = {Computational modeling of nano-structured glass fibers},
  author = {Alavinasab, A. and Jha, R. and Ahmadi, G. and Cetinkaya, C. and Sokolov, I.},
  date = {2008-12-01},
  journaltitle = {Comput Mater Sci},
  volume = {44},
  number = {2},
  pages = {622--627},
  doi = {10.1016/j.commatsci.2008.05.004},
  langid = {english}
}

@article{2008sauer,
  title = {An atomistically enriched continuum model for nanoscale contact mechanics and its application to contact scaling},
  author = {Sauer, Roger A. and Li, S.},
  date = {2008-07},
  journaltitle = {J. Nanosci. Nanotechnol.},
  volume = {8},
  number = {7},
  eprint = {19051933},
  eprinttype = {pubmed},
  pages = {3757--3773},
  doi = {PMID: 19051933},
  langid = {english}
}

@article{2009smith,
  title = {{{ABAQUS}}/{{Standard User}}'s {{Manual}}, {{Version}} 6.9},
  author = {Smith, Michael},
  date = {2009},
  publisher = {Dassault Systèmes Simulia Corp},
  url = {https://www.research.manchester.ac.uk/portal/en/publications/abaqusstandard-users-manual-version-69(0b112d0e-5eba-4b7f-9768-cfe1d818872e)/export.html},
  urldate = {2020-06-22},
  langid = {english}
}

@book{2011israelachvili,
  title = {Intermolecular and surface forces - 3rd {{Edition}}},
  author = {Israelachvili, J. N.},
  date = {2011},
  publisher = {Academic Press},
  url = {https://www.elsevier.com/books/intermolecular-and-surface-forces/israelachvili/978-0-12-391927-4},
  urldate = {2021-08-03}
}

@article{2011qualmann,
  title = {Let's go bananas: revisiting the endocytic {{BAR}} code},
  shorttitle = {Let's go bananas},
  author = {Qualmann, B. and Koch, D. and Kessels, M.M.},
  date = {2011-08-31},
  journaltitle = {EMBO J},
  volume = {30},
  number = {17},
  eprint = {21878992},
  eprinttype = {pubmed},
  pages = {3501--3515},
  doi = {10.1038/emboj.2011.266},
  pmcid = {PMC3181480}
}

@article{2012murrell,
  title = {F-actin buckling coordinates contractility and severing in a biomimetic actomyosin cortex},
  author = {Murrell, M. P. and Gardel, M. L.},
  date = {2012-12-18},
  journaltitle = {PNAS},
  volume = {109},
  number = {51},
  eprint = {23213249},
  eprinttype = {pubmed},
  pages = {20820--20825},
  publisher = {National Academy of Sciences},
  doi = {10.1073/pnas.1214753109},
  langid = {english}
}

@article{2012yoo,
  title = {Scalable fabrication of silicon nanotubes and their application to energy storage},
  author = {Yoo, J.-K. and Kim, J. and Jung, Y.S. and Kang, K.},
  date = {2012},
  journaltitle = {Adv Mater},
  volume = {24},
  number = {40},
  pages = {5452--5456},
  doi = {10.1002/adma.201201601}
}

@article{2014sauerb,
  title = {A geometrically exact finite beam element formulation for thin film adhesion and debonding},
  author = {Sauer, Roger A. and Mergel, Janine C.},
  date = {2014-09-01},
  journaltitle = {Finite Elem. Anal. Des.},
  volume = {86},
  pages = {120--135},
  doi = {10.1016/j.finel.2014.03.009},
  langid = {english}
}

@article{2016fan,
  title = {A three-dimensional surface stress tensor formulation for simulation of adhesive contact in finite deformation},
  author = {Fan, H. and Li, S.},
  date = {2016},
  journaltitle = {Int. J. Numer. Methods Eng.},
  volume = {107},
  number = {3},
  pages = {252--270},
  doi = {10.1002/nme.5169},
  langid = {english}
}

@article{2017ambrosetti,
  title = {Physical adsorption at the nanoscale: {{Towards}} controllable scaling of the substrate-adsorbate van der {{Waals}} interaction},
  shorttitle = {Physical adsorption at the nanoscale},
  author = {Ambrosetti, Alberto and Silvestrelli, Pier Luigi and Tkatchenko, Alexandre},
  date = {2017-06-12},
  journaltitle = {Phys. Rev. B},
  volume = {95},
  number = {23},
  pages = {235417},
  publisher = {American Physical Society},
  doi = {10.1103/PhysRevB.95.235417}
}

@article{2017islama,
  title = {Morphology and mechanics of fungal mycelium},
  author = {Islam, M. R. and Tudryn, G. and Bucinell, R. and Schadler, L. and Picu, R. C.},
  date = {2017-10-12},
  journaltitle = {Sci Rep},
  volume = {7},
  number = {1},
  pages = {13070},
  publisher = {Nature Publishing Group},
  doi = {10.1038/s41598-017-13295-2},
  issue = {1},
  langid = {english}
}

@article{2017venkataram,
  title = {Unifying {{Microscopic}} and {{Continuum Treatments}} of van der {{Waals}} and {{Casimir Interactions}}},
  author = {Venkataram, Prashanth S. and Hermann, Jan and Tkatchenko, Alexandre and Rodriguez, Alejandro W.},
  date = {2017-06-29},
  journaltitle = {Phys. Rev. Lett.},
  volume = {118},
  number = {26},
  pages = {266802},
  publisher = {American Physical Society},
  doi = {10.1103/PhysRevLett.118.266802}
}

@article{2018franquelim,
  title = {Membrane sculpting by curved {{DNA}} origami scaffolds},
  author = {Franquelim, H. G. and Khmelinskaia, A. and Sobczak, J.-P. and Dietz, H. and Schwille, P.},
  date = {2018-02-23},
  journaltitle = {Nat Commun},
  volume = {9},
  number = {1},
  eprint = {29476101},
  eprinttype = {pubmed},
  pages = {811},
  doi = {10.1038/s41467-018-03198-9},
  langid = {english},
  pmcid = {PMC5824810}
}

@article{2018nishiyama,
  title = {Molecular interactions in nanocellulose assembly},
  author = {Nishiyama, Y.},
  date = {2018-02-13},
  journaltitle = {Philos Trans Royal Soc A},
  volume = {376},
  number = {2112},
  pages = {20170047},
  publisher = {Royal Society},
  doi = {10.1098/rsta.2017.0047}
}

@article{2019borkovica,
  title = {Rotation-free isogeometric dynamic analysis of an arbitrarily curved plane {{Bernoulli-Euler}} beam},
  author = {Borković, A. and Kovačević, S. and Radenković, G. and Milovanović, S. and Majstorović, D.},
  date = {2019-02},
  journaltitle = {Eng. Struct.},
  volume = {181},
  pages = {192--215},
  doi = {10.1016/j.engstruct.2018.12.003},
  langid = {english}
}

@article{2020grill,
  title = {A computational model for molecular interactions between curved slender fibers undergoing large {{3D}} deformations with a focus on electrostatic, van der {{Waals}}, and repulsive steric forces},
  author = {Grill, Maximilian J. and Wall, Wolfgang A. and Meier, C.},
  date = {2020},
  journaltitle = {Int. J. Numer. Methods Eng.},
  volume = {121},
  number = {10},
  pages = {2285--2330},
  doi = {10.1002/nme.6309},
  langid = {english}
}

@article{2021čanadija,
  title = {Deep learning framework for carbon nanotubes: mechanical properties and modeling strategies},
  shorttitle = {Deep learning framework for carbon nanotubes},
  author = {Čanadija, M.},
  date = {2021-09-06},
  journaltitle = {Carbon},
  doi = {10.1016/j.carbon.2021.08.091},
  langid = {english}
}

@article{2021grilla,
  title = {Investigation of the peeling and pull-off behavior of adhesive elastic fibers via a novel computational beam interaction model},
  author = {Grill, Maximilian J. and Meier, C. and Wall, Wolfgang A.},
  date = {2021-06-11},
  journaltitle = {J. Adhes.},
  volume = {97},
  number = {8},
  pages = {730--759},
  publisher = {Taylor \& Francis},
  doi = {10.1080/00218464.2019.1699795}
}

@article{2021slepukhin,
  title = {Topological defects produce kinks in biopolymer filament bundles},
  author = {Slepukhin, V. M. and Grill, Maximilian J. and Hu, Q. and Botvinick, E.L. and Wall, W.A. and Levine, A.J.},
  date = {2021-04-13},
  journaltitle = {PNAS},
  volume = {118},
  number = {15},
  eprint = {33876768},
  eprinttype = {pubmed},
  publisher = {National Academy of Sciences},
  doi = {10.1073/pnas.2024362118},
  langid = {english}
}

@thesis{2022roy,
  type = {Thesis},
  title = {Quasi-{{Static}} and {{Implicit-Dynamic Finite Element Solution}} of {{Large Deformation Elastic Adhesive Contacts Using}} a {{Volumetric Interaction Scheme}}},
  author = {Roy, Suprateek},
  date = {2022-04-22T04:56:39Z},
  url = {https://etd.iisc.ac.in/handle/2005/5699},
  urldate = {2025-05-19},
  langid = {american}
}

@article{2023borković,
  title = {Geometrically exact isogeometric {{Bernoulli}}–{{Euler}} beam based on the {{Frenet}}–{{Serret}} frame},
  author = {Borković, A. and Gfrerer, M. H. and Marussig, B.},
  date = {2023-02-15},
  journaltitle = {Comput. Methods Appl. Mech. Eng.},
  volume = {405},
  pages = {115848},
  doi = {10.1016/j.cma.2022.115848},
  langid = {english}
}

@article{2023grill,
  title = {Analytical disk–cylinder interaction potential laws for the computational modeling of adhesive, deformable (nano)fibers},
  author = {Grill, Maximilian J. and Wall, Wolfgang A. and Meier, Christoph},
  date = {2023-05-01},
  journaltitle = {Int. J. Solids Struct.},
  volume = {269},
  pages = {112175},
  doi = {10.1016/j.ijsolstr.2023.112175},
  langid = {english}
}

@article{2023meier,
  title = {Generalized section-section interaction potentials in the geometrically exact beam theory: {{Modeling}} of intermolecular forces, asymptotic limit as strain-energy function, and formulation of rotational constraints},
  shorttitle = {Generalized section-section interaction potentials in the geometrically exact beam theory},
  author = {Meier, Christoph and Grill, Maximilian J. and Wall, Wolfgang A.},
  date = {2023-05-08},
  journaltitle = {Int. J. Solids Struct.},
  pages = {112255},
  doi = {10.1016/j.ijsolstr.2023.112255},
  langid = {english}
}

@article{2023roy,
  title = {A custom arc-length {{Finite Element}} solver for large deformation adhesive contacts using a k-d tree accelerated volumetric interaction scheme},
  author = {Roy, Suprateek and Sundaram, Narayan K.},
  date = {2023},
  journaltitle = {Int. J. Numer. Methods Eng.},
  volume = {124},
  number = {11},
  pages = {2393--2422},
  doi = {10.1002/nme.7215},
  langid = {english}
}

@article{2024borkovićb,
  title = {A novel section–section potential for short-range interactions between plane beams},
  author = {Borković, A. and Gfrerer, M. H. and Sauer, R. A. and Marussig, B. and Bui, T. Q.},
  date = {2024-09-01},
  journaltitle = {Comput. Methods Appl. Mech. Eng.},
  volume = {429},
  pages = {117143},
  doi = {10.1016/j.cma.2024.117143}
}

@inproceedings{2024borkovićg,
  title = {A note on beam-to-beam contact dynamics},
  booktitle = {Int. {{Conf}}. {{Contemp}}. {{Theory Pract}}. {{Constr}}.},
  author = {Borković, Aleksandar and Jočković, Miloš and Tatar, Dijana and Milovanović, Snježana},
  date = {2024-06-12},
  volume = {16},
  number = {1},
  pages = {337--350},
  publisher = {{University of Banja Luka, Faculty of Architecture, Civil Engineering and Geodesy}},
  doi = {10.61892/stp202401080B},
  langid = {english}
}

@article{2024grill,
  title = {Asymptotically consistent and computationally efficient modeling of short-ranged molecular interactions between curved slender fibers undergoing large {{3D}} deformations},
  author = {Grill, Maximilian J. and Wall, Wolfgang A. and Meier, Christoph},
  date = {2024-04-15},
  journaltitle = {Adv. Model. Simul. Eng. Sci.},
  volume = {11},
  number = {1},
  pages = {7},
  doi = {10.1186/s40323-023-00257-9}
}

@article{2024khan,
  title = {{{ANN}} based optimization of nano-beam oscillations with intermolecular forces and geometric nonlinearity},
  author = {Khan, Naveed Ahmad and Sulaiman, Muhammad and Lu, Benzhou},
  date = {2024-11-01},
  journaltitle = {International Journal of Solids and Structures},
  volume = {304},
  pages = {113054},
  doi = {10.1016/j.ijsolstr.2024.113054}
}

@article{2024manikandan,
  title = {A constitutive model for predicting the time-dependent behavior of multi-material {{4D}} printed structures},
  author = {Manikandan, N. and Rajesh, P. K.},
  date = {2024-02-01},
  journaltitle = {Prog Addit Manuf},
  volume = {9},
  number = {1},
  pages = {27--35},
  doi = {10.1007/s40964-023-00408-9},
  langid = {english}
}

@article{2024mokhalingam,
  title = {Continuum contact model for friction between graphene sheets that accounts for surface anisotropy and curvature},
  author = {Mokhalingam, Aningi and Gupta, Shakti S. and Sauer, Roger A.},
  date = {2024},
  journaltitle = {Phys. Rev. B},
  volume = {109},
  pages = {035435},
  doi = {10.1103/PhysRevB.109.035435}
}

@online{2024wang,
  title = {Analytical {{Interaction Potential}} for {{Lennard-Jones Rods}}},
  author = {Wang, Junwen and Seidel, Gary and Cheng, Shengfeng},
  date = {2024-05-06},
  eprint = {2405.03941},
  eprinttype = {arXiv},
  eprintclass = {cond-mat},
  doi = {10.48550/arXiv.2405.03941},
  pubstate = {prepublished}
}

@article{2025borković,
  title = {New analytical laws and applications of interaction potentials with a focus on van der {{Waals}} attraction},
  author = {Borković, A. and Gfrerer, M. H. and Sauer, R. A.},
  date = {2025-03-24},
  journaltitle = {Applied Mathematical Modelling},
  pages = {116100},
  doi = {10.1016/j.apm.2025.116100}
}

@software{2025borkovićb,
  title = {Supplementary notebooks for the manuscript regarding the new analytical laws and applications of interaction potentials with a focus on van der {{Waals}} attraction},
  author = {Borković, Aleksandar and Gfrerer, M. H.},
  date = {2025-03-06},
  doi = {10.3217/81zxn-1np78},
  organization = {Graz University of Technology}
}

\end{document}